\documentclass[traditabstract,longauth]{aa} 

\usepackage[nonamebreak]{natbib}
\usepackage{amsmath}
\usepackage{txfonts}
\usepackage{graphicx}
\usepackage{epstopdf}
\usepackage[breaklinks, colorlinks, citecolor=blue]{hyperref} 
\usepackage{ifthen}
\usepackage[usenames,dvipsnames,table]{xcolor}
\usepackage{fixltx2e}
\usepackage{placeins}

\bibpunct{(}{)}{;}{a}{}{,} 

\providecommand{\sorthelp}[1]{}

\def\setsymbol#1#2{\expandafter\def\csname #1\endcsname{#2}}
\def\getsymbol#1{\csname #1\endcsname}

\def\Planck{\textit{Planck}}

\def\alltwentyfifteenresultspapers{\nocite{planck2014-a01, planck2014-a03, planck2014-a04, planck2014-a05, planck2014-a06, planck2014-a07, planck2014-a08, planck2014-a09, planck2014-a11, planck2014-a12, planck2014-a13, planck2014-a14, planck2014-a15, planck2014-a16, planck2014-a17, planck2014-a18, planck2014-a19, planck2014-a20, planck2014-a22, planck2014-a24, planck2014-a26, planck2014-a28, planck2014-a29, planck2014-a30, planck2014-a31, planck2014-a35, planck2014-a36, planck2014-a37, planck2014-ES}}

\newbox\tablebox    \newdimen\tablewidth
\def\leaderfil{\leaders\hbox to 5pt{\hss.\hss}\hfil}
\def\endPlancktablewide{\tablewidth=\textwidth 
    $$\hss\copy\tablebox\hss$$
    \vskip-\lastskip\vskip -2pt}
\def\tablenote#1 #2\par{\begingroup \parindent=0.8em
    \abovedisplayshortskip=0pt\belowdisplayshortskip=0pt
    \noindent
    $$\hss\vbox{\hsize\tablewidth \hangindent=\parindent \hangafter=1 \noindent
    \hbox to \parindent{$^#1$\hss}\strut#2\strut\par}\hss$$
    \endgroup}
\def\doubleline{\vskip 3pt\hrule \vskip 1.5pt \hrule \vskip 5pt}

\def\L2{\ifmmode L_2\else $L_2$\fi}

\def\DeltaT{\ifmmode \Delta T\else $\Delta T$\fi}
\def\deltat{\ifmmode \Delta t\else $\Delta t$\fi}
\def\fknee{\ifmmode f_{\rm knee}\else $f_{\rm knee}$\fi}
\def\Fmax{\ifmmode F_{\rm max}\else $F_{\rm max}$\fi}
\def\solar{\ifmmode{\rm M}_{\mathord\odot}\else${\rm M}_{\mathord\odot}$\fi}
\def\Msolar{\ifmmode{\rm M}_{\mathord\odot}\else${\rm M}_{\mathord\odot}$\fi}
\def\Lsolar{\ifmmode{\rm L}_{\mathord\odot}\else${\rm L}_{\mathord\odot}$\fi}
\def\inv{\ifmmode^{-1}\else$^{-1}$\fi}
\def\mo{\ifmmode^{-1}\else$^{-1}$\fi}
\def\sup#1{\ifmmode ^{\rm #1}\else $^{\rm #1}$\fi}
\def\expo#1{\ifmmode \times 10^{#1}\else $\times 10^{#1}$\fi}
\def\,{\thinspace}
\def\lsim{\mathrel{\raise .4ex\hbox{\rlap{$<$}\lower 1.2ex\hbox{$\sim$}}}}
\def\gsim{\mathrel{\raise .4ex\hbox{\rlap{$>$}\lower 1.2ex\hbox{$\sim$}}}}

\def\simprop{\mathrel{\raise .4ex\hbox{\rlap{$\propto$}\lower 1.2ex\hbox{$\sim$}}}}
\def\deg{\ifmmode^\circ\else$^\circ$\fi}
\def\pdeg{\ifmmode $\setbox0=\hbox{$^{\circ}$}\rlap{\hskip.11\wd0 .}$^{\circ}
          \else \setbox0=\hbox{$^{\circ}$}\rlap{\hskip.11\wd0 .}$^{\circ}$\fi}
\def\arcs{\ifmmode {^{\scriptstyle\prime\prime}}
          \else $^{\scriptstyle\prime\prime}$\fi}
\def\arcm{\ifmmode {^{\scriptstyle\prime}}
          \else $^{\scriptstyle\prime}$\fi}
\newdimen\sa  \newdimen\sb
\def\parcs{\sa=.07em \sb=.03em
     \ifmmode \hbox{\rlap{.}}^{\scriptstyle\prime\kern -\sb\prime}\hbox{\kern -\sa}
     \else \rlap{.}$^{\scriptstyle\prime\kern -\sb\prime}$\kern -\sa\fi}
\def\parcm{\sa=.08em \sb=.03em
     \ifmmode \hbox{\rlap{.}\kern\sa}^{\scriptstyle\prime}\hbox{\kern-\sb}
     \else \rlap{.}\kern\sa$^{\scriptstyle\prime}$\kern-\sb\fi}
\def\ra[#1 #2 #3.#4]{#1\sup{h}#2\sup{m}#3\sup{s}\llap.#4}
\def\dec[#1 #2 #3.#4]{#1\deg#2\arcm#3\arcs\llap.#4}
\def\deco[#1 #2 #3]{#1\deg#2\arcm#3\arcs}
\def\rra[#1 #2]{#1\sup{h}#2\sup{m}}

\def\dots{\relax\ifmmode \ldots\else $\ldots$\fi}
\def\WHzsr{\ifmmode $W\,Hz\mo\,sr\mo$\else W\,Hz\mo\,sr\mo\fi}
\def\mHz{\ifmmode $\,mHz$\else \,mHz\fi}
\def\GHz{\ifmmode $\,GHz$\else \,GHz\fi}
\def\mKs{\ifmmode $\,mK\,s$^{1/2}\else \,mK\,s$^{1/2}$\fi}
\def\muKs{\ifmmode \,\mu$K\,s$^{1/2}\else \,$\mu$K\,s$^{1/2}$\fi}
\def\muKRJs{\ifmmode \,\mu$K$_{\rm RJ}$\,s$^{1/2}\else \,$\mu$K$_{\rm RJ}$\,s$^{1/2}$\fi}
\def\muKHz{\ifmmode \,\mu$K\,Hz$^{-1/2}\else \,$\mu$K\,Hz$^{-1/2}$\fi}
\def\MJysr{\ifmmode \,$MJy\,sr\mo$\else \,MJy\,sr\mo\fi}
\def\MJysrmK{\ifmmode \,$MJy\,sr\mo$\,mK$_{\rm CMB}\mo\else \,MJy\,sr\mo\,mK$_{\rm CMB}\mo$\fi}
\def\microns{\ifmmode \,\mu$m$\else \,$\mu$m\fi}

\def\muK{\ifmmode \,\mu$K$\else \,$\mu$\hbox{K}\fi}
\def\microK{\ifmmode \,\mu$K$\else \,$\mu$\hbox{K}\fi}
\def\muW{\ifmmode \,\mu$W$\else \,$\mu$\hbox{W}\fi}
\def\kms{\ifmmode $\,km\,s$^{-1}\else \,km\,s$^{-1}$\fi}
\def\kmsMpc{\ifmmode $\,\kms\,Mpc\mo$\else \,\kms\,Mpc\mo\fi}
\providecommand{\sorthelp}[1]{}

\graphicspath{{./Figures/}}
 
\newcommand{\Nside}{\ensuremath{N_{\mathrm{side}}}} 
\newcommand{\VEV}[1]{\langle#1\rangle}

\newcommand{\rms}{rms}

\newcommand{\healpix}{{\tt HEALPix}}
\newcommand{\polkapix}{{\tt polkapix}}
\newcommand{\bogopix}{{\tt bogopix}}

\newcommand{\WMAP}{\textit{WMAP\/}}
\newcommand{\COBE}{\textit{COBE\/}}

\begin{document}


\title{\Planck\ 2015 results. VIII. High Frequency Instrument data processing: Calibration and maps}
\titlerunning{\Planck\ 2015 results. VIII. HFI calibration \& maps}

\author{
\author{\small
Planck Collaboration: R.~Adam\inst{72}
\and
P.~A.~R.~Ade\inst{83}
\and
N.~Aghanim\inst{55}
\and
M.~Arnaud\inst{70}
\and
M.~Ashdown\inst{65, 5}
\and
J.~Aumont\inst{55}
\and
C.~Baccigalupi\inst{82}
\and
A.~J.~Banday\inst{91, 9}
\and
R.~B.~Barreiro\inst{60}
\and
N.~Bartolo\inst{28, 61}
% ...
}
\institute{\small
APC, AstroParticule et Cosmologie, Universit\'{e} Paris Diderot, CNRS/IN2P3, CEA/lrfu, Observatoire de Paris, Sorbonne Paris Cit\'{e}, 10, rue Alice Domon et L\'{e}onie Duquet, 75205 Paris Cedex 13, France\goodbreak
\and
African Institute for Mathematical Sciences, 6-8 Melrose Road, Muizenberg, Cape Town, South Africa\goodbreak
\and
Agenzia Spaziale Italiana Science Data Center, Via del Politecnico snc, 00133, Roma, Italy\goodbreak
\and
Aix Marseille Universit\'{e}, CNRS, LAM (Laboratoire d'Astrophysique de Marseille) UMR 7326, 13388, Marseille, France\goodbreak
\and
Astrophysics Group, Cavendish Laboratory, University of Cambridge, J J Thomson Avenue, Cambridge CB3 0HE, U.K.\goodbreak
\and
Astrophysics \& Cosmology Research Unit, School of Mathematics, Statistics \& Computer Science, University of KwaZulu-Natal, Westville Campus, Private Bag X54001, Durban 4000, South Africa\goodbreak
\and
Atacama Large Millimeter/submillimeter Array, ALMA Santiago Central Offices, Alonso de Cordova 3107, Vitacura, Casilla 763 0355, Santiago, Chile\goodbreak
\and
CITA, University of Toronto, 60 St. George St., Toronto, ON M5S 3H8, Canada\goodbreak
\and
CNRS, IRAP, 9 Av. colonel Roche, BP 44346, F-31028 Toulouse cedex 4, France\goodbreak
\and
CRANN, Trinity College, Dublin, Ireland\goodbreak
\and
California Institute of Technology, Pasadena, California, U.S.A.\goodbreak
\and
Centre for Theoretical Cosmology, DAMTP, University of Cambridge, Wilberforce Road, Cambridge CB3 0WA, U.K.\goodbreak
\and
Computational Cosmology Center, Lawrence Berkeley National Laboratory, Berkeley, California, U.S.A.\goodbreak
\and
DSM/Irfu/SPP, CEA-Saclay, F-91191 Gif-sur-Yvette Cedex, France\goodbreak
\and
DTU Space, National Space Institute, Technical University of Denmark, Elektrovej 327, DK-2800 Kgs. Lyngby, Denmark\goodbreak
\and
D\'{e}partement de Physique Th\'{e}orique, Universit\'{e} de Gen\`{e}ve, 24, Quai E. Ansermet,1211 Gen\`{e}ve 4, Switzerland\goodbreak
\and
Departamento de F\'{\i}sica, Universidad de Oviedo, Avda. Calvo Sotelo s/n, Oviedo, Spain\goodbreak
\and
Department of Astronomy and Astrophysics, University of Toronto, 50 Saint George Street, Toronto, Ontario, Canada\goodbreak
\and
Department of Astrophysics/IMAPP, Radboud University Nijmegen, P.O. Box 9010, 6500 GL Nijmegen, The Netherlands\goodbreak
\and
Department of Physics \& Astronomy, University of British Columbia, 6224 Agricultural Road, Vancouver, British Columbia, Canada\goodbreak
\and
Department of Physics and Astronomy, Dana and David Dornsife College of Letter, Arts and Sciences, University of Southern California, Los Angeles, CA 90089, U.S.A.\goodbreak
\and
Department of Physics and Astronomy, University College London, London WC1E 6BT, U.K.\goodbreak
\and
Department of Physics, Florida State University, Keen Physics Building, 77 Chieftan Way, Tallahassee, Florida, U.S.A.\goodbreak
\and
Department of Physics, Gustaf H\"{a}llstr\"{o}min katu 2a, University of Helsinki, Helsinki, Finland\goodbreak
\and
Department of Physics, Princeton University, Princeton, New Jersey, U.S.A.\goodbreak
\and
Department of Physics, University of California, Santa Barbara, California, U.S.A.\goodbreak
\and
Department of Physics, University of Illinois at Urbana-Champaign, 1110 West Green Street, Urbana, Illinois, U.S.A.\goodbreak
\and
Dipartimento di Fisica e Astronomia G. Galilei, Universit\`{a} degli Studi di Padova, via Marzolo 8, 35131 Padova, Italy\goodbreak
\and
Dipartimento di Fisica e Scienze della Terra, Universit\`{a} di Ferrara, Via Saragat 1, 44122 Ferrara, Italy\goodbreak
\and
Dipartimento di Fisica, Universit\`{a} La Sapienza, P. le A. Moro 2, Roma, Italy\goodbreak
\and
Dipartimento di Fisica, Universit\`{a} degli Studi di Milano, Via Celoria, 16, Milano, Italy\goodbreak
\and
Dipartimento di Matematica, Universit\`{a} di Roma Tor Vergata, Via della Ricerca Scientifica, 1, Roma, Italy\goodbreak
\and
Discovery Center, Niels Bohr Institute, Blegdamsvej 17, Copenhagen, Denmark\goodbreak
\and
European Southern Observatory, ESO Vitacura, Alonso de Cordova 3107, Vitacura, Casilla 19001, Santiago, Chile\goodbreak
\and
European Space Agency, ESAC, Planck Science Office, Camino bajo del Castillo, s/n, Urbanizaci\'{o}n Villafranca del Castillo, Villanueva de la Ca\~{n}ada, Madrid, Spain\goodbreak
\and
European Space Agency, ESTEC, Keplerlaan 1, 2201 AZ Noordwijk, The Netherlands\goodbreak
\and
Facolt\`{a} di Ingegneria, Universit\`{a} degli Studi e-Campus, Via Isimbardi 10, Novedrate (CO), 22060, Italy\goodbreak
\and
Gran Sasso Science Institute, INFN, viale F. Crispi 7, 67100 L'Aquila, Italy\goodbreak
\and
HGSFP and University of Heidelberg, Theoretical Physics Department, Philosophenweg 16, 69120, Heidelberg, Germany\goodbreak
\and
Helsinki Institute of Physics, Gustaf H\"{a}llstr\"{o}min katu 2, University of Helsinki, Helsinki, Finland\goodbreak
\and
INAF - Osservatorio Astronomico di Padova, Vicolo dell'Osservatorio 5, Padova, Italy\goodbreak
\and
INAF - Osservatorio Astronomico di Roma, via di Frascati 33, Monte Porzio Catone, Italy\goodbreak
\and
INAF - Osservatorio Astronomico di Trieste, Via G.B. Tiepolo 11, Trieste, Italy\goodbreak
\and
INAF/IASF Bologna, Via Gobetti 101, Bologna, Italy\goodbreak
\and
INAF/IASF Milano, Via E. Bassini 15, Milano, Italy\goodbreak
\and
INFN, Sezione di Bologna, Via Irnerio 46, I-40126, Bologna, Italy\goodbreak
\and
INFN, Sezione di Roma 1, Universit\`{a} di Roma Sapienza, Piazzale Aldo Moro 2, 00185, Roma, Italy\goodbreak
\and
INFN, Sezione di Roma 2, Universit\`{a} di Roma Tor Vergata, Via della Ricerca Scientifica, 1, Roma, Italy\goodbreak
\and
IPAG: Institut de Plan\'{e}tologie et d'Astrophysique de Grenoble, Universit\'{e} Grenoble Alpes, IPAG, F-38000 Grenoble, France, CNRS, IPAG, F-38000 Grenoble, France\goodbreak
\and
IUCAA, Post Bag 4, Ganeshkhind, Pune University Campus, Pune 411 007, India\goodbreak
\and
Imperial College London, Astrophysics group, Blackett Laboratory, Prince Consort Road, London, SW7 2AZ, U.K.\goodbreak
\and
Infrared Processing and Analysis Center, California Institute of Technology, Pasadena, CA 91125, U.S.A.\goodbreak
\and
Institut N\'{e}el, CNRS, Universit\'{e} Joseph Fourier Grenoble I, 25 rue des Martyrs, Grenoble, France\goodbreak
\and
Institut Universitaire de France, 103, bd Saint-Michel, 75005, Paris, France\goodbreak
\and
Institut d'Astrophysique Spatiale, CNRS (UMR8617) Universit\'{e} Paris-Sud 11, B\^{a}timent 121, Orsay, France\goodbreak
\and
Institut d'Astrophysique de Paris, CNRS (UMR7095), 98 bis Boulevard Arago, F-75014, Paris, France\goodbreak
\and
Institut f\"ur Theoretische Teilchenphysik und Kosmologie, RWTH Aachen University, D-52056 Aachen, Germany\goodbreak
\and
Institute of Astronomy, University of Cambridge, Madingley Road, Cambridge CB3 0HA, U.K.\goodbreak
\and
Institute of Theoretical Astrophysics, University of Oslo, Blindern, Oslo, Norway\goodbreak
\and
Instituto de F\'{\i}sica de Cantabria (CSIC-Universidad de Cantabria), Avda. de los Castros s/n, Santander, Spain\goodbreak
\and
Istituto Nazionale di Fisica Nucleare, Sezione di Padova, via Marzolo 8, I-35131 Padova, Italy\goodbreak
\and
Jet Propulsion Laboratory, California Institute of Technology, 4800 Oak Grove Drive, Pasadena, California, U.S.A.\goodbreak
\and
Jodrell Bank Centre for Astrophysics, Alan Turing Building, School of Physics and Astronomy, The University of Manchester, Oxford Road, Manchester, M13 9PL, U.K.\goodbreak
\and
Kavli Institute for Cosmological Physics, University of Chicago, Chicago, IL 60637, USA\goodbreak
\and
Kavli Institute for Cosmology Cambridge, Madingley Road, Cambridge, CB3 0HA, U.K.\goodbreak
\and
Kazan Federal University, 18 Kremlyovskaya St., Kazan, 420008, Russia\goodbreak
\and
LAL, Universit\'{e} Paris-Sud, CNRS/IN2P3, Orsay, France\goodbreak
\and
LERMA, CNRS, Observatoire de Paris, 61 Avenue de l'Observatoire, Paris, France\goodbreak
\and
LESIA, Observatoire de Paris, CNRS, UPMC, Universit\'{e} Paris-Diderot, 5 Place J. Janssen, 92195 Meudon, France\goodbreak
\and
Laboratoire AIM, IRFU/Service d'Astrophysique - CEA/DSM - CNRS - Universit\'{e} Paris Diderot, B\^{a}t. 709, CEA-Saclay, F-91191 Gif-sur-Yvette Cedex, France\goodbreak
\and
Laboratoire Traitement et Communication de l'Information, CNRS (UMR 5141) and T\'{e}l\'{e}com ParisTech, 46 rue Barrault F-75634 Paris Cedex 13, France\goodbreak
\and
Laboratoire de Physique Subatomique et Cosmologie, Universit\'{e} Grenoble-Alpes, CNRS/IN2P3, 53, rue des Martyrs, 38026 Grenoble Cedex, France\goodbreak
\and
Laboratoire de Physique Th\'{e}orique, Universit\'{e} Paris-Sud 11 \& CNRS, B\^{a}timent 210, 91405 Orsay, France\goodbreak
\and
Lawrence Berkeley National Laboratory, Berkeley, California, U.S.A.\goodbreak
\and
Lebedev Physical Institute of the Russian Academy of Sciences, Astro Space Centre, 84/32 Profsoyuznaya st., Moscow, GSP-7, 117997, Russia\goodbreak
\and
Max-Planck-Institut f\"{u}r Astrophysik, Karl-Schwarzschild-Str. 1, 85741 Garching, Germany\goodbreak
\and
McGill Physics, Ernest Rutherford Physics Building, McGill University, 3600 rue University, Montr\'{e}al, QC, H3A 2T8, Canada\goodbreak
\and
National University of Ireland, Department of Experimental Physics, Maynooth, Co. Kildare, Ireland\goodbreak
\and
Nicolaus Copernicus Astronomical Center, Bartycka 18, 00-716 Warsaw, Poland\goodbreak
\and
Niels Bohr Institute, Blegdamsvej 17, Copenhagen, Denmark\goodbreak
\and
Optical Science Laboratory, University College London, Gower Street, London, U.K.\goodbreak
\and
SISSA, Astrophysics Sector, via Bonomea 265, 34136, Trieste, Italy\goodbreak
\and
School of Physics and Astronomy, Cardiff University, Queens Buildings, The Parade, Cardiff, CF24 3AA, U.K.\goodbreak
\and
School of Physics and Astronomy, University of Nottingham, Nottingham NG7 2RD, U.K.\goodbreak
\and
Sorbonne Universit\'{e}-UPMC, UMR7095, Institut d'Astrophysique de Paris, 98 bis Boulevard Arago, F-75014, Paris, France\goodbreak
\and
Space Sciences Laboratory, University of California, Berkeley, California, U.S.A.\goodbreak
\and
Special Astrophysical Observatory, Russian Academy of Sciences, Nizhnij Arkhyz, Zelenchukskiy region, Karachai-Cherkessian Republic, 369167, Russia\goodbreak
\and
Sub-Department of Astrophysics, University of Oxford, Keble Road, Oxford OX1 3RH, U.K.\goodbreak
\and
Theory Division, PH-TH, CERN, CH-1211, Geneva 23, Switzerland\goodbreak
\and
UPMC Univ Paris 06, UMR7095, 98 bis Boulevard Arago, F-75014, Paris, France\goodbreak
\and
Universit\'{e} de Toulouse, UPS-OMP, IRAP, F-31028 Toulouse cedex 4, France\goodbreak
% ...
}
}

\authorrunning{Planck Collaboration}

\date{Submitted to A\&A February 5, 2015; Revised July 15, 2015}

\abstract {This paper describes the processing applied to the \Planck\ High Frequency Instrument (HFI) cleaned, time-ordered information to produce photometrically calibrated maps in temperature and (for the first time) in polarization. The data from the entire 2.5 year HFI mission include almost  five independent full-sky surveys. HFI observes the sky over a broad range of frequencies, from 100 to 857\GHz. To obtain the best accuracy on the calibration over such a large range, two different photometric calibration schemes have been used. The 545 and 857\GHz\ data are calibrated using models of planetary atmospheric emission. The lower frequencies (from 100 to 353\GHz) are calibrated using the time-variable cosmological microwave background dipole, which we call the "orbital dipole". This source of calibration only depends on the satellite velocity with respect to the solar system. Using a CMB temperature of $T_{CMB}=2.7255\pm0.0006$\,K, it permits an independent measurement of the amplitude of the CMB solar dipole ($3364.3 \pm 1.5 \ \mu\mathrm{K}$) which is approximatively 1$\sigma$ higher than the \WMAP\ measurement with a direction that is consistent between both experiments.

We describe the pipeline used to produce the maps of intensity and linear polarization from the HFI timelines, and the scheme used to set the zero level of the maps a posteriori. We also summarize the noise characteristics of the HFI maps in the 2015 \Planck\ data release and present some null tests to assess their quality. Finally, we discuss the major systematic effects and in particular the leakage induced by flux mismatch between the detectors that leads to spurious polarization signal.}

\maketitle

\alltwentyfifteenresultspapers
\defcitealias{planck2014-a08}{Paper~1}

\clearpage

\section{Introduction}

This paper, one of a set associated with the 2015 \Planck\ data release, is the second of two describing the processing of the data from the High Frequency Instrument (HFI). The HFI is one of the two instruments on board \Planck\footnote{\Planck\ (\url{http://www.esa.int/Planck}) is a project of the European Space Agency (ESA) with instruments provided by two scientific consortia funded by ESA member states and led by Principal Investigators from France and Italy, telescope reflectors provided through a collaboration between ESA and a scientific consortium led and funded by Denmark, and additional contributions from NASA (USA).}, the European Space Agency's satellite dedicated to precision measurements of the cosmic microwave background (CMB). The HFI uses cold optics (at 4\,K, 1.6\,K, and 100\,mK), filters, and 52 bolometers cooled to 100\,mK. Coupled to the \Planck\ telescope, it enables us to map the continuum emission of the sky in intensity and polarization at frequencies of 100, 143, 217, and 353\GHz, and in intensity at 545 and 857\GHz.
Paper~1~\citep{planck2014-a08} describes the processing of the data at the time-ordered level and the measurement of the beam. Paper~2 (this paper) describes the HFI photometric calibration and mapmaking.

\citetalias{planck2014-a08} describes how the TOI (time-ordered information) of each of the 52 bolometers is processed and flagged. Sampled at $5.544\,\mathrm{ms}$, the TOI is first corrected for the ADC nonlinearity, then it is demodulated and converted to the absorbed power with a simple nonlinear bolometric correction. Glitches are flagged and glitch tails are removed from the TOI. Thermal fluctuations are removed on the 1\,min timescale. Sharp lines in the temporal power spectrum of the TOI from the influence of the 4-K cooler are removed. Finally, the bolometer time response is deconvolved and the TOI is low-pass filtered. At this point, the TOIs are cleaned but not yet calibrated. The measurement of the beam is done using a combination of observations of planets for the main beam and GRASP calculations for the sidelobes. The focal plane geometry, or the relative position of bolometers in the sky, is deduced from Mars observations.

This paper describes how the prepared TOIs are used to make the calibrated maps for all \Planck\ HFI bands.
After a  summary of the photometric definitions (Sect.~\ref{sec:photometry}), this paper gives a description of the main steps of the mapmaking processing, focusing on the changes made since the previous \Planck\ releases \citep{planck2013-p03,planck2011-1.7}.
The major difference concerns the calibration (Sect.~\ref{sec:calibration}) which is now based on the orbital CMB dipole for the lower-frequency channels (100, 143, 217, and 353\GHz, also called "CMB channels") while the 545 and 857\GHz\ channels are photometrically calibrated using the signal from Uranus and Neptune.
Section~\ref{sec:mapmaking} explains the mapmaking upgrades, including the polarization treatment. Section~\ref{sec:maps} describes the maps, the solar dipole measurement, and the derivation of far sidelobes and zodiacal maps. 
Section~\ref{sec:noise} presents the noise characteristics and the null tests obtained by splitting the \Planck\ HFI dataset into different groups based on ring period, time period, or detector sets. Consistency checks are performed in order to assess the fidelity of the maps.
Finally, Sect.~\ref{sec:systematics} is dedicated to the description of systematic effects, in particular in polarization. The major systematic residuals in \Planck\ HFI data are due to the leakage from temperature to polarization induced by flux mismatch between associated bolometers. This is the result of either bandpass mismatch, or zero level uncertainty, or calibration uncertainty. We present a first attempt to correct the maps for the intensity-to-polarization leakage. At lower frequencies, even after correction, the residuals are still higher than the noise level and thus the maps cannot yet be used for cosmological analysis and are not released.

 \section{Photometric Equations}
 \label{sec:photometry}

The power absorbed by a given detector at time $t$ can be written as a sum of three terms corresponding to the first three Stokes parameters ($I_{p},\, Q_{p},\, U_{p}$) at the sampled pixel $p$ of the beam-convolved sky:
\begin{equation}
 	P_{t} = G \left[ I_{p} + \rho \left\{Q_{p}\cos{2(\psi_{t}+\alpha)}+U_{p}\sin{2(\psi_{t}+\alpha)}\right\} \right] + n_t \,,
        \label{eq:pabs}
\end{equation}
where $G$ encodes a photometric calibration factor, $\rho$ is the detector polarization efficiency, $\psi_{t}$ is the roll angle of the satellite, $\alpha$ is the detector polarization angle, and $n_t$ represents all the noise contributions to the absorbed power (photon noise, phonon noise, glitch residuals, etc.). The polarization efficiency is derived from the cross-polarization coefficient $\eta$ through $\rho=(1-\eta)/(1+\eta)$. It allows us to describe spider-web bolometers (SWB, $\rho\approx0$) as well as polarization-sensitive bolometers (PSB, $\rho\approx1$). According to the bolometer model and given the stability of the HFI operational conditions during the mission, the gain $G$ is expected to be constant over the whole mission (see section~1 of \citetalias{planck2014-a08}), once the bolometer nonlinearities have been corrected.

For an axisymmetric beam response, the ``smearing'' and ``pointing'' operations commute, and one can solve directly for the pixelized beam-convolved map:
\begin{eqnarray}
	P_{t} &=& G \left( 1, \rho \cos{2(\psi_{t}+\alpha)}, \rho \sin{2(\psi_{t}+\alpha)} \right)^\dag \cdot \left( I_p, Q_p, U_p \right) + n_t 
	\nonumber\\
	&\equiv& G \times A_{tp} T_p + n_t  \, ,
	\label{eq:pabs2}
\end{eqnarray}
where the direction of observation at the time $t$, $(\theta_t,\phi_t)$, falls into pixel $p$ and we define the map-pointing matrix $\tens{A}$ and the sky signal $\vec{T}=(I,Q,U)$.

\section{Calibration}
\label{sec:calibration}

The bolometer signal measured through current-biasing is proportional to the small variation in the incoming power from the sky. To express the measurement in sky temperature units, one has to determine a gain per detector based on a known source in the sky. For the HFI low frequency channels ($100$ to $353$\GHz), we use the CMB orbital dipole as a primary calibrator. This signal fills the entire beam and is almost insensitive to the beam profile and only marginally affected by pointing errors, while its signal-to-noise is high enough thanks to the full-sky coverage. Moreover, it is a stronger signal than CMB anisotropies (by a factor of around $10$), but not bright enough to cause nonlinearities in the detectors, and has the same electromagnetic spectrum as the anisotropies. At higher frequencies (545 and 857\GHz), calibration is performed on planets.

\subsection{CMB dipole conventions}
\label{ssec:cmb_dip_conv}

The CMB dipole is induced by the Doppler effect of the relative motion of the satellite with respect to the CMB frame,
\begin{equation}
	T_{\mathrm{Doppler}}(t,\hat{\vec{u}}) = \frac{T_{\mathrm{CMB}}}{\gamma_t (1-\vec{\beta}_t \cdot \hat{\vec{u}})}  \,, \label{eq:dipole}
\end{equation}
where $\vec{\beta}_t = \vec{v}_t/c$ and $\gamma_t = (1-\beta_t^2)^{-1/2}$. $\vec{v}_t$ is the satellite velocity at time $t$ and $\hat{\vec{u}}$ is the unit vector along the line of sight.

The solar system motion with respect to the CMB frame, giving rise to what is referred to as the ``solar dipole", is the dominant component of the satellite velocity. A residual contribution (called the``orbital dipole") is induced by the yearly motion of the satellite with respect to the solar system barycentre.
The solar dipole can be considered as sky-stationary during the observations and is thus projected onto the sky as an $\ell=1$ component with amplitude previously measured by COBE and \WMAP, $3355 \pm 8 \ \mu\mathrm{K}$ \citep{hinshaw2009}. Relativistic corrections to the solar dipole produce second order anisotropies at multipoles $\ell\ge1$, with amplitudes proportional to $\beta^\ell$, and more importantly couple the two dipole components, as will be discussed below. 
Though the orbital dipole velocity is typically an order of magnitude lower than the solar dipole, it is time dependent, and its time-variability is precisely determined by the satellite velocity which is known at the level of $10^{-4}\,\mathrm{km\ s}^{-1}$.
Finally, to calibrate in temperature, we only rely on an external measurement of the CMB absolute temperature. We use $T_{CMB} = 2.7255$\,K \citep{fixsen2009}.

The expansion of Eq.~\eqref{eq:dipole} in $\beta$ gives 
\begin{equation}
	\Delta(\hat{\vec{u}}) = \frac{T_{\mathrm{Doppler}}}{T_{\mathrm{CMB}}} - 1 \approx \vec{\beta}\cdot\hat{\vec{u}} -\frac{\beta^2}{2} + (\vec{\beta}\cdot\hat{\vec{u}})^2 + {\cal{O}}(\beta^3) \,.
\end{equation}
If we decompose the velocity into a \emph{solar} boost $\vec{\beta}_1$ and an \emph{orbital} boost $\vec{\beta}_2$ then
\begin{equation}
	\Delta = \Delta_1 + \Delta_2 - \vec{\beta}_1\cdot\vec{\beta}_2 + 2 (\vec{\beta}_1\cdot\hat{\vec{u}})(\vec{\beta}_2\cdot\hat{\vec{u}} ) + {\cal{O}}(\beta^3)
\end{equation}
where the first term corresponds to the solar dipole, the second term is the orbital dipole, and the third and fourth terms show the coupling between both due to relativistic corrections.

\subsection{Absolute calibration on orbital dipole}
\label{sec:orbdipcal}

The calibration algorithm takes advantage of the orbital dipole not being fixed on the sky, unlike the solar dipole (during the length of the mission). In practice, relativistic corrections and second order in the development of the conversion from $I_\nu$ to $T_{cmb}$ couple solar and orbital dipoles creating an additional non-stationary signal which also depends on the frequency.
We use as calibration reference signal, the total CMB dipole computed using Eq.~\eqref{eq:dipole}, assuming that it has the same scaling with frequency than the higher multipole CMB anisotropies. This approximation will leave in the HFI frequency maps the frequency dependent fraction of the kinematic quadrupole arising from the 2d order term in the $\beta$ expansion described in \citep{kamionkowski2003} and \citep{quartin2015}. 
The amplitude of the kinematic quadrupole is expected to be lower than 0.5\% on the CMB-calibrated \Planck-HFI channels. However, it is not directly correlated to the orbital dipole on which we calibrate the data. Thus we expect the systematic error induced on the gain estimation to be much smaller.

The orbital dipole signal is modulated on a one-year period. To take into account the time variation of this signal, we need to add a term in Eq.~\eqref{eq:pabs2}:
\begin{equation}
	\label{eq:cal}
	\vec{P} = G \times ( \tens{A}  \vec{T} + \vec{t}_{\mathrm{orb}} ) + \vec{n} \,,
\end{equation}
where $\vec{t}_\mathrm{orb}$ is the time-dependent orbital dipole signal, while in this formula the solar dipole is part of the sky signal $\vec{T}$. Note that we can arbitrarily set all or part of the solar dipole signal either in the calibration template or in the sky without changing the resulting  gain estimation.

Since $\vec{t}_\mathrm{orb}$ is known, we can solve Eq.~\eqref{eq:cal} for each bolometer independently, rewriting the system as
\begin{equation}
	\vec{P} = \tens{A} \tilde{\vec T} + G  \vec{t}_\mathrm{orb} + \vec{n} \label{eq:cal2}\,,
\end{equation}
where the unknowns are the sky-signal in absorbed power units $\tilde{\vec T}$ and the gain $G$. The calibration problem is thus linear and can be solved directly. 
The maximum-likelihood estimate of the gain $G$ is obtained by combining all available samples using the noise covariance matrix $\tens{N}=\VEV{n^\dag n}$ and marginalizing over $\tilde{\vec T}$, solving the equation
\begin{equation}
	\left( \vec{t}_{\mathrm{orb}}^\mathrm{T} \tens{N}^{-1} \tens{Z} \vec{t}_{\mathrm{orb}} \right) \times G = \vec{t}_{\mathrm{orb}}^\mathrm{T} \tens{N}^{-1} \tens{Z} \vec{d},
\label{eq:calibration}
\end{equation}
where $\mathbf{d}$ is the vector of input data $P_t$ and we define
\begin{equation}
	\tens{Z} = \tens{I} - \tens{A} \left(\tens{A}^\mathrm{T} \tens{N}^{-1} \tens{A}\right)^{-1} \tens{A}^\mathrm{T} \tens{N}^{-1}.
\label{eq:Z}
\end{equation}

In practice, the noise is treated by assuming that the destriping reduces the matrix $\tens{N}$ to a diagonal one once the data $\vec{d}$ have been corrected for a constant offset (see Sect.~\ref{sec:mapmethod}).
For the HFI, the mapmaking problem is degenerate for the reconstruction of polarization if we solve for a single detector at a time. Neglecting polarization in Eq.~\eqref{eq:calibration} for polarization-sensitive detectors biases the calibration solution. Moreover, we need a very accurate relative calibration between detectors that are combined to reconstruct polarization in order to minimize leakage from intensity to polarization (see Sect.~\ref{sec:leakage}). For this reason, for the \Planck\ 2015 release we have extended the algorithm described in~\cite{tristram2011} to perform a multi-detector gain estimation for all bolometers at a given frequency together with the offsets (see Sect~\ref{sec:mapmethod}).

\subsection{Long time-constant residuals}
\label{sec:vltc}

The observation strategy of Planck results in the path across a particular part of the sky being almost reversed 6 months later.
As described in~\citetalias{planck2014-a08}, we take advantage of this to derive the time transfer function below one second.
Nevertheless, longer time responses (larger than the second), even with low amplitudes, may bias the calibration estimation by distorting the dipole signal and causing some leakage of the solar into the orbital dipole. To take into account the systematic residuals after time-constant deconvolution, we built a simplified model describing a pure single-mode sinusoidal dipole signal (including solar and orbital dipole) convolved with an exponential decay in the time domain. In the frequency domain, this reads:
\begin{equation}
	{\mathcal T}_{\mathrm{dip}} = {\cal F}\left( t_{\mathrm{dip}} * B e^{-t/\tau} \right) = \frac{B}{1/\tau + 2i\pi \nu} \delta_{\nu,\nu_{\mathrm{spin}}} \,,
	\label{eq:vltc}
\end{equation}
where $B$ and $\tau$ are the amplitude and the time constant, and $\nu_{\mathrm{spin}}$ is the spin frequency. $\mathcal{T}_{\mathrm{dip}}$ is a complex coefficient, the real part of which corresponds to the relative change in the gain $G$, while the imaginary part corresponds to the amplitude of the dipole mode shifted by $90\deg$. 

In practice, prior to the absolute calibration, we solve for the amplitude of a shifted-dipole template using single detector calibration:
\begin{equation}
	\vec{P} = \tens{A} \tilde{\vec T} + G \vec{t}_\mathrm{dip} + C \vec{t}^{90\deg}_\mathrm{dip} + \vec{n}  \,,	
\label{eq:cal3}
\end{equation}
where $C$ gives the amplitude of the dipole mode shifted by $90\deg$. 
With this toy-model, we cannot reconstruct the amplitude and the time constant because we only fit for one coefficient $C$. But we trace the systematic effect on the dipoles due to the very long time  response.
Once the coefficient $C$ has been determined, we correct the data for $C \vec{t}^{90\deg}_\mathrm{dip}$ to account for this additional shifted mode coming from the residual time constant.

\begin{figure}[htbp]
	\centering
	\includegraphics[width=0.5\textwidth]{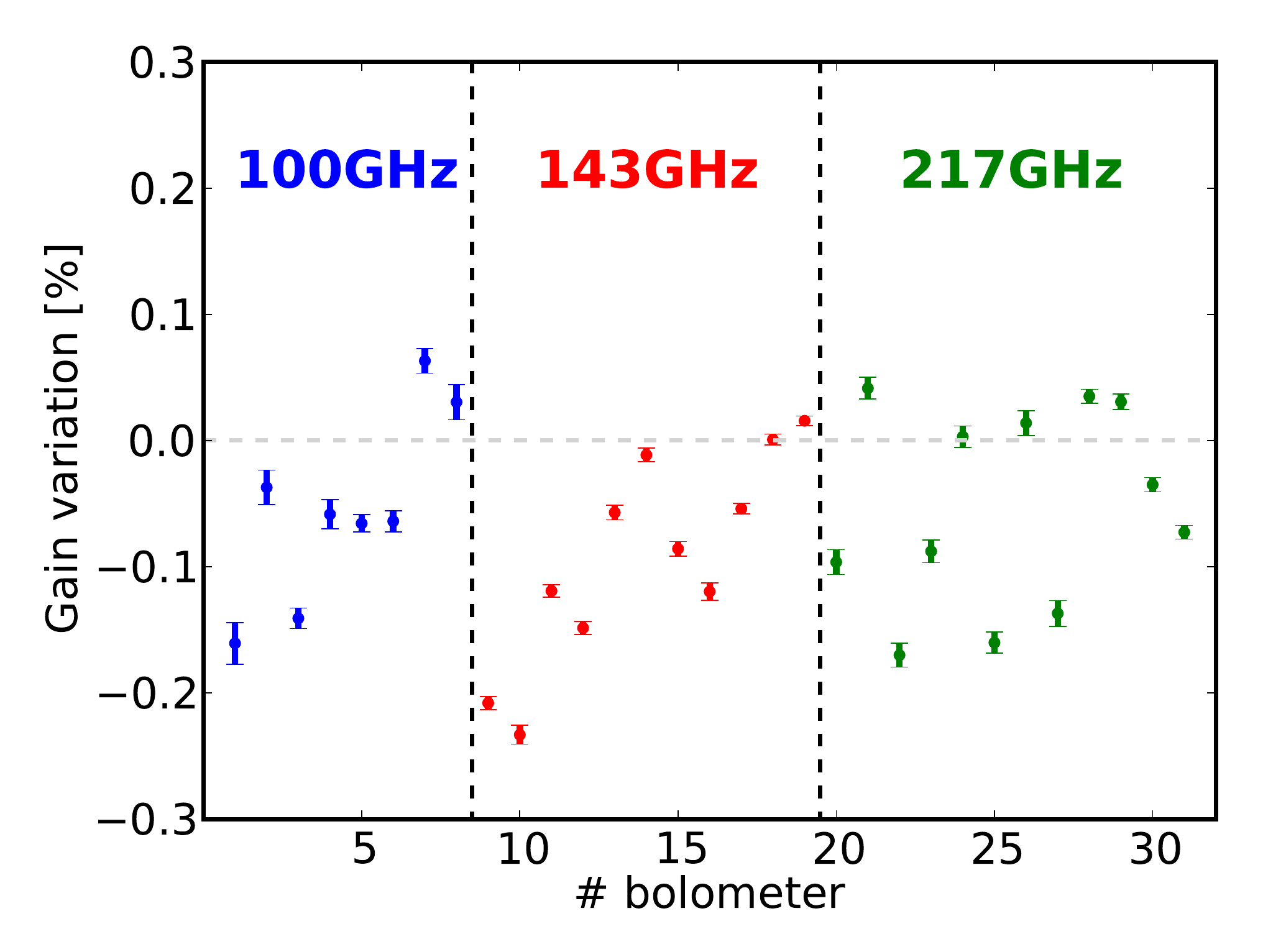}
	\caption{Effect on the gains when including the model for long time constant residuals as described in the text.}
	\label{fig:vltc_gain}
\end{figure}

The effect on the gain $G$ depends on the unknown value of the time constant $\tau$. Figure~\ref{fig:vltc_gain} shows the impact on the gains for each bolometer when including the shifted-dipole correction. At higher frequencies (353\GHz\ and above), the signal is no longer dominated by a dipole and cannot be approximated by the model described above. For the 2015 release, we do not include any correction for time constant residuals in the mapmaking for those channels.

\subsection{Submillimetre calibration}
\label{sec:planet}

Mars, Jupiter, Saturn, Uranus, and Neptune are all observed several times during the full length of the mission. The submillimetre channels of HFI (545 and 857\GHz) are calibrated on models of Uranus and Neptune. We do not use Jupiter and Saturn observations for calibration, since both planets have strong absorption features at those frequencies, which complicate comparison with broadband measurements. The flux from Jupiter also leads to detector saturation at the highest HFI frequencies. Similarly, we choose not to use Mars as a calibrator, because strong seasonal variations complicate the modelling.

Various methods are used to derive planet flux densities, including aperture photometry and point-spread function (PSF) fits. Planet measurements with \Planck\ are studied fully in~\cite{planck2014-a33}. We focus here on calibration using aperture photometry in the time-ordered data from the submillimetre channels. The simulation pipeline used for the main beam reconstruction computes the reconstruction bias and the error budget (the mean and the variance) for each planet observation. The comparison of flux measurements of Neptune and Uranus with up-to-date models provides the calibration factors at 545 and 857\GHz.

\subsubsection{Planet models}

We use the ESA~2 model for Uranus and the ESA~3 model for Neptune \citep{Moreno2010EHSC}. Both models quote absolute uncertainties of 5\,\%. Planet model spectra are produced from their modelled brightness temperatures using the planet solid angles at the time of observation and integrated under the individual 545 and  857\GHz\ bolometer bandpasses. Flux densities are colour-corrected to a reference spectrum defined by a constant $\nu I_\nu$ law, so as to be directly comparable to HFI flux density measurements.

\cite{planck2014-a33} give a detailed account of the ratio between the expected planet fluxes and the measured ones at all HFI frequencies and for the five observed planets (Mars, Jupiter, Saturn, Uranus, and Neptune). They all fall within the 5\,\% model uncertainty range. This is a validation of the models at  100--353\GHz. Hence the models can be used with some confidence to calibrate the 545 and 857\GHz\ channels. 

\subsubsection{Aperture photometry in the timelines}
\label{sec:ap_pho}

We select all samples in a $2^\circ \times 2^\circ$ box around the planet positions and build time-ordered vector objects (hereafter ``timelines'') for each bolometer and each planet scan. We use the first four scans of Neptune and Uranus (season 1 to 4). We build corresponding background timelines using all the samples in a $2^\circ \times 2^\circ$ box around the planet position when the planet is not there. The resulting background has a much higher spatial density than the planet timelines. We use this to build a background mini-map with $2' \times 2'$ pixels that can then be interpolated at each sample position in the planet timelines in order to remove a local background estimate (see appendix B in \citetalias{planck2014-a08} for details).

The aperture photometry measurement procedure applied to our planet timelines is an extension of the usual aperture photometry approach to irregularly gridded data. Flux is integrated in an aperture of radius 3 times the effective FWHM. Typically, aperture photometry is applied to maps of fixed-size pixels, which means integrating the flux in the aperture is equivalent to summing the pixel values. In our case, we have to take into account the inhomogenous spatial distribution of the samples. To do this we  assume that the beam products perfectly describe the spatial light profile of Neptune and Uranus, and compute a spatial sampling correction factor as the ratio of the integrated flux in a highly spatially oversampled beam and the integrated flux in a beam sampled on the planet timelines. This sampling correction has to be computed for each bolometer and planet crossing, because the spatial sampling varies between planet observations. We estimate the statistical uncertainty of the measurements as the standard deviation of the samples in an annulus of radius 3 to 5 times the effective FWHM of the beam. 

\begin{figure}
	\begin{center}
	\includegraphics[width=\columnwidth]{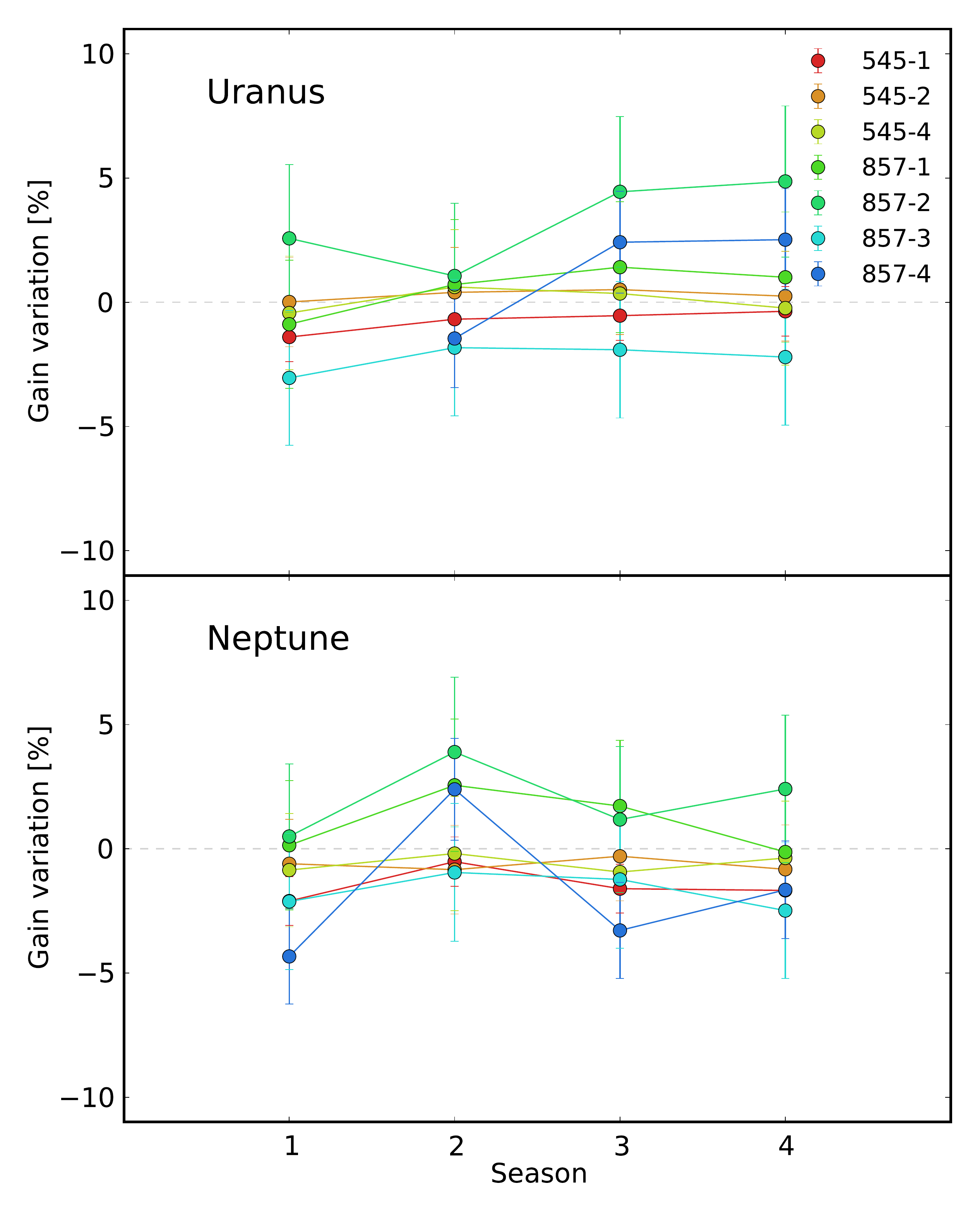}
	\caption{\label{fig:Dx11PlanetCalib} Dispersion of the planet-derived calibration factor per season around the planet calibration estimates for Uranus ({\it top}), and Neptune ({\it bottom}).}
	\end{center}
\end{figure}

We find large variations between the individual planet measurements in each detector and at each season of the full mission survey (Fig.~\ref{fig:Dx11PlanetCalib}) which we attribute to underestimation of the measurement uncertainty. Indeed, the signal from Neptune and Uranus is not expected to vary in time apart from the differences in solid angle which are very small and already taken into account. While accurately corrected by the timeline aperture photometry algorithm presented here, the limited available spatial sampling of the planet signal at these frequencies could explain part of the variations. We therefore decided to include the ``seasonal'' rms of the measurements (we have 4 observations per bolometer per planet) in the measurement uncertainty.

\begin{figure}
	\begin{center}
	\includegraphics[width=\columnwidth]{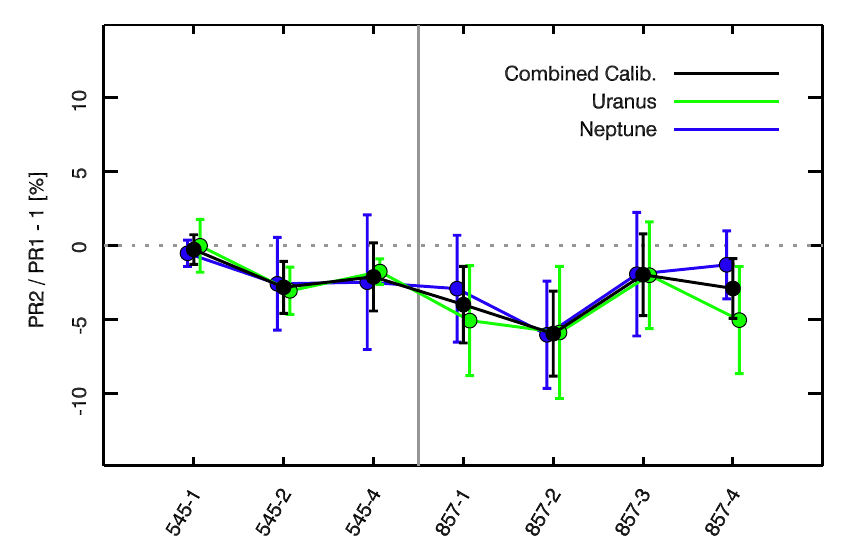}
	\caption{\label{fig:PlanetCalib_PR2_vs_PR1} Comparison of the 2015 \Planck\ release calibration to the one from the 2013 release. We show the relative difference in percent per bolometer for Uranus (green), Neptune (blue), and both calibrators combined (black).}
	\end{center}
\end{figure}

The averaged calibration factors for each detector for each planet are in very good agreement. The final calibration factors are the means of both planet estimates. We compare them to the 2013 \Planck\ release in Fig.~\ref{fig:PlanetCalib_PR2_vs_PR1}: the calibration factors changed by 1.9 and 4.1\,\% at 545 and 857\GHz, respectively, which is within the planet modelling uncertainty. Combined with other pipeline changes (such as the ADC corrections), the 2015 frequency maps have decreased in brightness by 1.8 and 3.3\,\% compared to 2013.

Calibration uncertainties are given in Table~\ref{table:Dx11PlanetCalib}. In order to produce the frequency maps, detectors are weighted by their inverse noise variance. We use the same weights to compute the corresponding calibration errors. We estimate combined statistical errors of 1.1\,\% and 1.4\,\% at 545 and 857\GHz\ respectively, to which one should add linearly (as it should be done for systematics), the 5\,\% uncertainty arising from the planet models. Errors on absolute calibration are therefore 6.1 and 6.4\,\% at 545 and 857\GHz, respectively. Since the reported relative uncertainty of the models is of the order of 2\,\%, we find the relative calibration between the two HFI highest frequency channels to be better than 3\,\%.

\begin{table}[tb]                 
\begingroup
\newdimen\tblskip \tblskip=5pt
\caption{Uncertainties in the planet-derived calibration factor for each bolometer. The systematic uncertainty is the absolute uncertainty of the planet model.}
\label{table:Dx11PlanetCalib}                          
\nointerlineskip
\vskip -3mm
\footnotesize
\setbox\tablebox=\vbox{
   \newdimen\digitwidth 
   \setbox0=\hbox{\rm 0} 
   \digitwidth=\wd0 
   \catcode`*=\active 
   \def*{\kern\digitwidth}
   \newdimen\signwidth 
   \setbox0=\hbox{-} 
   \signwidth=\wd0 
   \catcode`!=\active 
   \def!{\kern\signwidth}
\halign{\hbox to 2.5cm{#\leaderfil}\tabskip 1em&\hfil#\hfil \tabskip 2em&\hfil#\hfil \tabskip 0pt\cr                            
\noalign{\doubleline}
\omit\hfil Bolometer\hfil&\omit\hfil Uncertainty \hfil& \omit\hfil Uncertainty\hfil\cr
\omit\hfil &\omit\hfil Stat. \hfil&\omit\hfil Syst. \hfil \cr
\noalign{\vskip 3pt\hrule\vskip 5pt}
545-1&1.0\,\%&5.0\,\%\cr
545-2&1.8\,\%&5.0\,\%\cr
545-4&2.3\,\%&5.0\,\%\cr
857-1&2.6\,\%&5.0\,\%\cr
857-2&2.9\,\%&5.0\,\%\cr
857-3&2.8\,\%&5.0\,\%\cr
857-4&2.0\,\%&5.0\,\%\cr
\noalign{\vskip 5pt\hrule\vskip 3pt}
}
}
\endPlancktablewide                 
\endgroup
\end{table}                 

\section{Mapmaking}
\label{sec:mapmaking}

\subsection{Summary}
\label{sec:mapsummary}

Each data sample is calibrated in $\mathrm{K_{CMB}}$ for the 100, 143, 217, and 353\GHz\ channels, or \MJysr\ (assuming a constant $\nu I_{\nu}$ law) for the 545 and 857\GHz\ channels, using the calibration scheme described above. Unlike for the 2013 release, the bolometer gains are assumed to be constant throughout the mission. The \Planck\ total dipole (solar and orbital) is computed and subtracted from the data.

As in the \Planck\ 2013 release, we average the measurements for each detector in each pixel visited during a stable pointing period (hereafter called a ``ring'') while keeping track of the bolometer orientation on the sky. The subsequent calibration and mapmaking operations use this intermediate product as input. 
The calibrated TOIs are only modified by a single offset value per ring and per detector. The offsets are determined with a destriping method described in~\cite{tristram2011}. Here, the size of the pixels where the consistency of different rings is tested is 8\arcm\ ($\Nside=512$). Maps in intensity and polarization are used to assess the consistency of the destriper solution. The offsets are simultaneously determined for all bolometers at a given frequency using the full mission data. For all the maps produced using a given bolometer, the same offset per ring is used (except in the case of half-rings, see below).

The products of the HFI mapmaking pipelines are maps of ($I$, $Q$, $U$) together with their covariances ($II$, $IQ$, $IU$, $QQ$, $QU$, $UU$), pixelized  according to the \healpix\ scheme~\citep{gorski2005}. The map resolution is $\Nside=2048$, and the pixel size is 1\parcm7. The mapmaking method is a simple projection of each unflagged sample to the nearest grid pixel. In the case of polarization data with several detectors solved simultaneously, the polarization equation is inverted on a per-pixel basis (see Sect.~\ref{sec:mapmethod}). For each sky map, a hit count map is computed (number of samples per pixel; one sample has a duration of $5.544\,\mathrm{ms}$).

We provide maps from which the zodiacal light component, which varies in time, has been removed, based on templates fitted on the survey-difference maps (see Sect.~\ref{sec:zodi}). These templates are systematically subtracted from the data of each bolometer prior to the mapmaking. 

Unlike in the 2013 release, the far sidelobes (FSL) are not removed from the maps. At most, this leaves residuals of order 0.5--1.5\,$\mu\mathrm{K}$ in the 100--353\GHz\ maps, with uncertainties on the residuals of roughly 100\,\% \citep{planck2014-a08}. At higher frequencies, Galactic pick-up from the FSL produces significant residuals in about half of the detectors of the order of 0.03\MJysr\ at 545 and 0.3\MJysr\ at 857\GHz\  \citep[see][]{planck2014-a12}. The total solid angle in the spillover is a better known quantity, and we describe in Sect.~\ref{sec:fsls} the effect of the FSL on the calibration.

\subsection{Mapmaking method}
\label{sec:mapmethod}

In the same way as in \citet{planck2013-p03f}, we use a destriping algorithm to deal with the HFI low-frequency noise. In this approach, the noise in a ring $r$ is represented by an offset, denoted $o_r$, and a white noise part $n$, which is uncorrelated with the low-frequency noise.
For a given bolometer, we can write Eq.~\eqref{eq:cal} as:
\begin{equation}
	P_t = G \times ( A_{tp} \, T_p + t_{\mathrm{orb}}) + \Gamma_{tr} \, o_r + n_t \,, 
\label{eq:MM2}
\end{equation}
where $\Gamma_{tr}$ is the ring-pointing matrix (which associates the data sample $t$ with the ring number $r$). The unknowns are the gain $G$, the offsets for each ring $o_r$, and the sky signal for each pixel $T_p = (I_p,Q_p,U_p)$. As there is a degeneracy between the average of the offsets and the zero level of the maps, we impose the constraint $\langle {o_r} \rangle = 0$. The absolute zero level of the maps is determined as described in Sect.~\ref{sec:monopole}.

For the production of the maps for the 2015 HFI data release, we first build the rings for all detectors. We apply the following frequency-dependent processing to these compressed data sets. 

For CMB frequencies (100, 143, and 217\GHz) channels:
\begin{enumerate}
	\item we estimate a first approximation of the orbital dipole gain together with the offsets and the amplitude of the long-time-constant residuals for each bolometer independently, neglecting the polarization signal;
	\item we then derive the gains and offsets for a fixed amplitude of the long-time-constant residuals using the multi-bolometer algorithm.
\end{enumerate}

At 353\GHz, the long-time-constant residuals are more difficult to constrain. They are driven more by Galactic emission drifts than by the dipole, which dominates at lower frequencies, so that the model described in Sect.~\ref{sec:vltc} is not relevant. Hence for this frequency we use a simpler pipeline without a long-time-constant residuals template:
\begin{enumerate}
	\item we estimate the orbital dipole gain together with the offsets for each bolometer independently, neglecting the polarization signal;
	\item we then estimate the final offsets using a destriping procedure for all bolometers at this frequency.
\end{enumerate}

For the two highest frequencies, at 545 and 857\GHz, the pipeline is considerably different, because we use the planets (Uranus and Neptune) as calibration sources:
\begin{enumerate}
	\item we estimate a first approximation of the offsets for each bolometer independently;
	\item we derive the gains from the planet flux comparison;
	\item we then estimate the final offsets with a destriping procedure for all bolometers at a given frequency.
\end{enumerate}

Finally, for each data set, using the pre-computed gains and offsets, we project the ring data onto maps. For polarization, we invert the $3\times 3$ system derived from Eq.~\eqref{eq:MM2} for each pixel independently with a criterion that the condition number be lower than $10^3$:
\begin{equation}
	\vec{T} = \left(\tens{A}^\mathrm{T} {\tens N}^{-1} {\tens A}\right)^{-1} \tens{A}^\mathrm{T} \tens{N}^{-1} \vec{d}\,,
	\label{eq:destMap}
\end{equation}
where $\vec{d}$ are the calibrated, cleaned ring data (offsets and orbital dipole removed)  $\vec{d} = \left(\vec{P} - \tens{\Gamma} \, \vec{o} \right) /G - \vec{t}_\mathrm{orb}$.
Note that we use \healpix\ (not IAU) conventions for the sign and normalization of the $Q$ and $U$ Stokes parameters.

For destriping, we use the same tool as before, \polkapix, which was thoroughly validated in~\cite{tristram2011}. Maps are built by simple co-addition in each pixel of the destriped, calibrated, and time-varying component-subtracted signal. We subtract the CMB dipole as measured by \Planck\ (see Sect.~\ref{sec:soldip}).

We introduced the following modifications with respect to~\citet{planck2013-p03f}:
\begin{itemize}
	\item we include polarization in the destriping for the channels that include PSBs;
	\item we enlarged the masked fraction of the sky, from 10 to 15\,\%, based on Galactic emission, to avoid signal gradients leaking into offsets;
	\item to improve the offset accuracy, we compute one set of offsets combining all detectors, using full-mission data, and use them for all derived maps involving these detectors.
\end{itemize}
This last change induces a small noise correlation between detector-set maps (see Sect.~\ref{sec:detset-correlation}). In 2013 we computed independent offsets for each detector or detector set, including the full mission.

\subsection{Map products}
\label{sec:mapsplit}

\begin{figure*}
	\begin{center}
	\includegraphics[width=\textwidth]{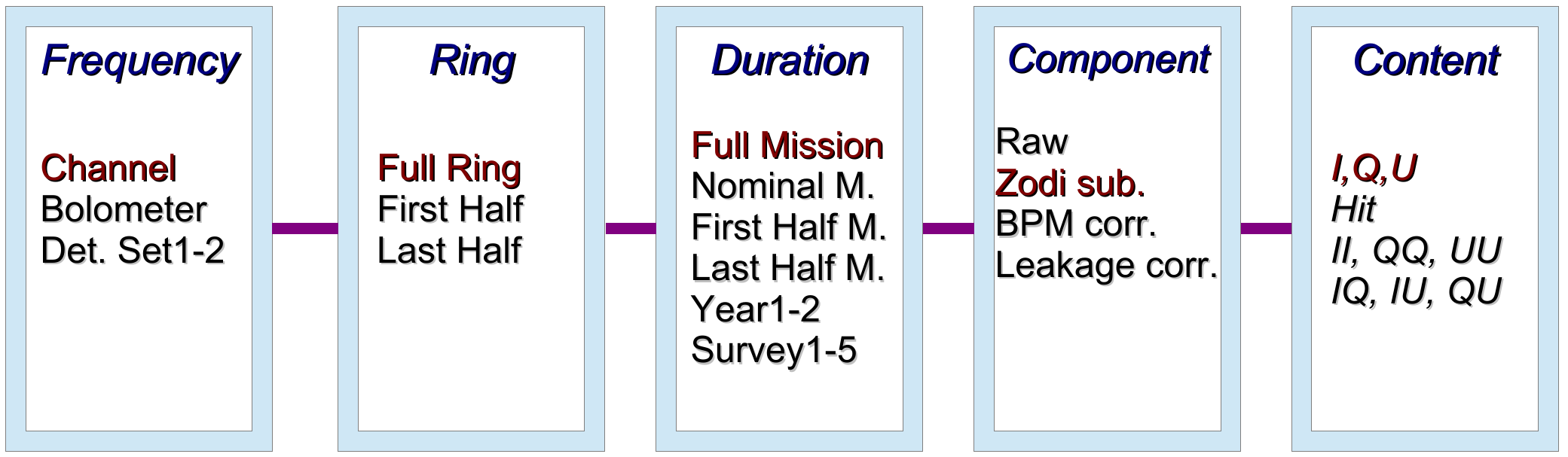}
	\caption{\label{fig:MapMatrix} Map matrix. The HFI maps are released in different flavours. Not all combinations are released but any map will correspond to a choice of lines in each box. The \emph{Frequency} box is related to the use of all detectors at a given frequency (Channel), or individual bolometers or sets of detectors as defined in Table~\ref{tab:detset}. The \emph{Ring} box is a way of splitting (or not) the data in equal halves at the ring level. The \emph{Duration} box indicates the different ways of splitting data between surveys, years, full or nominal mission, first half mission, or last half mission. The \emph{Component} box indicates the systematic corrections that can be applied at the map level. The recommended first choice map is highlighted in red. See Appendix~\ref{annex:official} for details.}
	\end{center}
\end{figure*}

The principal HFI final product consists of six maps that cover the six frequencies (100--857\GHz) for the full mission in intensity at high resolution ($\Nside=2048$). However, many more maps are needed to assess the noise and the consistency of the data. Figure~\ref{fig:MapMatrix} summarizes the various splits produced. The branches that are delivered in the 2015 release are described in Appendix~\ref{annex:official}.

Maps from different halves of each ring period (first and last) are computed independently of each other, including the offset per half-ring. Thus, half-ring half-difference maps can give a quick account of the noise level in the maps.

For each frequency, we also produce temperature and polarization maps using detector sets (each set including four polarization sensitive bolometers). In addition, we produce a temperature map for each spider-web bolometer (SWB).

\Planck's scanning strategy samples almost all the sky pixels every six months, with alternating scan directions in successive 6-month periods. The full cold HFI mission encompasses five surveys, each covering a large fraction of the sky. Surveys 1--2 and 3--4 are paired to produce Year~1 and Year~2 maps (see Table~\ref{tab:timesplit}).
Maps are produced for the full-mission dataset together with the survey, year, and half-mission maps. With each map is associated a hit-count map and variance maps ($II$; and $QQ$, $UU$,  $IQ$, $IU$, and $QU$ when polarization is reconstructed). Overall, a total of about 6500 sky maps have been produced. We have used this data set to evaluate the performance of the photometric calibration by examining difference maps (see Sect.~\ref{sec:jackknives}).

\subsection{Zero levels} 
\label{sec:monopole}

\Planck-HFI cannot measure the absolute sky background. The mapmaking procedure does not change the mean value of the input TOI. We therefore adjust the monopole on the maps a posteriori, in a similar manner to the method used in~\citet{planck2013-p03f} that relies on external datasets. To achieve this, we need to take into account two major components of the monopole:
\begin{enumerate}
	\item Galactic dust emission: we estimate the brightness in the HFI single-detector maps that corresponds to zero gas column-density (i.e., zero Galactic dust emission). To do so, we use the \ion{H}{i} column density, which is assumed to be a reliable tracer of the Galactic gas column-density in very diffuse areas \citep[see][Sect.~5.1]{planck2013-p03f}. The offsets derived are then subtracted from each detector's data in the processing.
	\item Extragalactic emission: the cosmic infrared background (CIB) monopole is taken into account by adding the levels from \cite{bethermin2012} to the maps (see Table~\ref{tab:summary}).
\end{enumerate}

The sum of the two offsets is appropriate for total emission analysis. For Galactic studies, only the Galactic zero level has to be set which can be achieved by subtracting the CIB levels (Table~\ref{tab:summary}) from the released maps. Unlike for the previous release, in the 2015 maps the zero level correction (both CIB and Galactic) has been applied.

Zodiacal light has not been accounted for in this procedure. The offsets that have to be removed at each frequency to set the Galactic zero level using zodiacal-light-corrected maps are smaller than those needed for the total maps. The correction to be applied to the released maps are given in Table~\ref{tab:summary}.

\subsection{Polarization efficiency and orientation}
\label{sec:pol_eff}
The calibration parameters for the PSBs were measured on the ground  before launch. \cite{rosset2010} have reported pre-flight measurements of the polarization efficiency of the HFI PSBs with an accuracy of 0.3\,\%. The absolute orientation of the focal plane has been measured at a level better than $0\pdeg3$. The relative orientation between PSBs is known with an accuracy better than $0\pdeg9$.

The SWBs are much less sensitive to polarization. Nonetheless, we take into account their polarization efficiency which is between 1 and 9\,\%, although their orientations have been less accurately determined (errors can be up to a few percent), as described in~\cite{rosset2010}.

\section{HFI temperature and polarization maps}
\label{sec:maps}

\subsection{Solar dipole measurement}
\label{sec:soldip}

\begin{table*}[!htb]
	\begingroup
	\newdimen\tblskip \tblskip=5pt
	\caption{CMB solar dipole measurements for the 100 and 143\GHz\ channels estimated for different sky coverage levels (37,~50, and~58\,\%) corresponding to three thresholds in 857\GHz\ amplitude (2,~3, and~4\,\MJysr). Uncertainties include only statistical errors. Systematic errors are 0.8\muK\ for the amplitude, and (0\pdeg024, 0\pdeg0034) in Galactic (longitude, latitude).}
	\label{tab:dipole_measure}
\nointerlineskip
\vskip -3mm
\footnotesize
\setbox\tablebox=\vbox{
   \newdimen\digitwidth 
   \setbox0=\hbox{\rm 0} 
   \digitwidth=\wd0 
   \catcode`*=\active 
   \def*{\kern\digitwidth}
   \newdimen\signwidth 
   \setbox0=\hbox{-} 
   \signwidth=\wd0 
   \catcode`!=\active 
   \def!{\kern\signwidth}

\halign{\hbox to 2cm{#\leaderfil}\tabskip 1em&\hfil# \tabskip 2em\hfil&\hfil# \tabskip 2em&\hfil# \tabskip 2em&\hfil# \tabskip 0pt&\hfil#\tabskip  0pt\cr       
\noalign{\doubleline}
		\omit\hfil Frequency \hfil & \omit Threshold \hfil & \omit\hfil $d$ \hfil & \omit\hfil $\mathrm{lon}$ \hfil & \omit\hfil $\mathrm{lat}$\hfil\cr
		\omit\hfil [\GHz] \hfil &  \omit\hfil [\MJysr] \hfil& \omit\hfil [\muK] \hfil & \omit\hfil [$\deg$] \hfil  & \omit\hfil [$\deg$] \hfil\cr
	\noalign{\vskip 3pt\hrule\vskip 5pt}
100 & 2 &$3364.81 \pm 0.06$ & $263.921 \pm 0.002$ & $48.2642 \pm 0.0008$\cr
100 & 3 & $3364.76 \pm 0.05$ & $263.922 \pm 0.002$ & $48.2640 \pm 0.0006$\cr
100 & 4 & $3364.99 \pm 0.04$ & $263.928 \pm 0.002$ & $48.2631 \pm 0.0006$\cr
143 & 2 & $3364.05 \pm 0.03$ & $263.908 \pm 0.001$ & $48.2641 \pm 0.0004$\cr
143 & 3 & $3363.72 \pm 0.02$ & $263.903 \pm 0.001$ & $48.2653 \pm 0.0003$\cr
143 & 4 & $3363.39 \pm 0.02$ & $263.905 \pm 0.001$ & $48.2668 \pm 0.0003$\cr
\noalign{\vskip 5pt\hrule\vskip 3pt}
}
}
\endPlancktablewide                 
\endgroup
\end{table*}

The $\ell=1$ mode of  CMB anisotropy is unique in that its amplitude is dominated by a large component associated with our motion with respect to the CMB rest frame. In this section, we present the CMB solar dipole results based on \Planck-HFI maps at the two lowest frequencies, 100 and 143\GHz. Low-frequency maps are dominated by CMB over a large fraction of the sky. Nevertheless, the inhomogeneous nature of the dust emission can bias CMB solar dipole estimates. 

We cleaned the Galactic emission from the HFI maps using a local correlation with the 857\GHz\ map. We model each HFI map $I_\nu$ as
\begin{equation}
	I_\nu - C = q I_{857} + D_{\mathrm{res}}
\end{equation}
where $C$ is the CMB anisotropy (here we use the {\tt SMICA} map, \citealt{planck2014-a11}, from which we remove any residual dipole component) and $I_{857}$ is the \Planck\ 857\GHz\ map that is assumed to have a negligible contribution from the solar dipole.\footnote{The amplitude of the solar dipole at 857\GHz\ is 0.0076 \MJysr. At least 90\,\% of this is removed in the mapmaking process, leaving a residual that is well below the noise level and any systematic effects.}
The term $D_{\mathrm{res}}$ includes the dipole and any systematic effects from both $I_{857}$ and $I_\nu$. In bright regions of the sky, $D_{\mathrm{res}}$ also contains extra emission that is uncorrelated or only partially correlated with $I_{857}$, for instance free-free emission or CO.

In order to capture any spatial variations of the dust SED, we estimated $q$ and $D_{\mathrm{res}}$ on an $\Nside=64$ grid. For each $\Nside=64$ pixel, we performed a linear regression of the $\Nside=2048$ pixels of $I_\nu$ vs $I_{857}$, assuming a constant dust SED over a 55\arcm\ area.
We then fit for the dipole amplitude and direction in $D_{\mathrm{res}}$ using sky pixels where $I_{857} < 2$, 3, or 4\,\MJysr\ (corresponding to 37, 50, or 58\,\% of the sky respectively) to limit the effect of Galactic-emission residuals (CO, free-free emission, and small-scale dust SED variations). 

The results are given in Table~\ref{tab:dipole_measure}. We measure a Solar System peculiar velocity of $370.06 \pm 0.09\ \mathrm{km\ s}^{-1}$ with respect to the CMB rest frame. We use the CMB temperature from \citet{fixsen2009} ($2.7255\pm0.0006$\,K) to convert that measurement into a CMB dipole measurement.

The error bars here only include statistical uncertainties, which are very low thanks to the \Planck-HFI signal-to-noise ratio. 
We evaluate the additional systematic uncertainties from the variation of the results between independent bolometer maps. For the amplitude, the peak-to-peak variation between bolometers and combined maps is $\pm0.8$\muK\ at 100 and 143\GHz. Variations with sky coverage are of the same order. Note that the uncertainty from the FIRAS temperature should be added to the budget ($\pm0.74$\muK). For the coordinates, we found variations of $\pm0\pdeg013$ in longitude and $\pm0\pdeg0019$ in latitude. These differences are observed when comparing results at different frequencies, and are likely to result from uncertainties in the foreground subtraction. This is also consistent with the magnitude of the direction shifts we observe when changing the sky fraction.

As an independent check, we also produce a cleaned CMB map using an internal linear combination (ILC) method. We used the HFI maps at 100, 143, and 217\GHz\ smoothed with a 1\deg~FWHM Gaussian kernel. Note that smoothing the data with a 1\deg\ kernel reduces the solar dipole in the maps by 0.005\,\%, i.e., 0.2\muK, which we corrected for afterwards. We then estimate the solar dipole amplitude and direction using a Galactic mask that removes less than 15\,\% of the sky to avoid the inner Galactic plane where the residuals are most intense. The measurement is compatible with the results in Table~\ref{tab:dipole_measure}.

At the end, the amplitude ($d$) and direction (Galactic longitude, latitude) of the solar dipole measured by \Planck-HFI is
\begin{eqnarray*}
	d &=& 3364.29 \pm 0.02 \mathrm{(stat)} \pm 0.8\mathrm{(sys)} \pm 0.74\mathrm{(FIRAS)} \ \mu\mathrm{K}\\
	\left(
	\begin{array}{c}
		\mathrm{lon}\\
		\mathrm{lat}\\
	\end{array} \right) 
	&=& \left(
	\begin{array}{lll}
		263\pdeg914  &\pm 0\pdeg001\phantom{0} \mathrm{(stat)} &\pm 0\pdeg013\phantom{0} \mathrm{(sys)}\\
		\phantom{0}48\pdeg2646 &\pm 0\pdeg0003 \mathrm{(stat)} &\pm 0\pdeg0019 \mathrm{(sys)}
	\end{array}
	\right)
\end{eqnarray*}

This is to be compared to the official \Planck\ solar dipole measurement obtained in combination with the \Planck-LFI:
\begin{eqnarray*}
	d &=& 3364.5  \pm 2.0 \ \mu{\mathrm{K}} \\	
 	\left(
	\begin{array}{c}
		\mathrm{lon}\\
		\mathrm{lat}\\
	\end{array} \right) 
	&=& \left(
	\begin{array}{c}
		264\pdeg00 \pm 0\pdeg03\\
		\ \ 48\pdeg24 \pm 0\pdeg02
	\end{array}
	\right)\\
\end{eqnarray*}

Compared to the \WMAP\ five-year results ($d, \mathrm{lon}, \mathrm{lat}) = (3355 \pm 8 \muK, 263\pdeg99 \pm 0\pdeg14, 48\pdeg26 \pm 0\pdeg03$; \citealt{hinshaw2009}), this is 9.3\muK\ (0.28\,\%) higher in amplitude while shifted by $(0\parcm6,1\parcm2)$ in longitude and latitude. Part of the difference (0.6\muK) is due to the revised CMB monopole temperature compared to \citet{mather1999} ($2.725$\,K). This total dipole (solar, orbital, relativistic, and interactions thereof) is removed from the calibrated TOI before final mapmaking.

\subsection{\Planck-HFI maps}
\begin{figure*}[htbp]
	\centering
	\vspace{-0.4in}
	\includegraphics[width=0.80\textwidth]{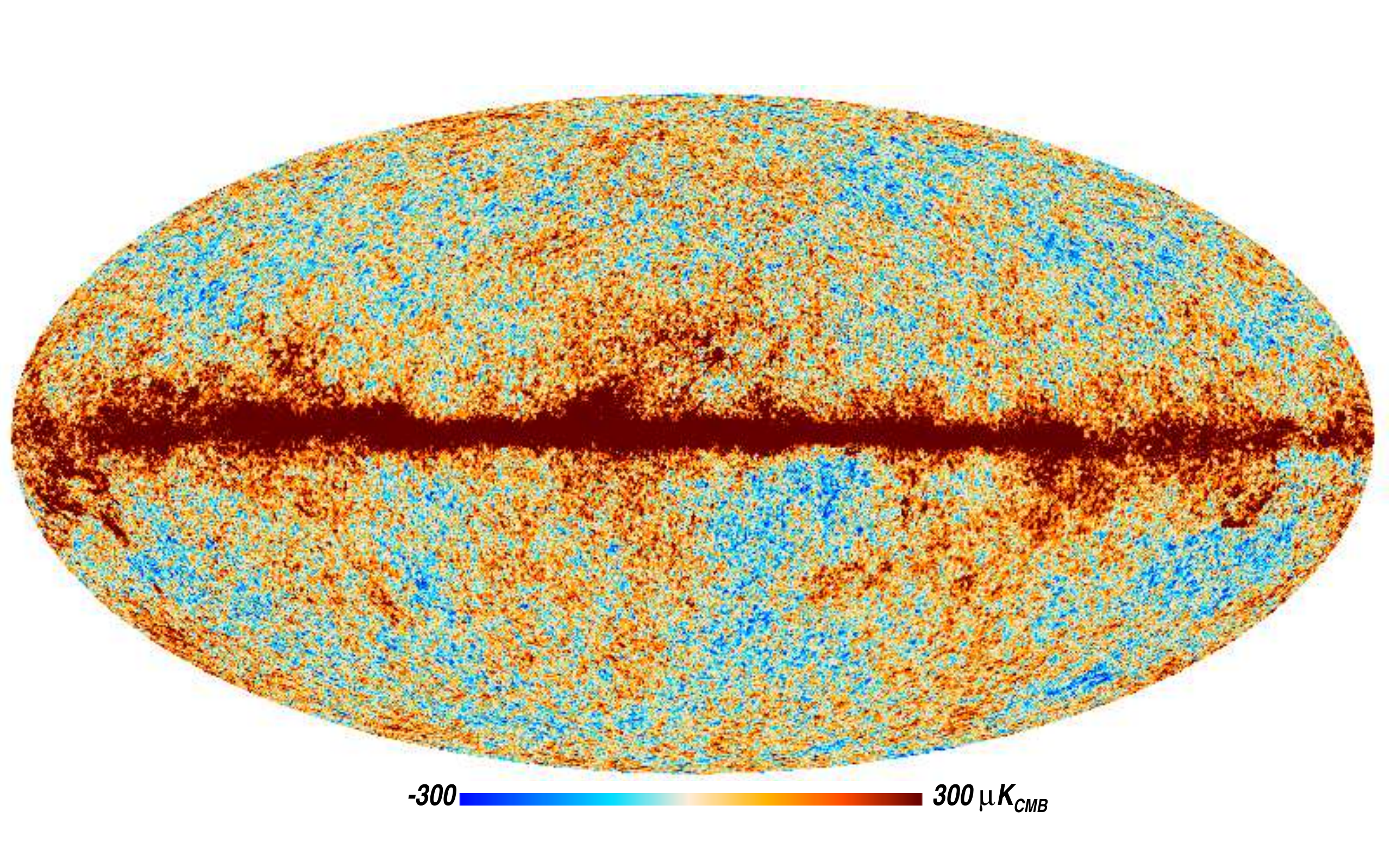} \\
	\vspace{-0.4in}
	\includegraphics[width=0.80\textwidth]{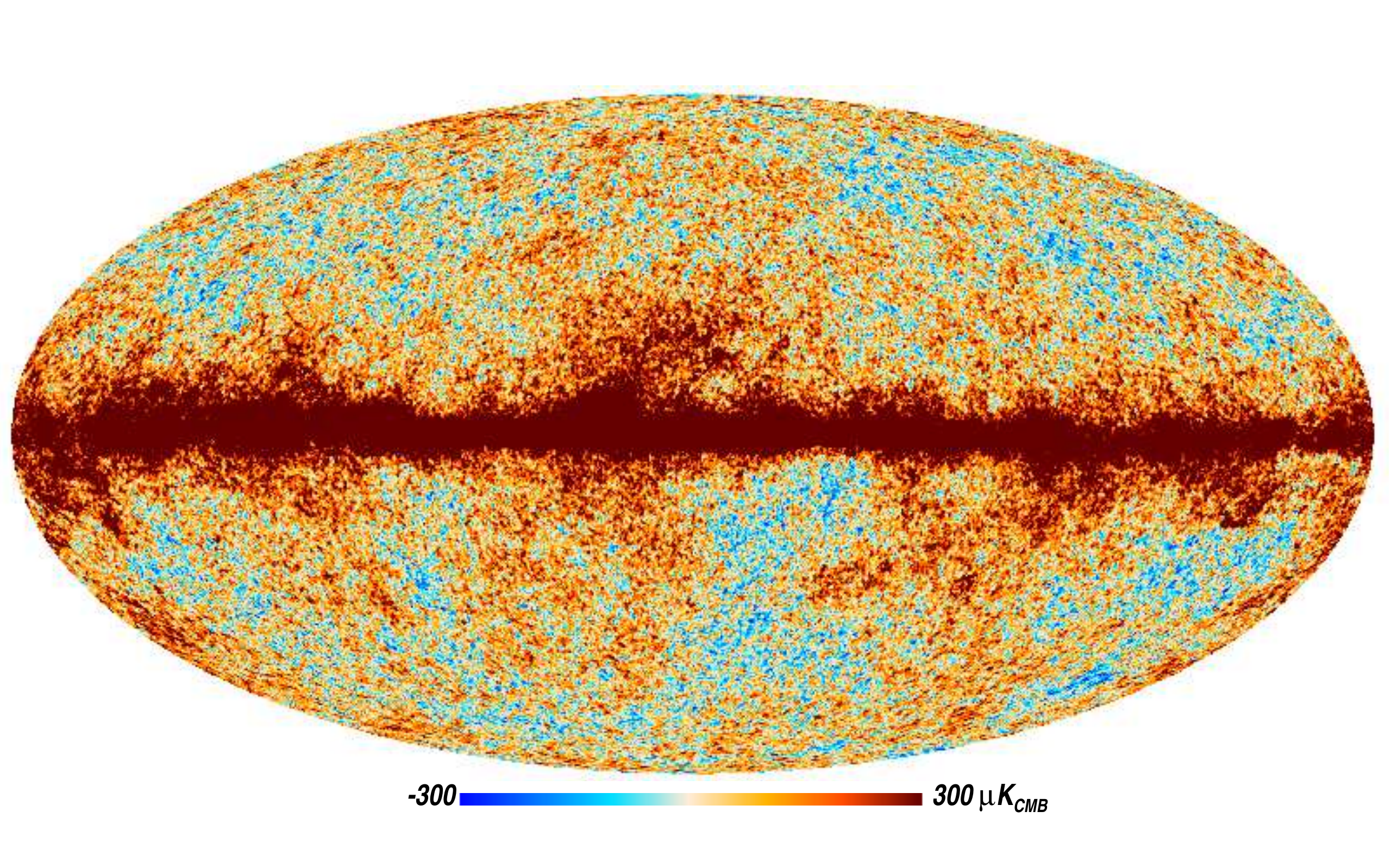} \\
	\vspace{-0.4in}
	\includegraphics[width=0.80\textwidth]{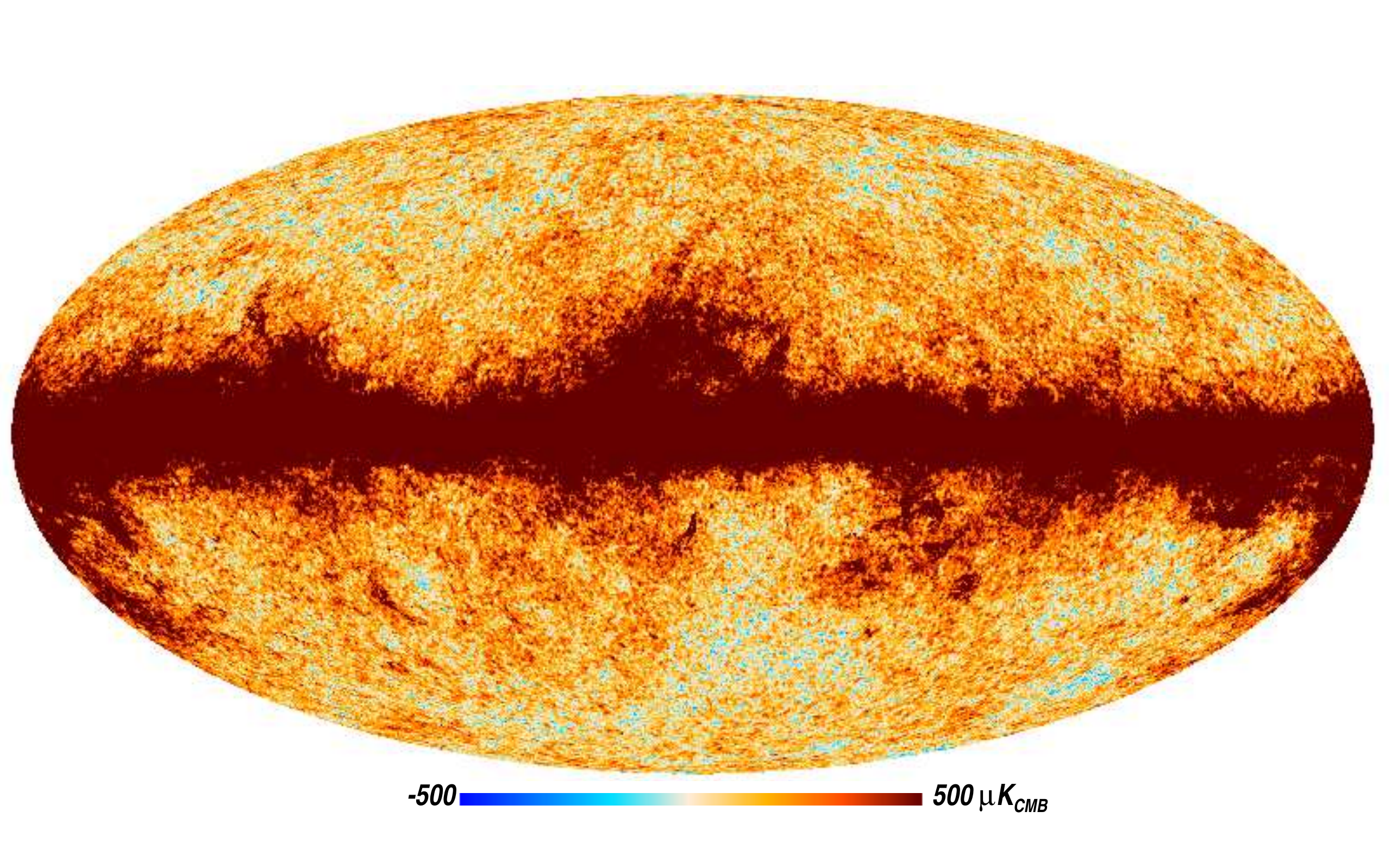} \\
	\caption{\Planck-HFI full mission channel intensity maps at 100, 143, and 217\GHz\ ({\it from top to bottom}) after removal of zodiacal emission.}
	\label{fig:maps_hfi_low}
\end{figure*}

\begin{figure*}[htbp]
	\centering
	\vspace{-0.4in}
	\includegraphics[width=0.80\textwidth]{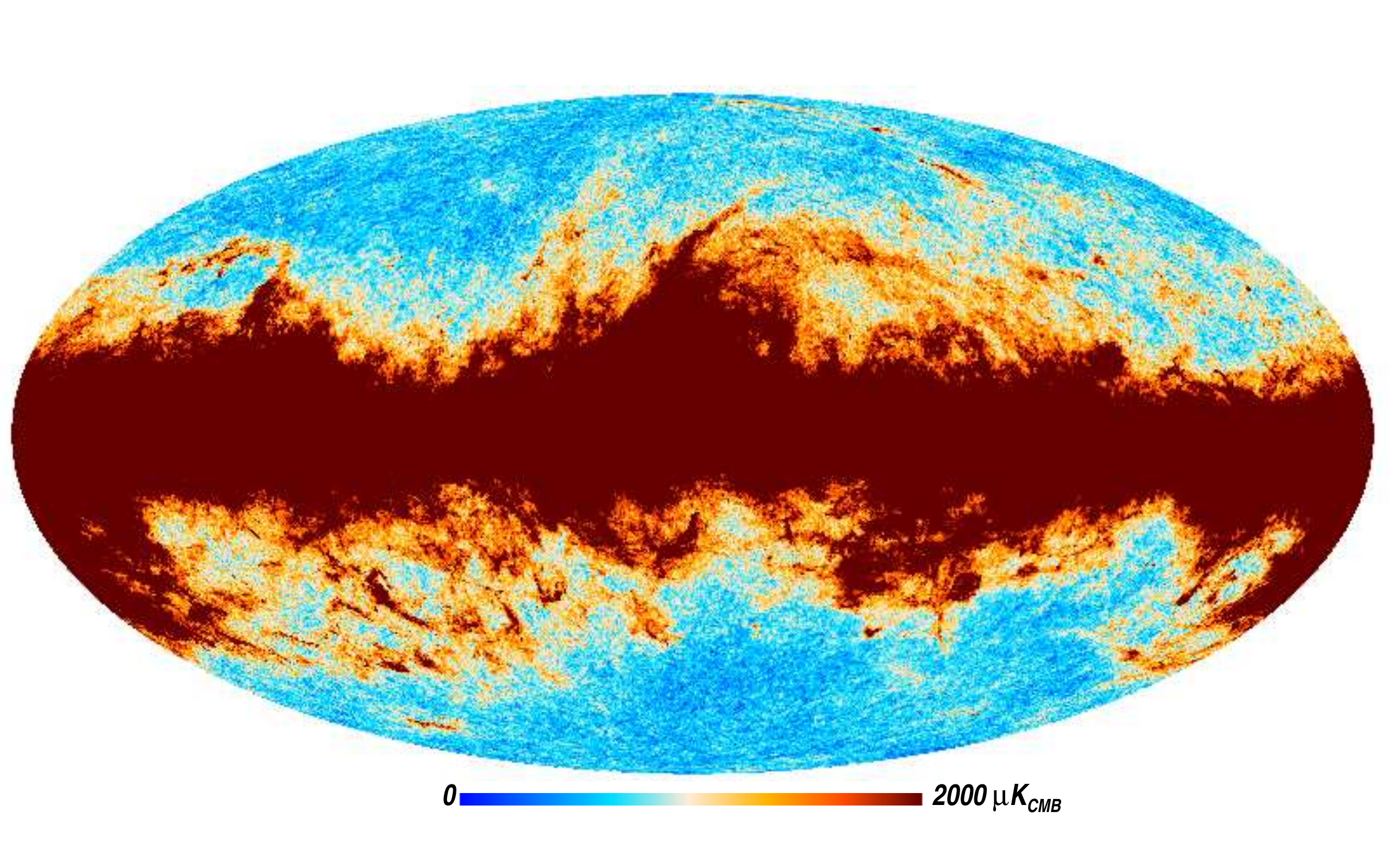} \\
	\vspace{-0.4in}
	\includegraphics[width=0.80\textwidth]{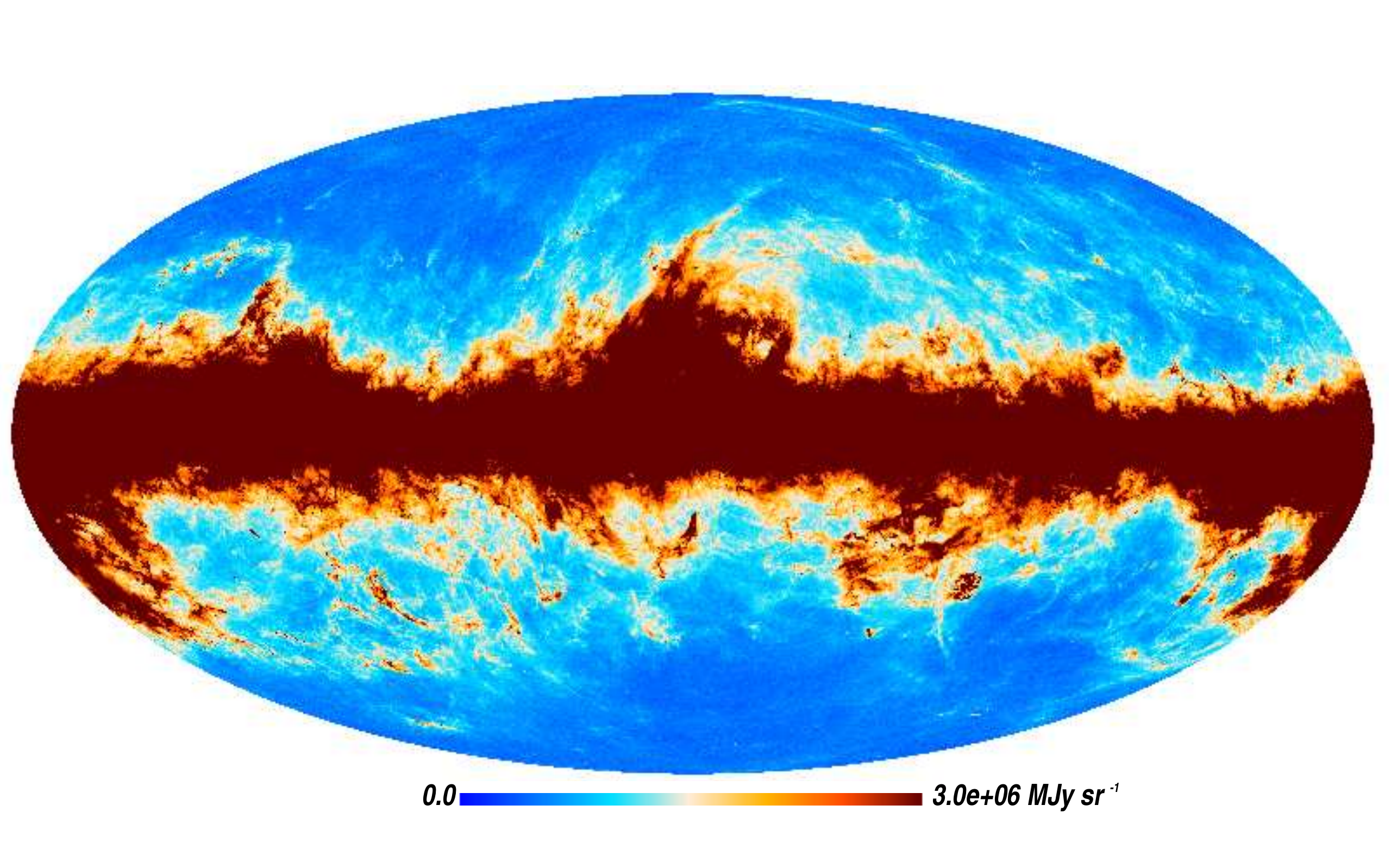} \\
	\vspace{-0.4in}
	\includegraphics[width=0.80\textwidth]{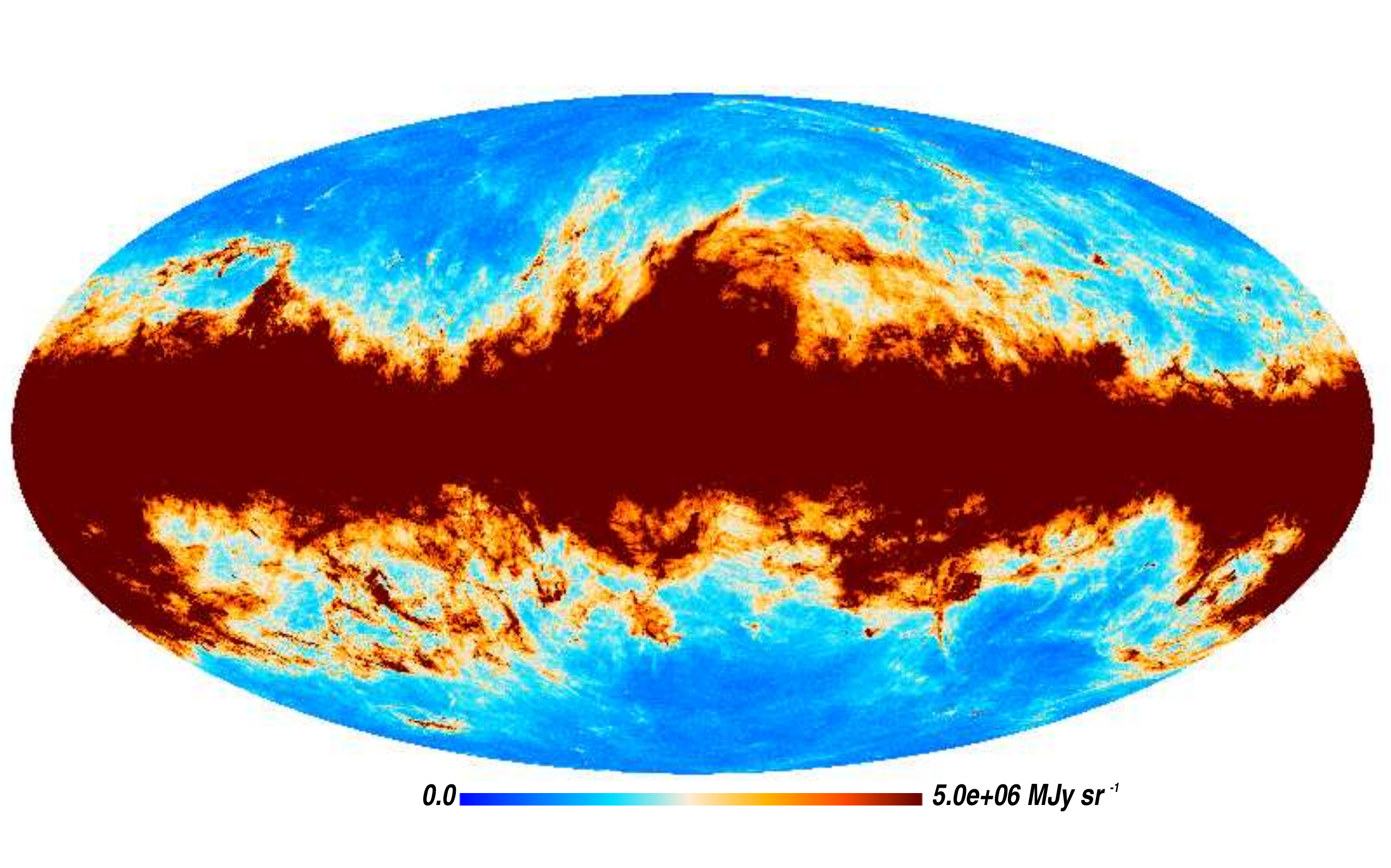}
	\caption{\Planck-HFI full mission channel intensity maps at 353, 545, and 857\GHz\ ({\it from top to bottom}) after removal of zodiacal emission.}
	\label{fig:maps_hfi_mm}
\end{figure*}

\begin{figure*}[ht!]
	\centering
	\includegraphics[width=0.49\textwidth]{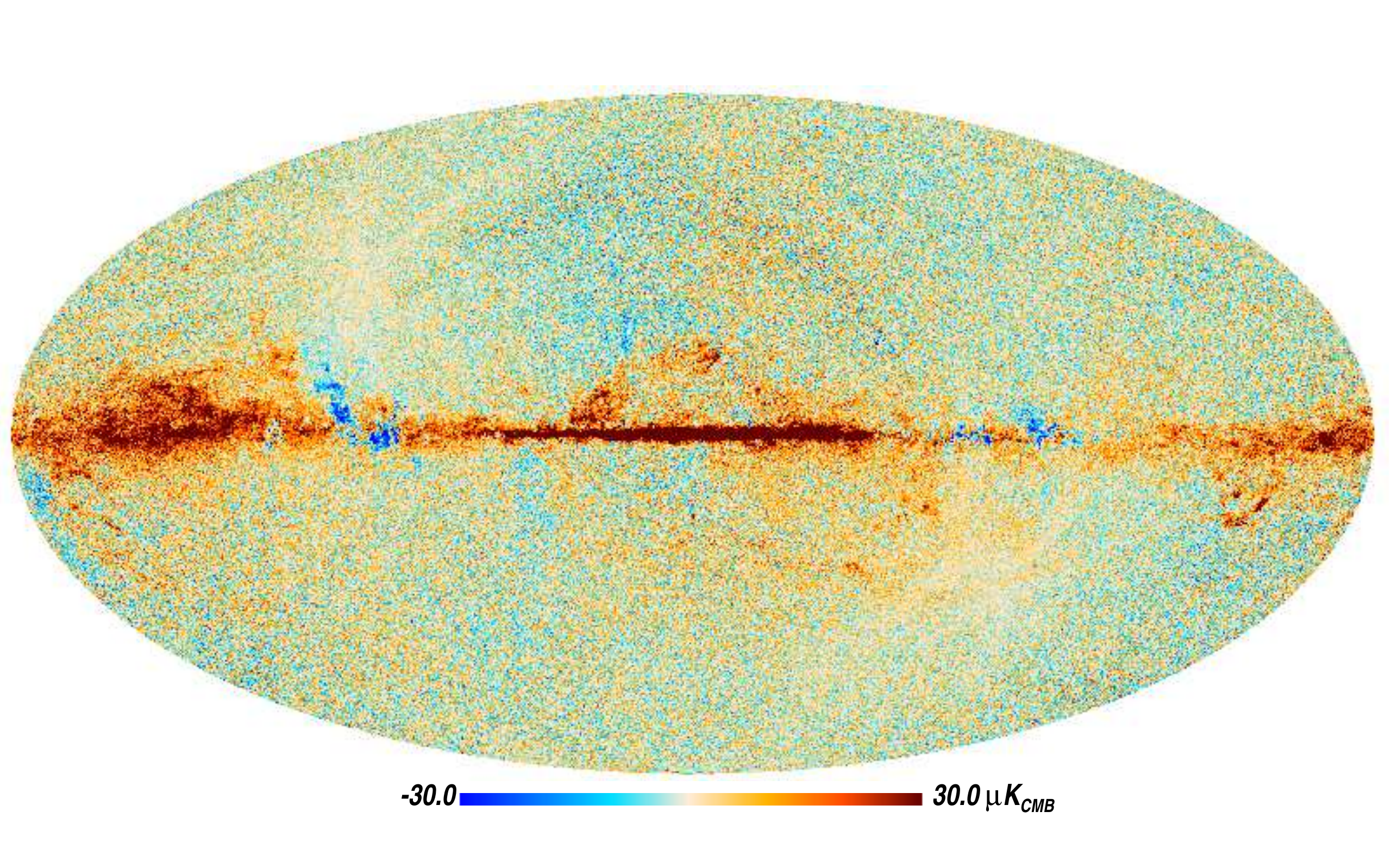} 
	\includegraphics[width=0.49\textwidth]{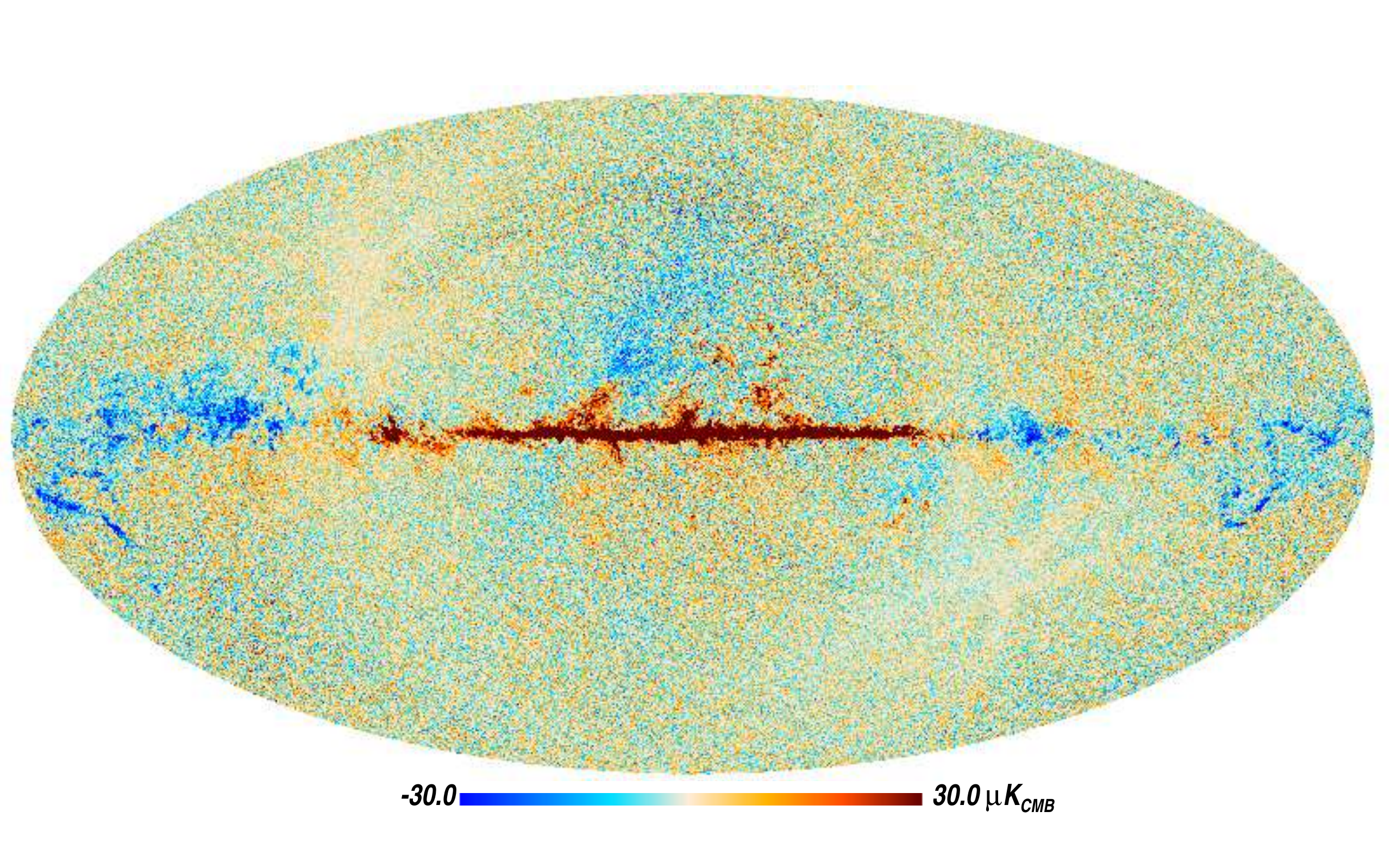} \\
	\includegraphics[width=0.49\textwidth]{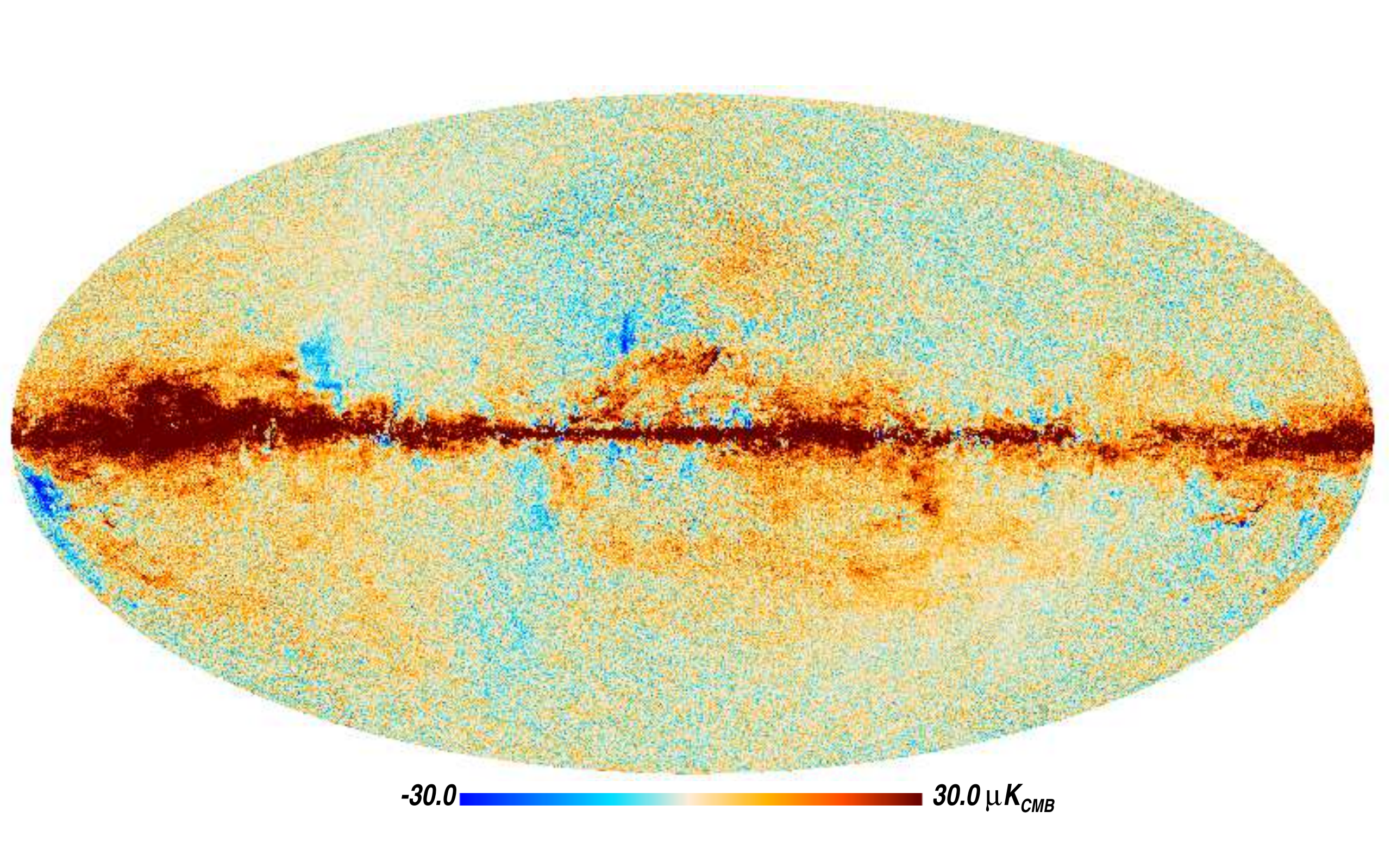}  
	\includegraphics[width=0.49\textwidth]{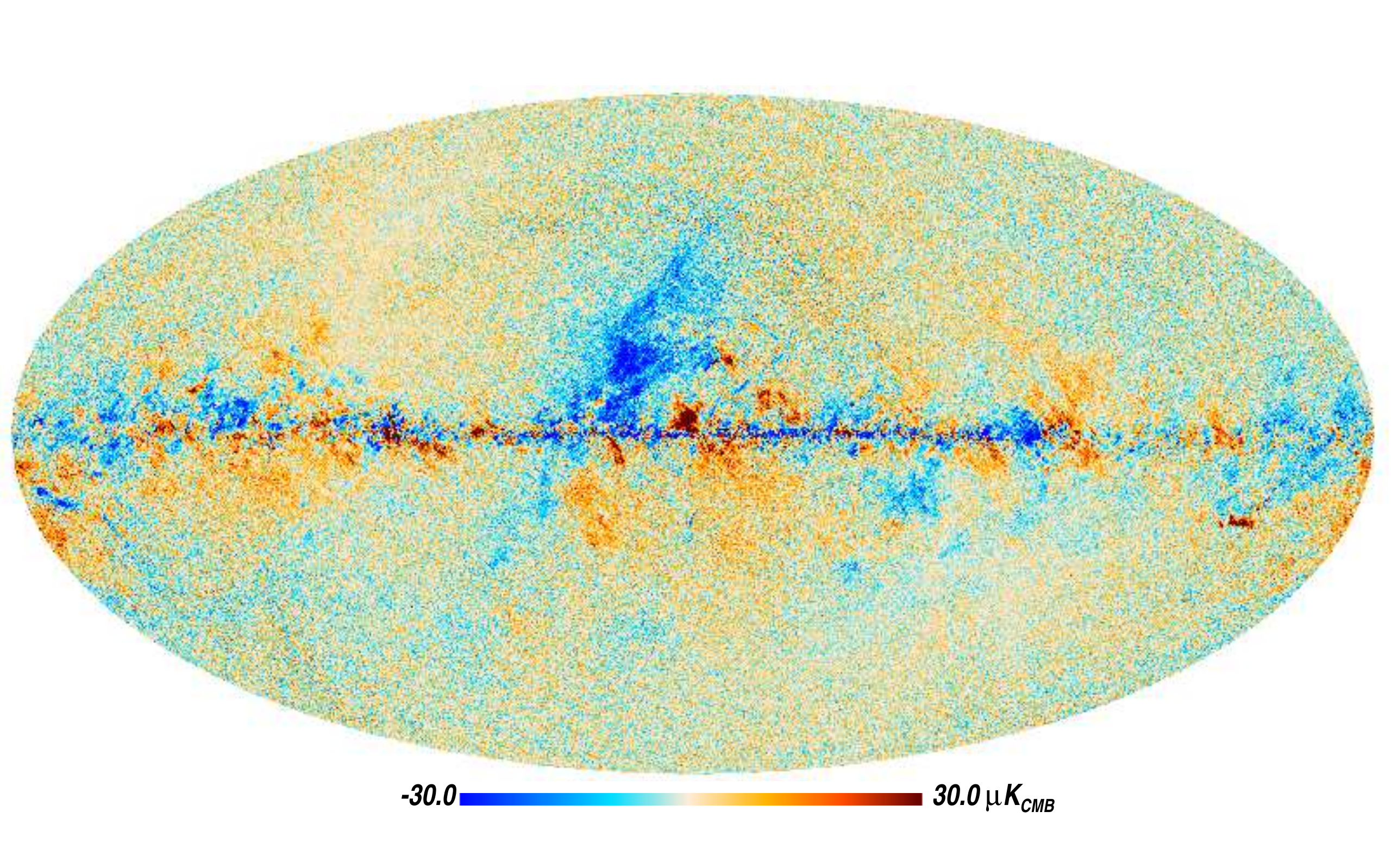} \\
	\includegraphics[width=0.49\textwidth]{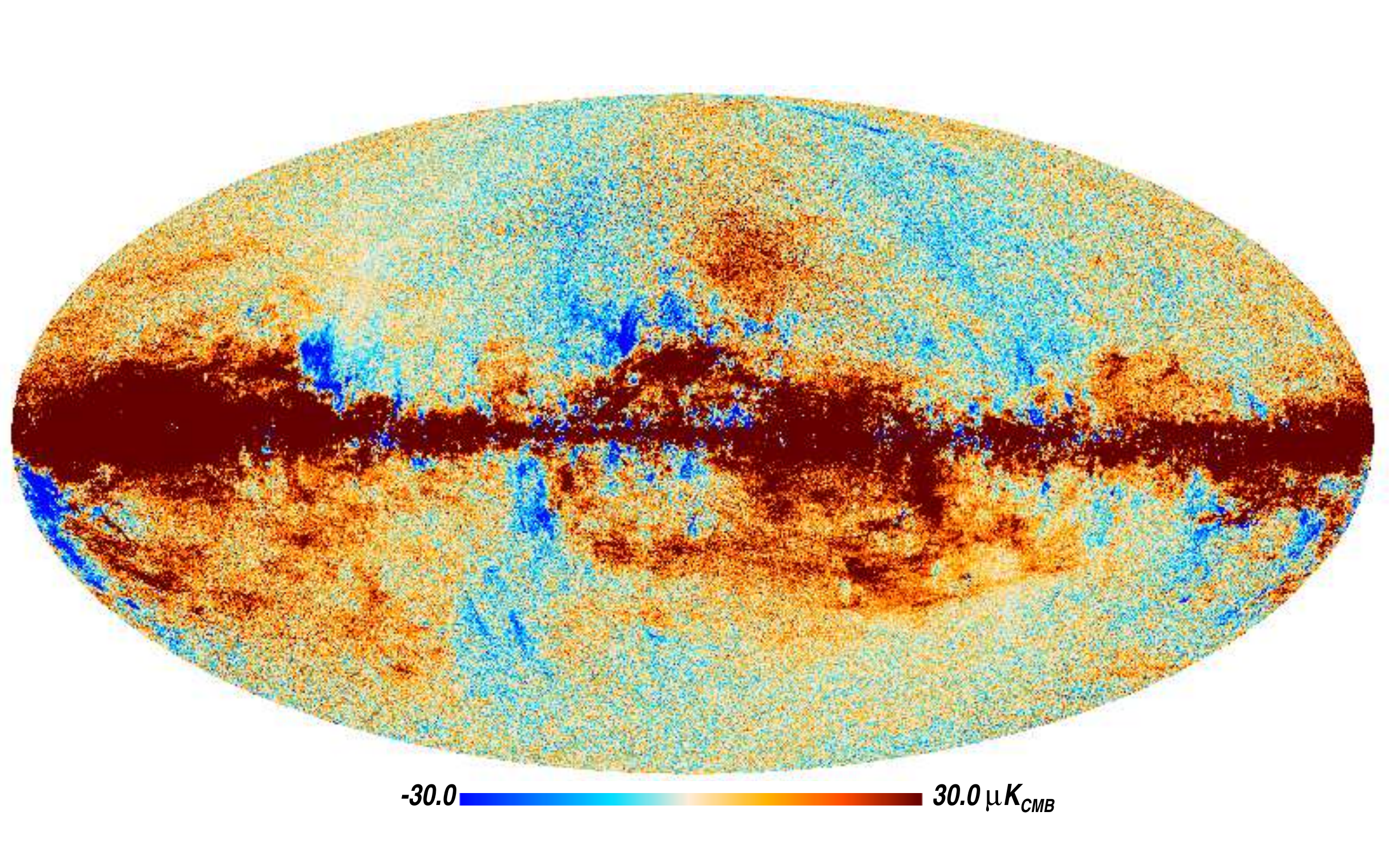}  
	\includegraphics[width=0.49\textwidth]{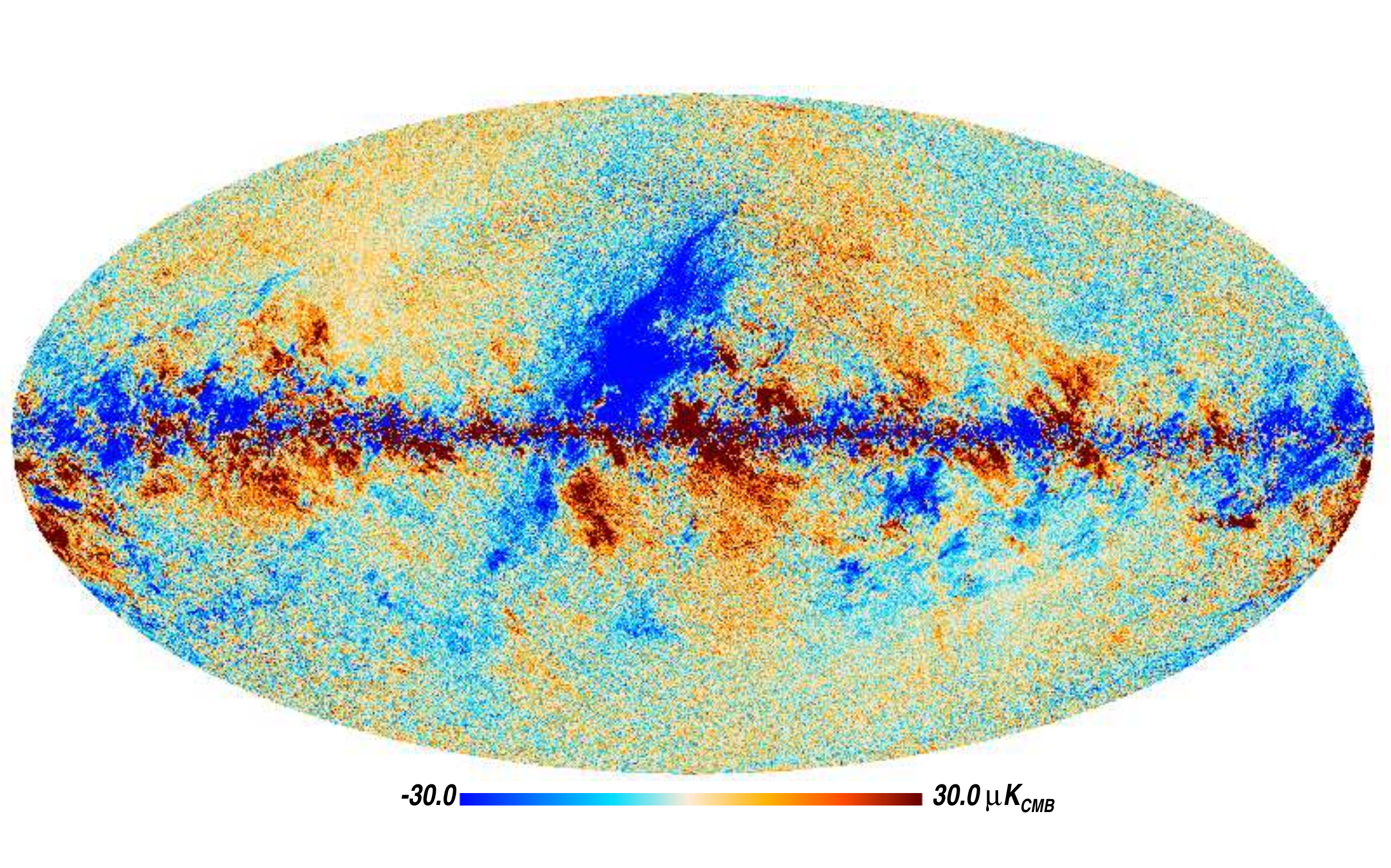} \\
	\includegraphics[width=0.49\textwidth]{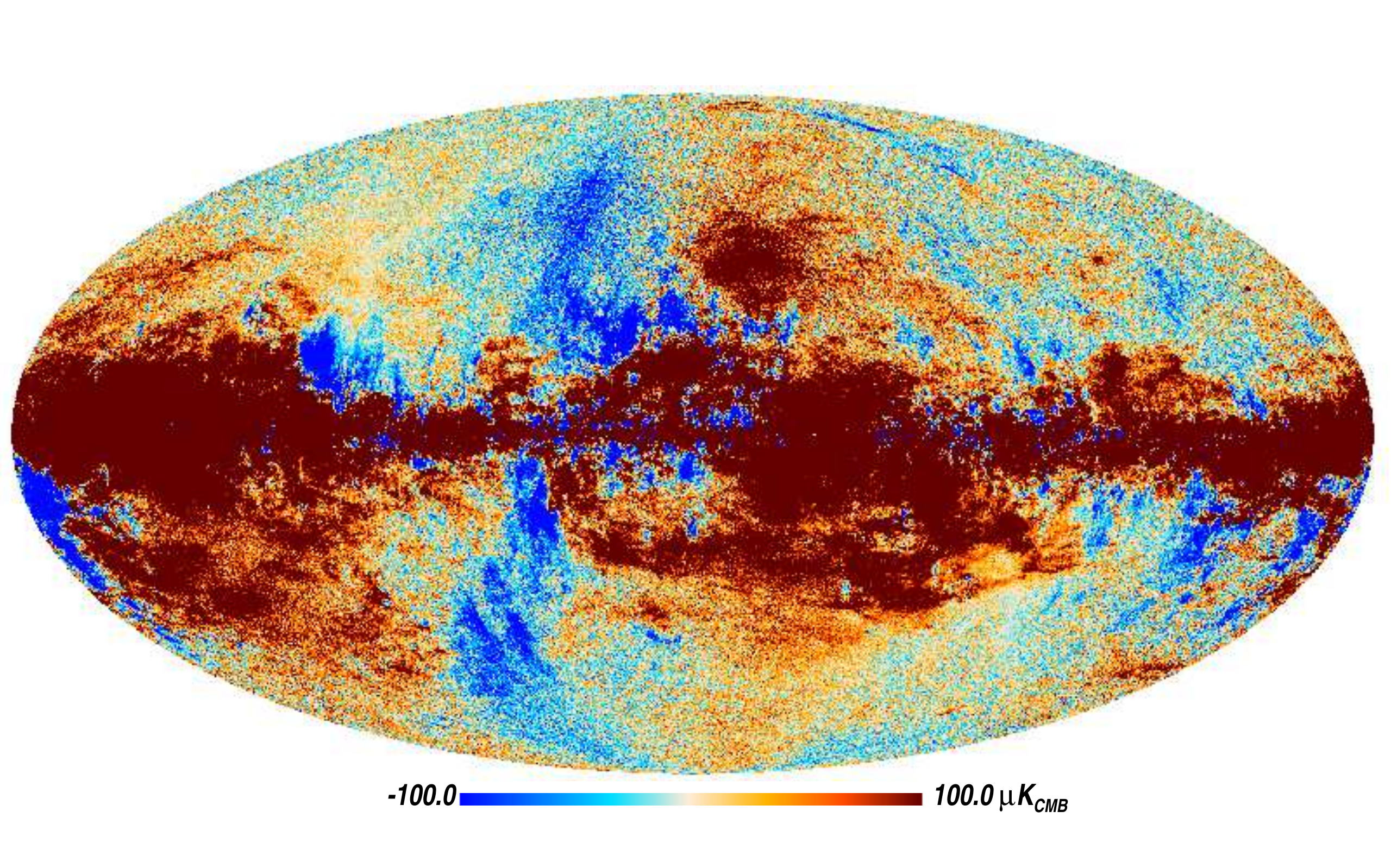}  
	\includegraphics[width=0.49\textwidth]{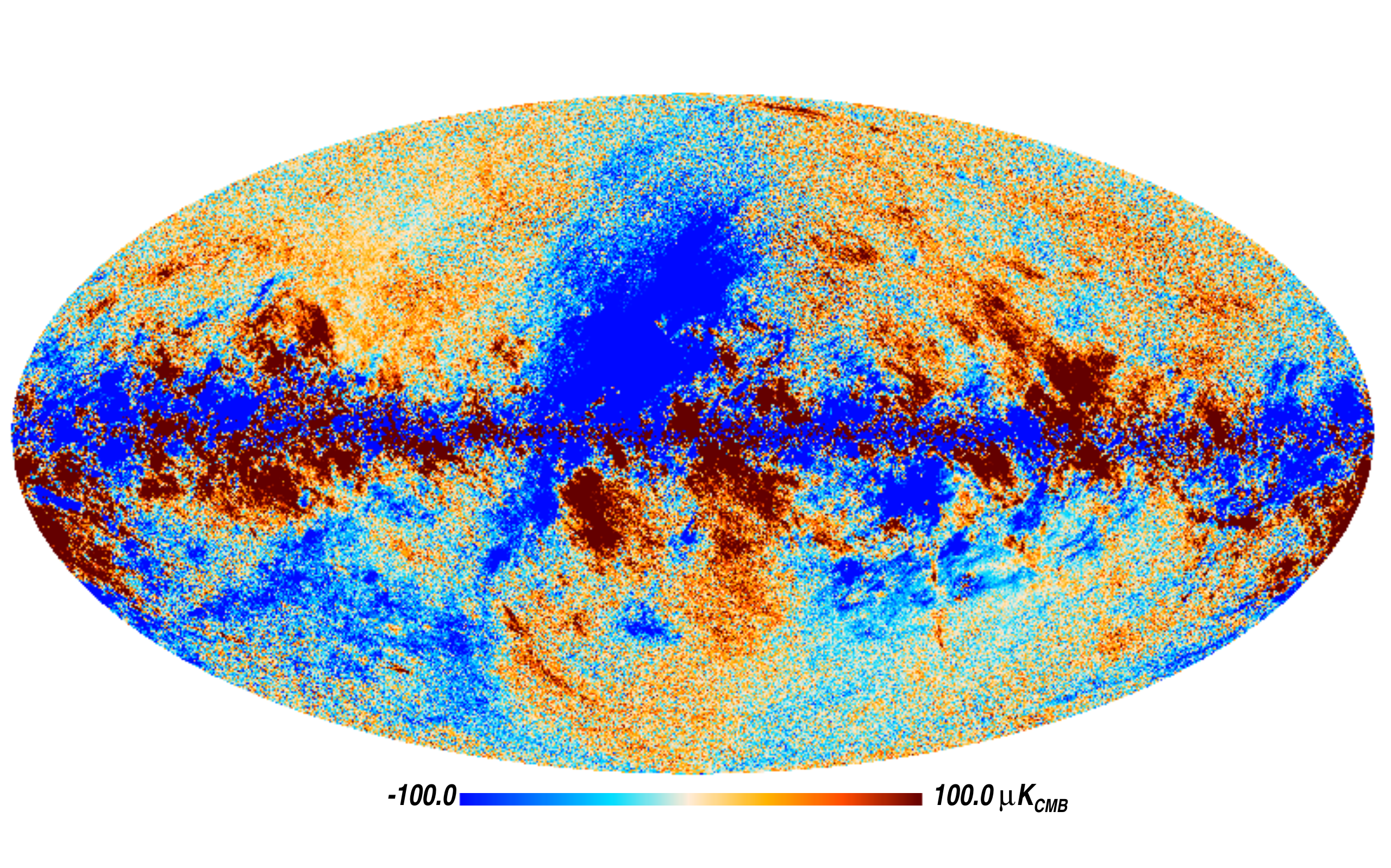}
	\caption{\Planck-HFI full mission $Q$ ({\it left}) and $U$ ({\it right}) polarization maps corrected from bandpass leakage (see Sect.~\ref{sec:bpm}). {\it from top to bottom}: 100\GHz, 143\GHz, 217\GHz, and 353\GHz}
	\label{fig:maps_polar}
\end{figure*}

Frequency maps have been produced using inverse noise weighting. In Figs.~\ref{fig:maps_hfi_low} and \ref{fig:maps_hfi_mm} we show the six intensity frequency maps from 100 to 857\GHz\ at full resolution ($\Nside=2048$). 
Figure~\ref{fig:maps_polar} presents polarization maps at the four first frequencies (100, 143, 217, and 353\GHz), degraded to lower resolution ($\Nside=256$) in order to enhance the signal-to-noise ratio. Those maps have been corrected from bandpass leakage as will be discussed in Sect.~\ref{sec:bpm}.
In both intensity and polarization, we clearly see the emission from the Galactic dust increasing with frequency. 
In intensity, CMB anisotropies are visible at high latitude in the low-frequency channels (between 100 and 217\GHz).
In polarization, the 100\GHz\ maps are contaminated in the Galactic plane by residual CO leakage coming from bandpass mismatch between bolometers.

\subsection{Far sidelobes}
\label{sec:fsls}

As noted in~\cite{planck2014-a08}, far sidelobes (FSLs) affect the response of the instrument to large-scale structure. In addition, the FSLs also affect the HFI calibration. 

At low frequencies, HFI calibrates by fitting to the sinusoidal signal created by the dipole modulated by the \Planck\ circular scanning strategy. As outlined in appendix B of~\citet{planck2013-p01a}, this effectively weights different parts of the beam in general, and the sidelobes in particular, by their angle from the spin-axis. For example, far-sidelobe contributions close to the spin-axis actually affect the calibration very little. Similarly, since we are calibrating with signals that are ``in phase'' with the known phase of the main beam as it scans, the further a sidelobe contribution is in angle {\em around} the spin-axis from the main lobe, the less it contributes to the calibration. So, a sidelobe contribution that is 90\deg\ in scan phase from the main lobe, for example, would not contribute to the HFI calibration, while something close to the main beam would potentially have a large effect. 
The change of the gain due to the far sidelobes is calculated by fitting the dipole to full timeline simulations of the dipole convolved by the FSLs. The factors are 0.09\,\% at 100\GHz, 0.05\,\% at 143\GHz, 0.04\,\% at 217\GHz, and negligible at 353\GHz. The delivered $100$--$217$\GHz\ maps have been scaled by these gain changes. It should be noted that these numbers are uncertain at the $20$--$30\,\%$ level, depending on a multitude of details, such as how the telescope is modelled.

For the planet photometry, some level of knowledge of the amplitude of the FSLs is needed to correctly compare the reconstructed flux with the planet brightness. However, the relative FSL power is lower than 0.3\,\%~\citep{tauber2010} for all HFI frequencies, which is well below the systematic uncertainties of the planet emission models we are using, which are around 5\,\% (see Sect.~\ref{sec:planet}). Therefore FSLs can safely be ignored in the 545 and 857\GHz\ calibration.

\subsection{Zodiacal emission}
\label{sec:zodi}

\begin{figure}[!t]
	\centering
	\includegraphics[width=88mm]{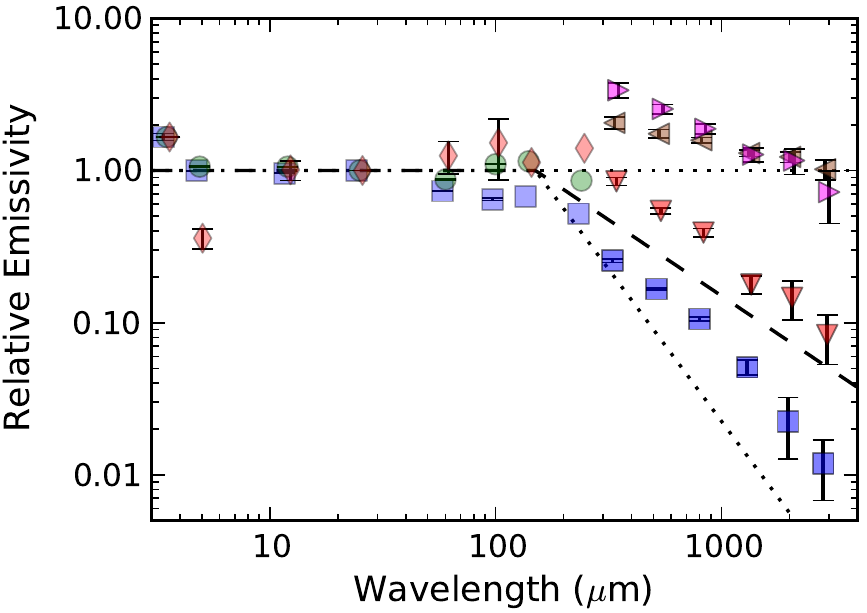}
	\caption{Zodiacal emissivities. Data on the left, at wavelengths shorter than about $300\,\mu$m, are from COBE/DIRBE \citep{kelsall1998}. Data for wavelengths greater than about $300\,\mu$m are from \Planck\ (Table~\ref{tab:zodi_emis}). In both cases, the blue squares represent the emissivity of the Diffuse Cloud. For DIRBE, the red diamonds represent the fitted emissivity for all three IRAS Bands, and the green circles show the values for the Circumsolar Ring and Trailing Blob. For \Planck: the pink, right-pointing triangles are for IRAS Band 3; the brown, left-pointing triangles are for IRAS Band 1; and the red, downward-pointing triangles are for IRAS Band 2. For reference, the lines mark emissivities that are unity at wavelengths less than 250\,$\mu$m, but that are proportional to $\lambda^{-2}$, $\lambda^{-1}$, and $\lambda^{0}$ at longer wavelengths.}
	\label{fig:Zodi_Emis}
\end{figure}

Zodiacal emission is reconstructed and subtracted in the same fashion as that used for the 2013 \Planck\ results~\citep{planck2013-pip88}. The basic procedure for characterizing and removing zodiacal emission from the \Planck\ maps is to:
\begin{itemize}
	\item make frequency maps for each horn and survey as described in previous sections;
	\item make survey difference maps for each horn and year;
	\item find the date ranges over which each $\Nside=256$ pixel was observed, and veto those pixels that were observed over a time-span of more than one week;
	\item use the \COBE\ model \citep{kelsall1998} to recreate the different zodiacal emission components, assuming blackbody emissivities;
	\item fit the components to the survey difference maps for each horn and year to extract the actual emissivities;
	\item use the average of the fitted emissivities to reconstruct the implied zodiacal emission seen during each pointing period, for each horn, and remove these from each detector.
\end{itemize}
The emissivities for each zodiacal component at each of the HFI frequencies are given in Table~\ref{tab:zodi_emis} and are plotted in Fig.~\ref{fig:Zodi_Emis}. As noted in~\cite{planck2013-pip88}, there seems to be a jump between the emissivities for the bands at DIRBE wavelengths and the emissivities of Bands~1 and~3 at \Planck\ wavelengths. This is being investigated, but is assumed to be a consequence of the assumption in the DIRBE analysis that all three bands have the same emissivities, while the \Planck\ analysis allows them to be different. For the \Planck\ cosmological studies this should be irrelevant, since the zodiacal analysis is being used only to remove the interplanetary dust contamination -- the overall amplitudes of the emissivities, which are completely degenerate with the assumed particle density in the bands, are not being interpreted physically.

\begin{table}[!t]
\begingroup
\newdimen\tblskip \tblskip=5pt
\caption{Frequency-averaged zodiacal emissivity values for the Diffuse Cloud and the three {\em IRAS} bands. These are also shown in Fig.~\ref{fig:Zodi_Emis}.}
\label{tab:zodi_emis}
\nointerlineskip
\vskip -3mm
\footnotesize
\setbox\tablebox=\vbox{
   \newdimen\digitwidth 
   \setbox0=\hbox{\rm 0} 
   \digitwidth=\wd0 
   \catcode`*=\active 
   \def*{\kern\digitwidth}
   \newdimen\signwidth 
   \setbox0=\hbox{-} 
   \signwidth=\wd0 
   \catcode`!=\active 
   \def!{\kern\signwidth}
\halign{\hbox to 1.5cm{#\leaderfil}\tabskip 0.5em&\hfil# \tabskip 0.5em&\hfil# \tabskip 0.5em&\hfil# \tabskip 0.5em&\hfil#\tabskip  0pt\cr       
\noalign{\doubleline}
	\omit Frequency \hfil & \omit\hfil Cloud \hfil& \omit\hfil Band 1 \hfil& \omit\hfil Band 2 \hfil& \omit\hfil Band 3 \hfil\cr
	 \omit $[$GHz$]$ &\cr
	\noalign{\vskip 3pt\hrule\vskip 5pt}
857 & 0.256 $\pm$ 0.007 & 2.06 $\pm$ 0.19 & 0.85 $\pm$ 0.05 & 3.37 $\pm$ 0.38\cr
545 & 0.167 $\pm$ 0.002 & 1.74 $\pm$ 0.11 & 0.54 $\pm$ 0.03 & 2.54 $\pm$ 0.18\cr
353 & 0.106 $\pm$ 0.003 & 1.58 $\pm$ 0.07 & 0.39 $\pm$ 0.02 & 1.88 $\pm$ 0.14\cr
217 & 0.051 $\pm$ 0.006 & 1.30 $\pm$ 0.07 & 0.15 $\pm$ 0.02 & 1.27 $\pm$ 0.14\cr
143 & 0.022 $\pm$ 0.010 & 1.23 $\pm$ 0.10 & 0.15 $\pm$ 0.04 & 1.16 $\pm$ 0.22\cr
100 & 0.012 $\pm$ 0.005 & 1.02 $\pm$ 0.16 & 0.08 $\pm$ 0.03 & 0.72 $\pm$ 0.27\cr
\noalign{\vskip 5pt\hrule\vskip 3pt}
}
}
\endPlancktablewide                 
\endgroup
\end{table}

\section{Noise description and subset differences}
\label{sec:noise}

\subsection{Map variance}
\label{sec:mapvariance}

As demonstrated in \citetalias{planck2014-a08}, the noise spectra for the \Planck-HFI bolometers show significant deviation from white noise, resulting in correlations between pixels after map projection. At large scales, the correlations are dominated by low-frequency noise, while at high resolution neighbouring pixels are correlated due to time-response deconvolution and filtering. The \Planck\ 2015 release does not provide a pixel-pixel correlation matrix; only the variance per pixel is given for each delivered map. At first order, the variance maps are proportional to $1/N_{\mathrm{hit}}$, where $N_{\mathrm{hit}}$ is the number of samples per pixel.

The half-difference half-ring maps, projected using the same gain, but destriped independently, are a good representation of the noise variance in the HFI maps. Indeed, in the difference between the first and the second half of a ring, the sky signal vanishes almost perfectly. Moreover, most of the HFI systematics are scan-synchronous and thus also vanish in the difference. 

Table~\ref{tab:noise_in_maps} compares the noise per sample from three estimators: (a) the mean value of the pre-whitened variance map (i.e., scaled using the hit counts); (b) the variance of the pre-whitened half-ring half-difference map; (c) the average of the half-ring map power spectra in the $\ell$ range $100$--$5000$ (see Sect.~\ref{sec:jackknives}). For polarization, the numbers are averages over $Q$ and $U$ for the maps and $E$ and $B$ for the spectra. The different estimators are sensitive to different kinds of systematic effects, such as time-response residuals and signal gradient in pixels. Nevertheless, the three noise estimators give very consistent results.

\begin{table}[h]
\begingroup
\newdimen\tblskip \tblskip=5pt
\caption{Estimation of the noise per sample for intensity ($I$) and polarization ($P$) estimated from: (a) the variance maps; (b) the half-ring difference maps; (c) the pseudo-spectra. Units are \muK$_{\mathrm{CMB}}$ for 100 to 353\GHz, and \MJysr\ for the submm channels.}
\label{tab:noise_in_maps}
\nointerlineskip
\vskip -3mm
\footnotesize
\setbox\tablebox=\vbox{
   \newdimen\digitwidth 
   \setbox0=\hbox{\rm 0} 
   \digitwidth=\wd0 
   \catcode`*=\active 
   \def*{\kern\digitwidth}
   \newdimen\signwidth 
   \setbox0=\hbox{-} 
   \signwidth=\wd0 
   \catcode`!=\active 
   \def!{\kern\signwidth}
\halign{\hbox to 2cm{#\leaderfil}\tabskip 1em&\hfil# \tabskip 1em\hfil &\hfil# \tabskip 1em\hfil&\hfil# \tabskip 0pt\hfil&\hfil#\tabskip  0pt\cr                            
\noalign{\doubleline}
	\omit Frequency\hfil & \omit\hfil Variance maps\hfil & \omit\hfil Diff. maps\hfil & \omit\hfil Pseudo-spectra\hfil\cr
	 \omit $[$GHz$]$\hfil & \omit\hfil (a)\hfil & \omit\hfil (b)\hfil & \omit\hfil (c)\hfil\cr
\noalign{\vskip 3pt\hrule\vskip 5pt}
100I &  1538 &  1531 &  1410\cr
100P &  2346 &  2344 &  2131\cr
143I &   769 &   758 &   759\cr
143P &  1631 &  1618 &  1611\cr
217I &  1105 &  1098 &  1141\cr
217P &  2512 &  2486 &  2440\cr
353I &  3692 &  3459 &  3780\cr
353P & 10615 & 10141 & 10181\cr
545I & 0.612 & 0.619 & 0.779\cr
857I & 0.660 & 0.866 & 0.860\cr
\noalign{\vskip 5pt\hrule\vskip 3pt}
}
}
\endPlancktablewide                 
\endgroup
\end{table}

\begin{figure*}[!h]
	\centering
	\includegraphics[width=\textwidth]{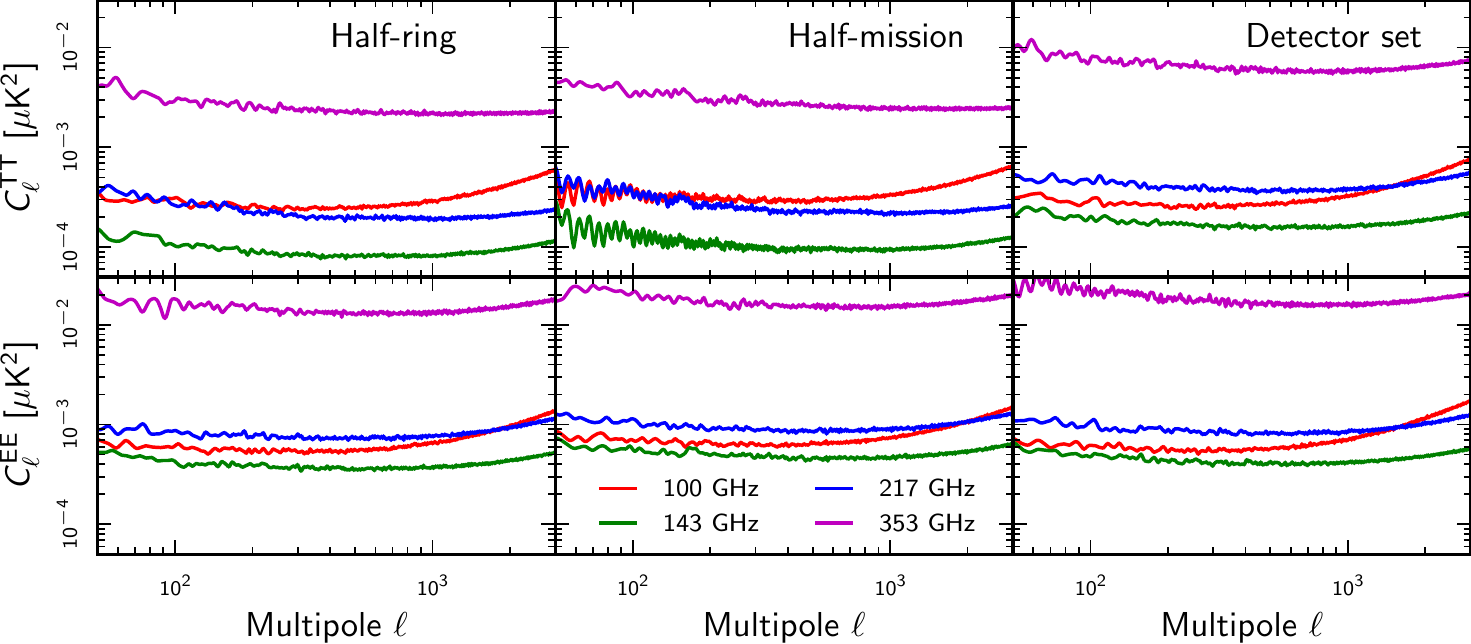}
	\caption{$TT$ and $EE$ power spectra reconstructed from the half-difference between data subset maps for the dipole-calibrated channels.}
	\label{fig:halfmap_spectra}
\end{figure*}

The variance maps ($I$--$I$, $Q$--$Q$, and $U$--$U$) are quite inhomogenous, owing to the \Planck\ scanning (which have negligible wobbling), the relative position of the detectors in the focal plane, and the rejection of some rings or groups of rings (see \citetalias{planck2014-a08}).
Moreover, the typical \healpix\ pixel size is about $1\parcm7$ at $\Nside=2048$ resolution while \Planck\ scans the sky on roughly (but not exactly) ecliptic
meridians separated by 2\parcm5 (near the ecliptic equator). As a
consequence, for single survey maps, lines of empty pixels appear between the scanning trajectories around $(l,b)=(0\deg,\pm45\deg)$.
Even when surveys are combined, inhomogeneities arise from the \healpix\ pixels being elongated parallelograms. The axis of their elongation changes at the boundaries between the 12 primary \healpix\ pixels. In the same regions of the sky and in Galactic coordinates, these elongations are parallel with the scanning trajectories, which induces moir\'e patterns in the coverage maps.

The degree of correlation between the Stokes parameters within each pixel reflects the distribution of the detector orientations, which results from the scanning strategy. The $I$--$Q$ and $I$--$U$ correlations are about 14, 9, 6, and 12\,\% at 100, 143, 217, and 353\GHz. The $Q$--$U$ correlation is about 11, 2, 3, and 8\,\% at 100,143, 217, and 353\GHz). In Appendix~\ref{annex:IQUcorrel} we show the sky distribution of these correlations.

\subsection{Map differences} 
\label{sec:jackknives}

The redundancy of the \Planck\ scanning history and focal plane layout provides numerous ways to check data consistency. We can create differences between maps built using data splits, as described in Sect.~\ref{sec:mapsplit}. In the limit that the signal is the same in each data subset, the difference map should contain only noise. The TOI processing includes several operations that introduce correlations on various time scales; these are discussed below. 

In the dipole-calibrated frequency channels (100--353\GHz), the signal differences are small enough that the data-split map differences can be evaluated at a spectral level, giving insight into the residual systematic errors. For high-frequency channels, we discuss the residuals in the map domain.

Given maps of two subsets of the data, $M_A$ and $M_B$, we construct the half-difference as $\Delta M = \left( M_A - M_B\right)/2$. We compute the power spectrum in temperature and polarization of the half-difference, masking the sky with the \Planck\ point source mask as well as the galaxy masks used in \cite{planck2014-a13}, i.e., leaving 65, 59, 48, and 32\,\% of the sky unmasked at 100, 143, 217, and 353\GHz, respectively.

In order to use this half-difference map to assess noise in the full maps, we account for widely varying integration time in the two subsets using a pixel-by-pixel weight map, which is multiplied by the half-difference map $\Delta M$ prior to computing the angular power spectrum. The weight is constructed as $W = 2 / \sqrt{\left(1/n_A + 1/n_B \right) \left( n_A + n_B \right)}$ where $n_A$ is the hit count map for $M_A$ and $n_B$ is the hit count map for $M_B$. In the limit that the half-difference map consists entirely of white noise, this exactly accounts for the differences in the hit counts. 
The $TT$ and $EE$ spectra of the difference maps are plotted in Fig.~\ref{fig:halfmap_spectra} and are described in the sections below. The $BB$ spectra are nearly the same as the $EE$ spectra and are not shown.

\subsubsection{Half-ring map differences} 
The half-ring difference is sensitive to high-frequency noise, since most low-frequency modes (on time scales longer than 1\,h) are common to both data sets and thus vanish. In the harmonic domain, the noise is nearly white with an amplitude compatible with the noise estimated in the map domain (see Sect.~\ref{sec:mapvariance}). At large multipoles, the noise blows up due to the time transfer function deconvolution, before being cut off by the low-pass filter. At lower multipoles, half-ring differences show low-frequency noise residuals due to the destriping. Indeed, the destriping is performed independently for each half, or essentially half the data are used to solve the offsets for the full ring maps. The residuals from the offset determination are therefore expected to be twice as large as in the full-mission map.

In addition, the deglitching operation performed during the TOI processing uses the full data set to estimate the signal in each ring, thereby introducing some correlation between the two halves of each ring. Taking the difference between the two half-rings in fact removes the correlated portion of the noise at the few percent level.

\subsubsection{Half-mission map differences} 
With half-mission differences, we can check for long-time-scale variations and for apparent gain variation with time due to ADC nonlinearities. 
Moreover, due to slightly shifted pointing between the first and second halves of the mission, the effect of a signal gradient within a pixel (especially on the Galactic plane where the signal is strong) is larger than for the half-ring map differences.

Because the number of observations in a given pixel can be very different between the two half-mission maps, using a weighting as described above is essential. Including the weighting, the half-mission differences give a power that is 10--20\,\% higher than the corresponding half-ring difference. This fraction of additional power is nearly the same in all the channels 100--353\GHz, and is the same in both temperature and polarization. 
We understand this small additional power to be due to effects from the TOI processing that introduce correlations in the noise between the subsets. The half-ring maps, as stated above, have correlations introduced by the deglitcher that are subsequently removed by the differencing. These correlations are not present between the half-mission data sets, so their difference shows a higher noise power.

\subsubsection{Detector-set map differences}
This difference probes systematic effects that are bolometer-dependent. Note that, in the case of 143, 217, and 353\GHz, the detector-set split excludes the unpolarized detectors, and the noise in $TT$ is correspondingly higher than in the half-mission and half-ring split. The 100\GHz\ channel has only polarization-sensitive bolometers and the $TT$ spectrum of the difference is much closer to the spectrum seen for the other data splits.

There are several other effects that make the power spectrum of the detector-set difference stand apart. A unique time response function is deconvolved from each bolometer. In the half-ring and half-mission data splits, the deconvolved function is identical between the two halves. However with the detector-sets, the time response is in general slightly different between the two halves. This effect leads to a tilt in the spectrum of the detector-set difference maps relative to the half-ring or half-mission split.
Moreover, at 353\GHz, signal residuals are larger due to relative calibration uncertainties between detectors.

\subsubsection{Map differences at low-$\ell$}

\begin{figure}[!t]
	\centering
	\includegraphics[width=\columnwidth]{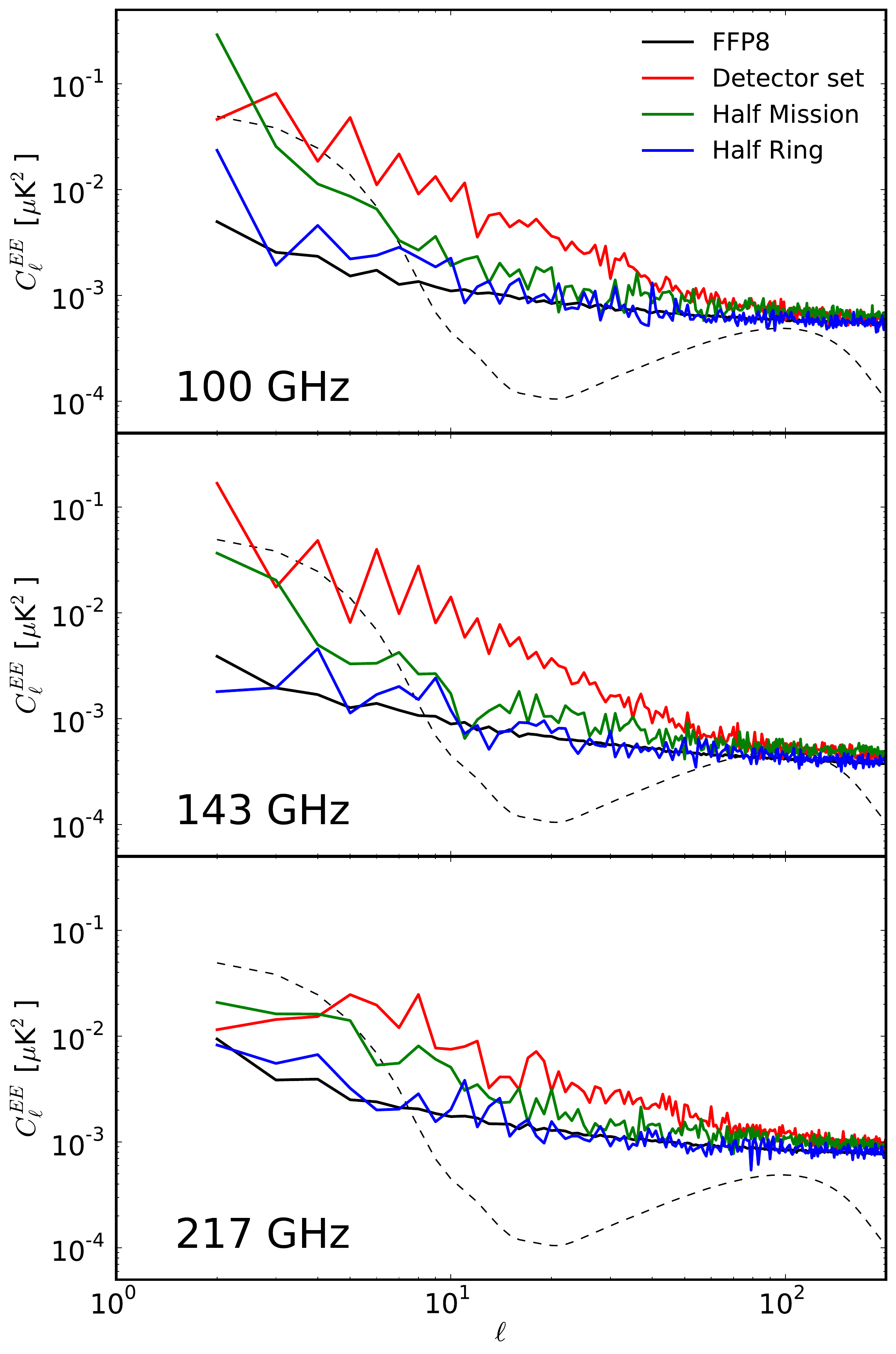}
	\caption{$EE$ power spectra reconstructed from the half-difference between data subset maps for the dipole-calibrated channels at low multipoles compared to the noise estimation from the FFP8 simulations. CMB signal from \Planck~2015 is plotted in dashed lines.}
	\label{fig:jack_lowl}
\end{figure}

At low multipoles, despite the huge progress in the control of the systematics, data are still contaminated by systematic residuals. Figure~\ref{fig:jack_lowl} shows the $EE$ power spectra from the half-difference maps at 100, 143, and 217\GHz\ and compared to the noise power spectrum from FPP8 simulations.  The half-ring differences are compatible with noise while, at multipoles typically lower than 50, detector-set and half-mission differences are dominated by excess power which is larger than the $EE$ CMB signal. The origin of the excess power will be explored in a forthcoming publication.

\subsubsection{High frequency channels}
\begin{figure}[!t]
	\centering
	\includegraphics[width=\columnwidth]{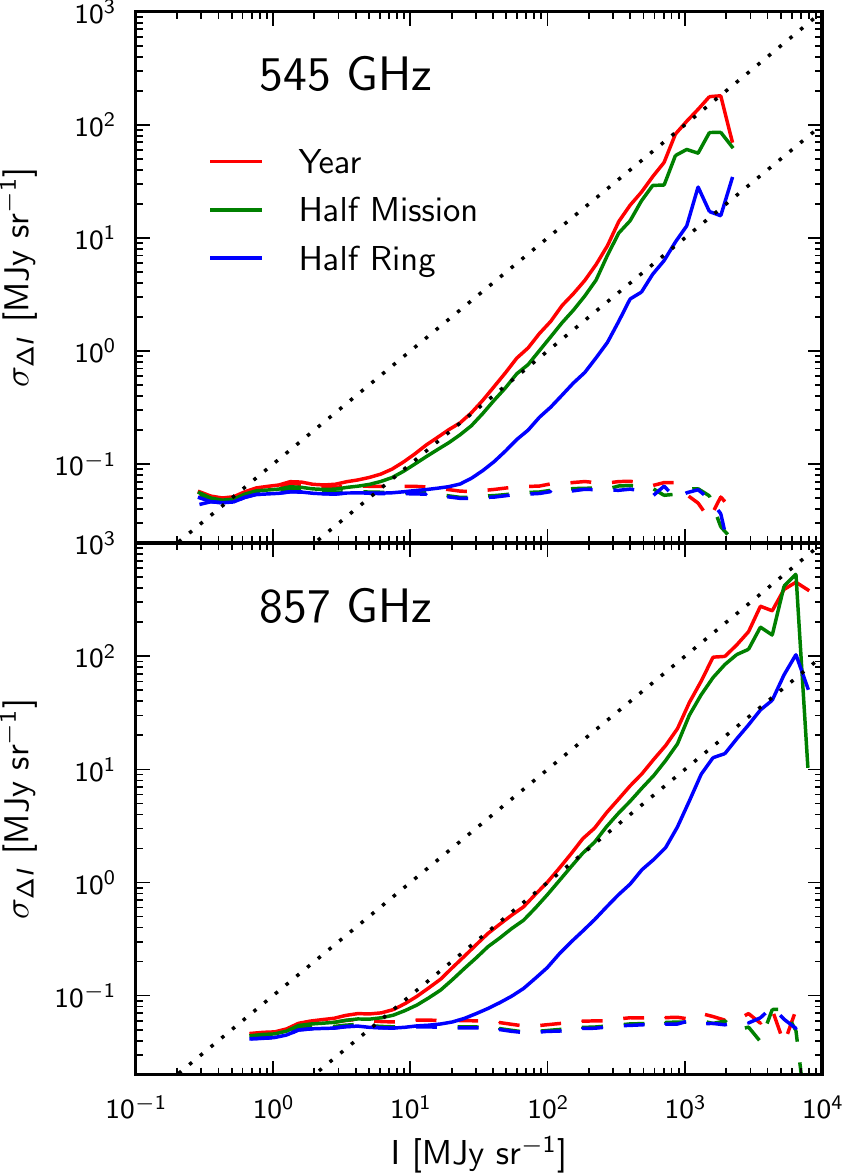}
	\caption{Rms of the residual signal in difference maps at 545 and 857\GHz, as a function of signal level in the full map. The solid coloured curves show the \rms\ of the data, while dashed coloured curves show the \rms\ of a simulated noise map. The diagonal dotted lines indicate 1\,\% and 10\,\% of the signal.}
	\label{fig:submmdifferencerms}
\end{figure}

For the highest frequency channels (545 and 857\GHz) the data-split map differences are dominated by residual signal.
Figure~\ref{fig:submmdifferencerms} shows the rms of the differences of intensity maps at 545 and 857\GHz\ for half-ring, half-mission, and year data splits compared to the same data split performed on a simulated noise map. At low signal, the difference is consistent with instrumental noise. At high signal levels, an additional residual appears in the difference map that is roughly proportional to the signal level. Part of this is due to pointing errors. For year and half-mission, the effect is enhanced by the combination of residual gain variations and the relative difference of pointing between the two splits. Over most of the sky, the signal is reproducible to better than 1\,\% for these frequencies.
Bolometer map differences (not shown here) are, in addition, sensitive to the relative calibration error.

\subsection{Noise cross-correlation}

\begin{figure*}[htbp]
	\centering
	\includegraphics[width=0.45\textwidth]{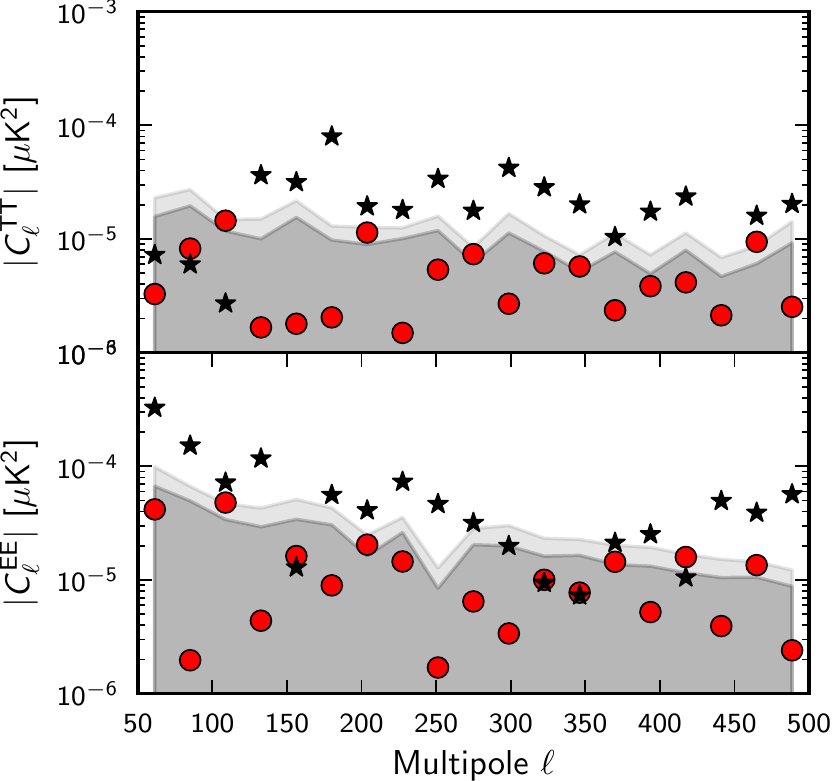}
	\includegraphics[width=0.45\textwidth]{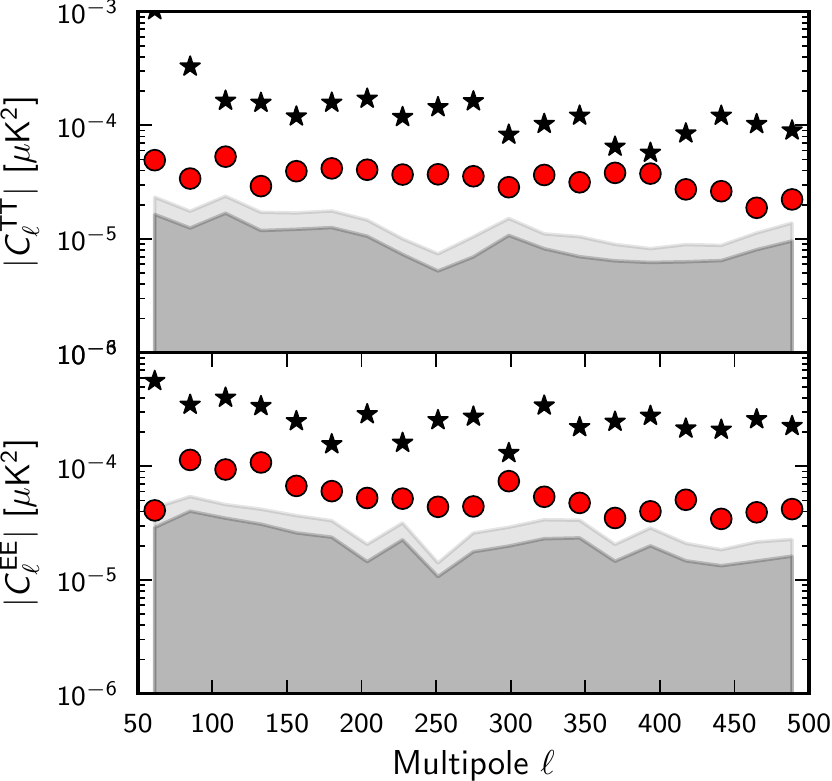}
	\caption{100\GHz\ difference map cross spectra. {\it Left}: half-mission (HM) correlation of half-ring differences (HR). {\it Right}: half-ring (HR) correlation of half-mission difference (HM). The real data are red dots. The end-to-end simulation are black stars. One and two sigma contours from ten FFP8 noise realizations are shaded grey.}
	\label{fig:100correlation_spectra}
\end{figure*}

\begin{figure*}[htbp]
	\centering
	\includegraphics[width=0.45\textwidth]{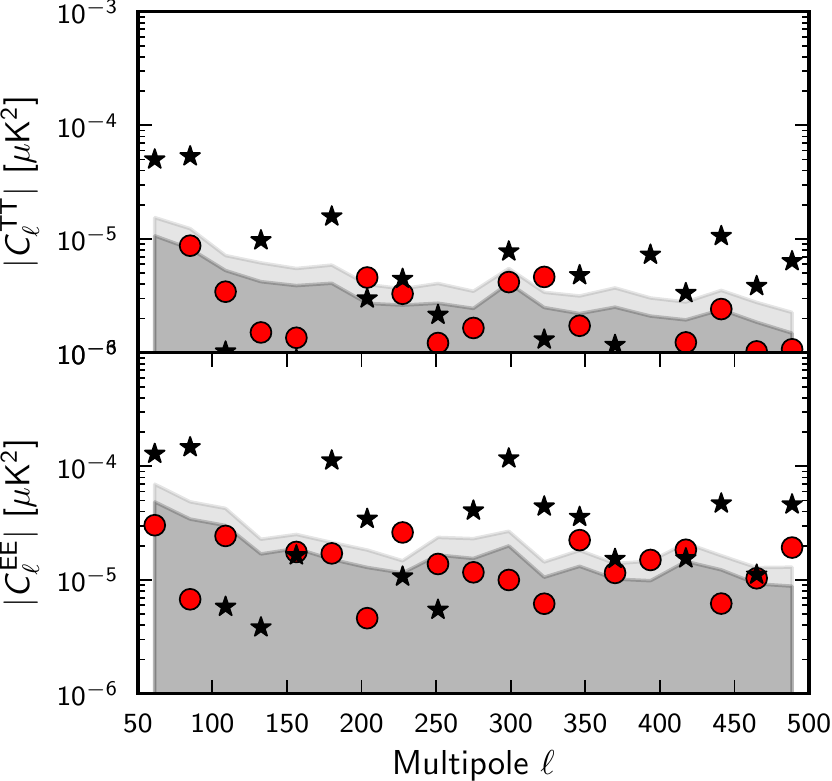}
	\includegraphics[width=0.45\textwidth]{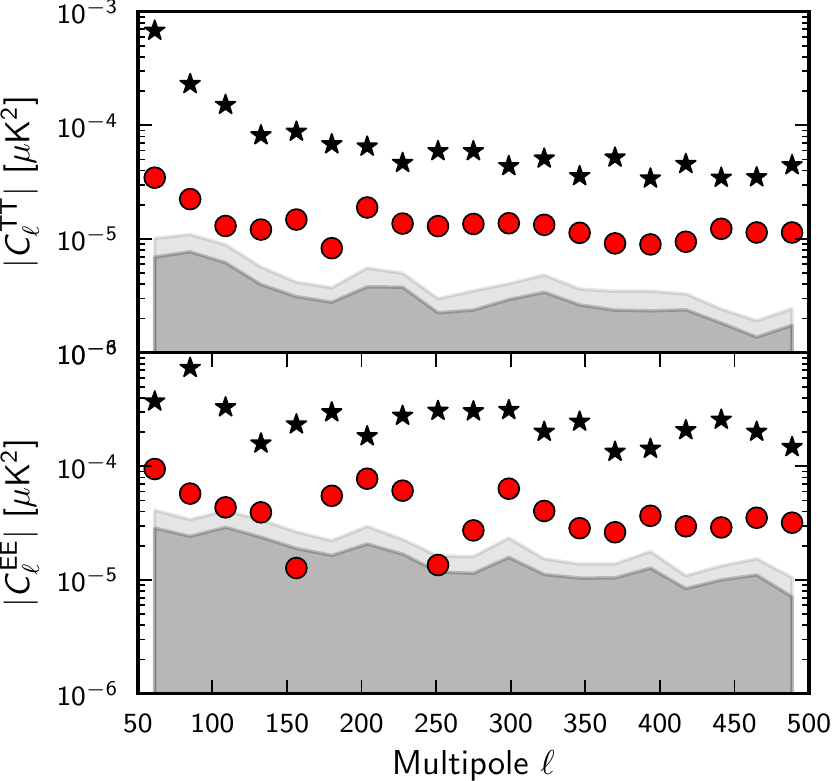}
	\caption{Same as Fig.~\ref{fig:100correlation_spectra} for 143\GHz\ difference map cross spectra.}
	\label{fig:143correlation_spectra}
\end{figure*}

\begin{figure*}[htbp]
	\centering
	\includegraphics[width=0.45\textwidth]{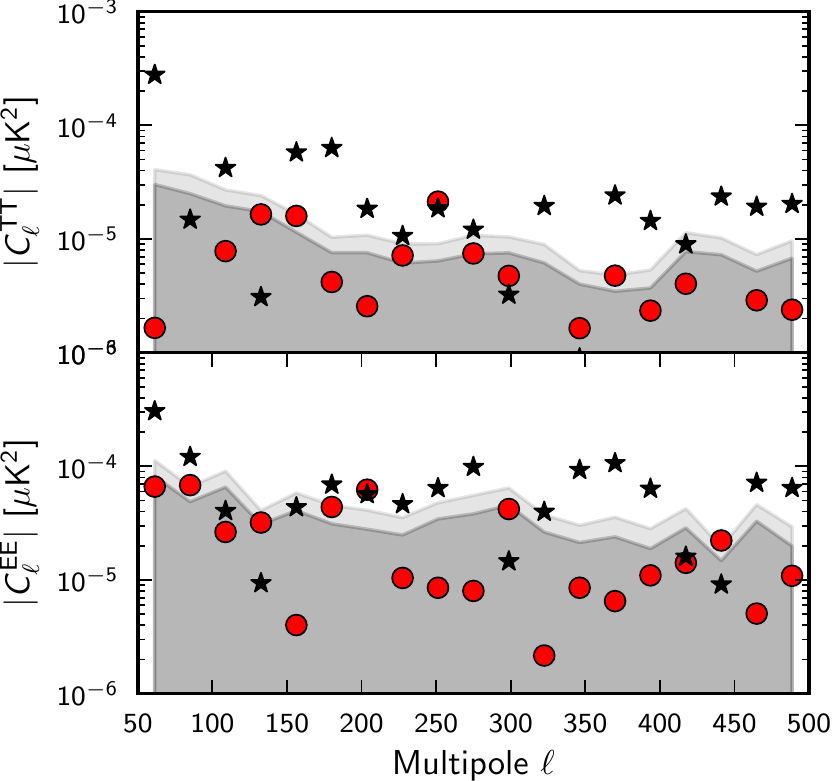}
	\includegraphics[width=0.45\textwidth]{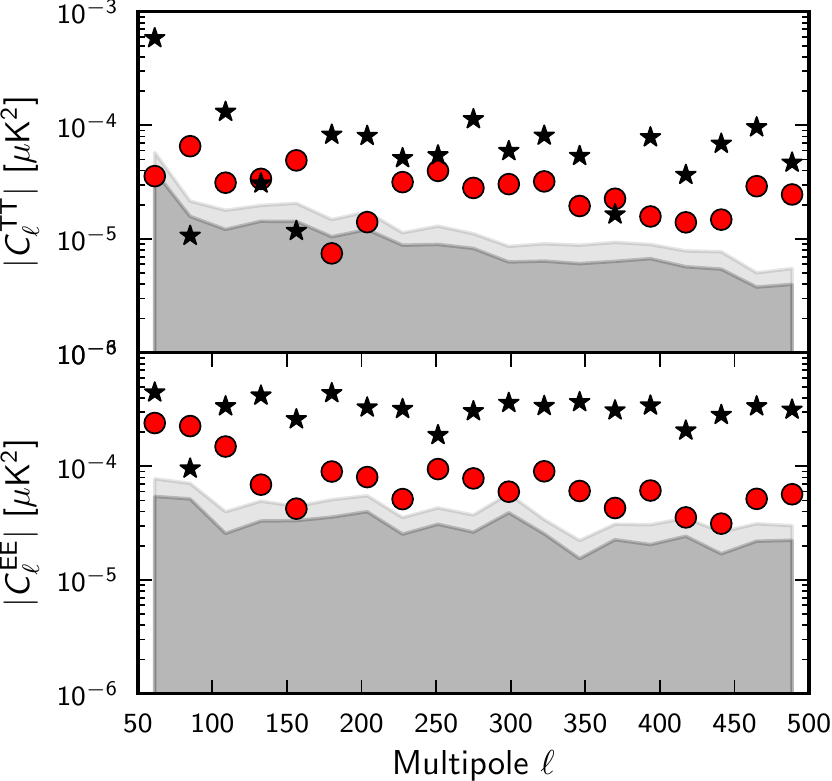}
	\caption{Same as Fig.~\ref{fig:100correlation_spectra} for 217\GHz\ difference map cross spectra.}
	\label{fig:217correlation_spectra}
\end{figure*}

Here we check for correlations in the noise by computing cross-spectra between the difference maps described earlier. We look at 100 (Fig.~\ref{fig:100correlation_spectra}), 143 (Fig.~\ref{fig:143correlation_spectra}), and 217\GHz\ (Fig.~\ref{fig:217correlation_spectra}) in comparison with the expectations from projecting noise realizations on the sky (using the FFP8 noise realizations described in~\citealt{planck2014-a14} and the end-to-end simulations described in \citetalias{planck2014-a08}).

When the half-mission cross-spectra of half-ring differences are computed, the results are roughly consistent with the FFP8 noise simulations. At 143\GHz\ in temperature, the end-to-end simulation produces a slight rise in power at low multipoles that is not seen in the data.

Large correlations are seen in the half-ring cross-spectra of half-mission differences. These are at least partially induced by our processing since the
end-to-end simulations also show correlations that are not as large in amplitude as those seen in the data, but show a similar spectral shape. These correlations are mainly due to the deglitcher, as described above.

\subsection{Temperature-polarization cross-variance}

\begin{figure}[htbp]
	\centering
	\includegraphics[width=.45\textwidth]{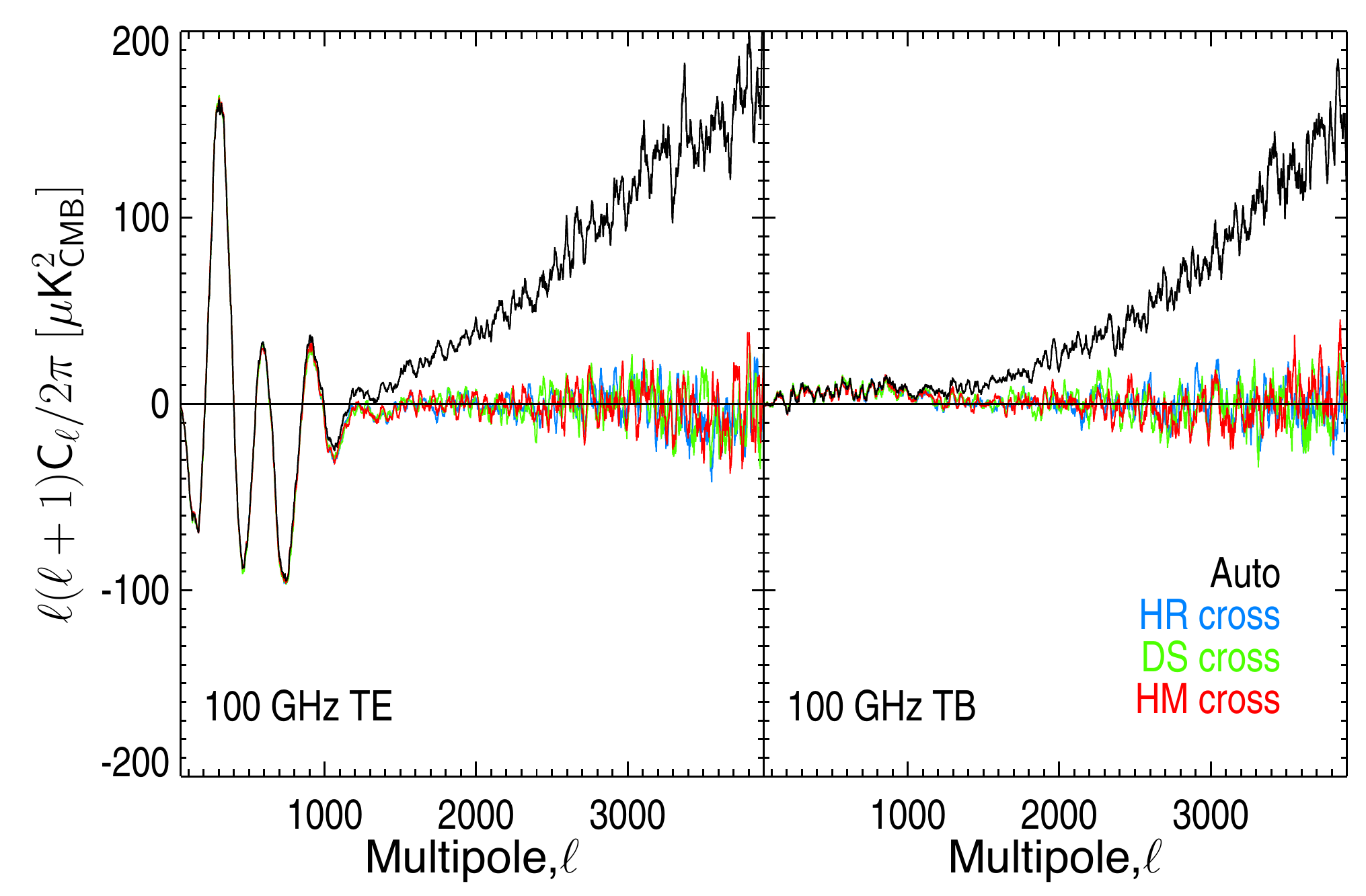}\\
	\includegraphics[width=.45\textwidth]{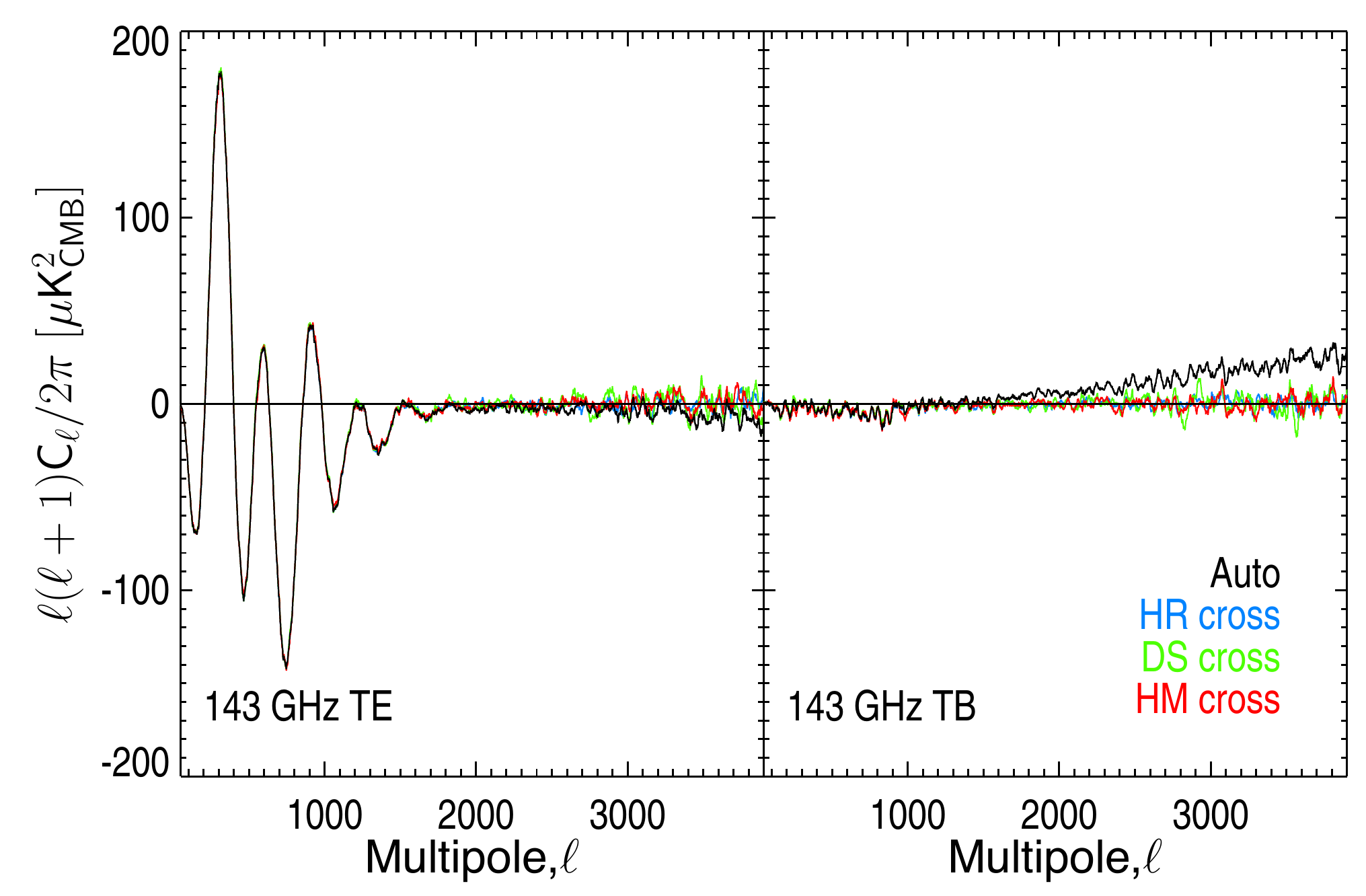}\\
	\includegraphics[width=.45\textwidth]{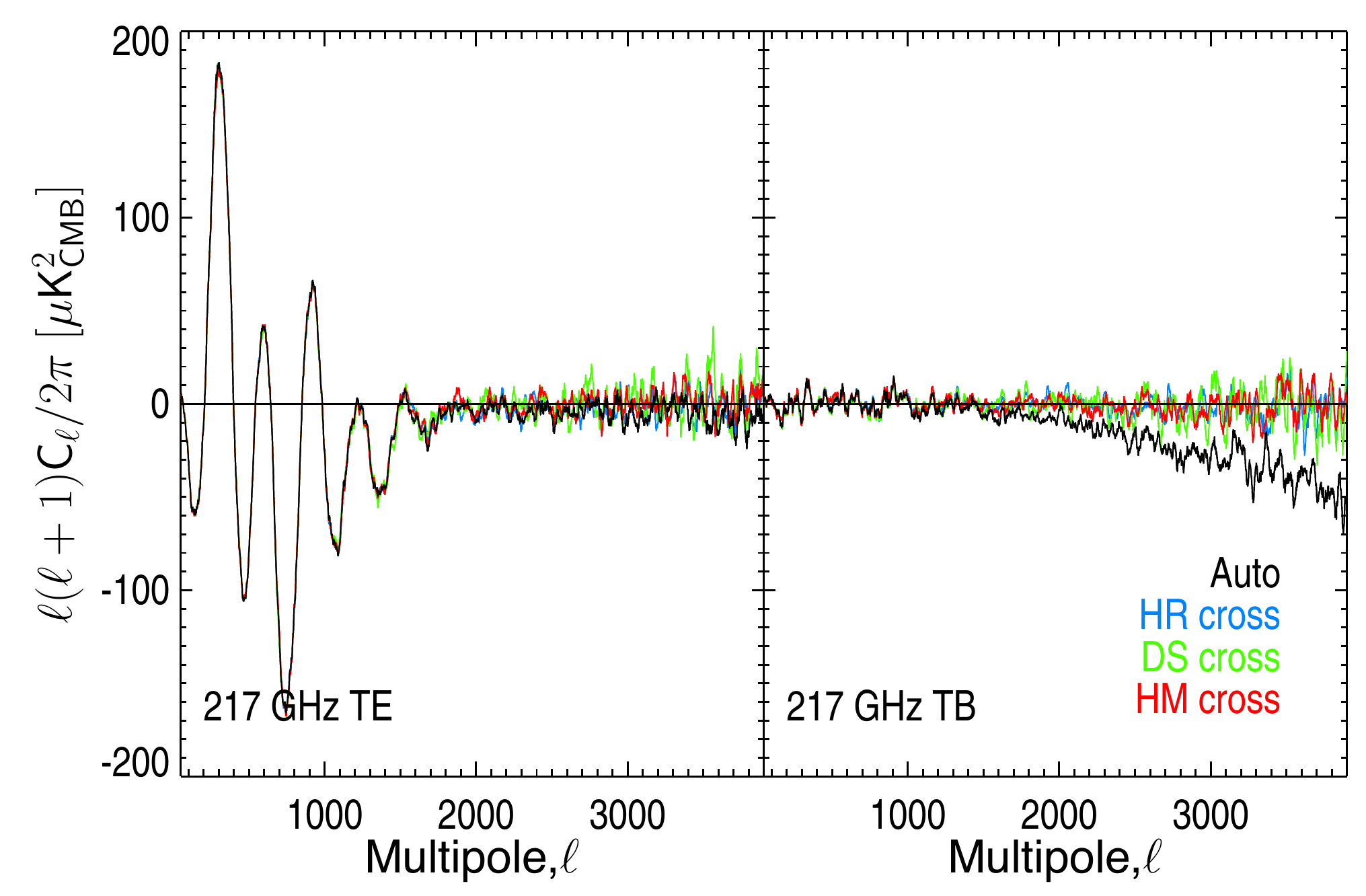}\\
	\includegraphics[width=.45\textwidth]{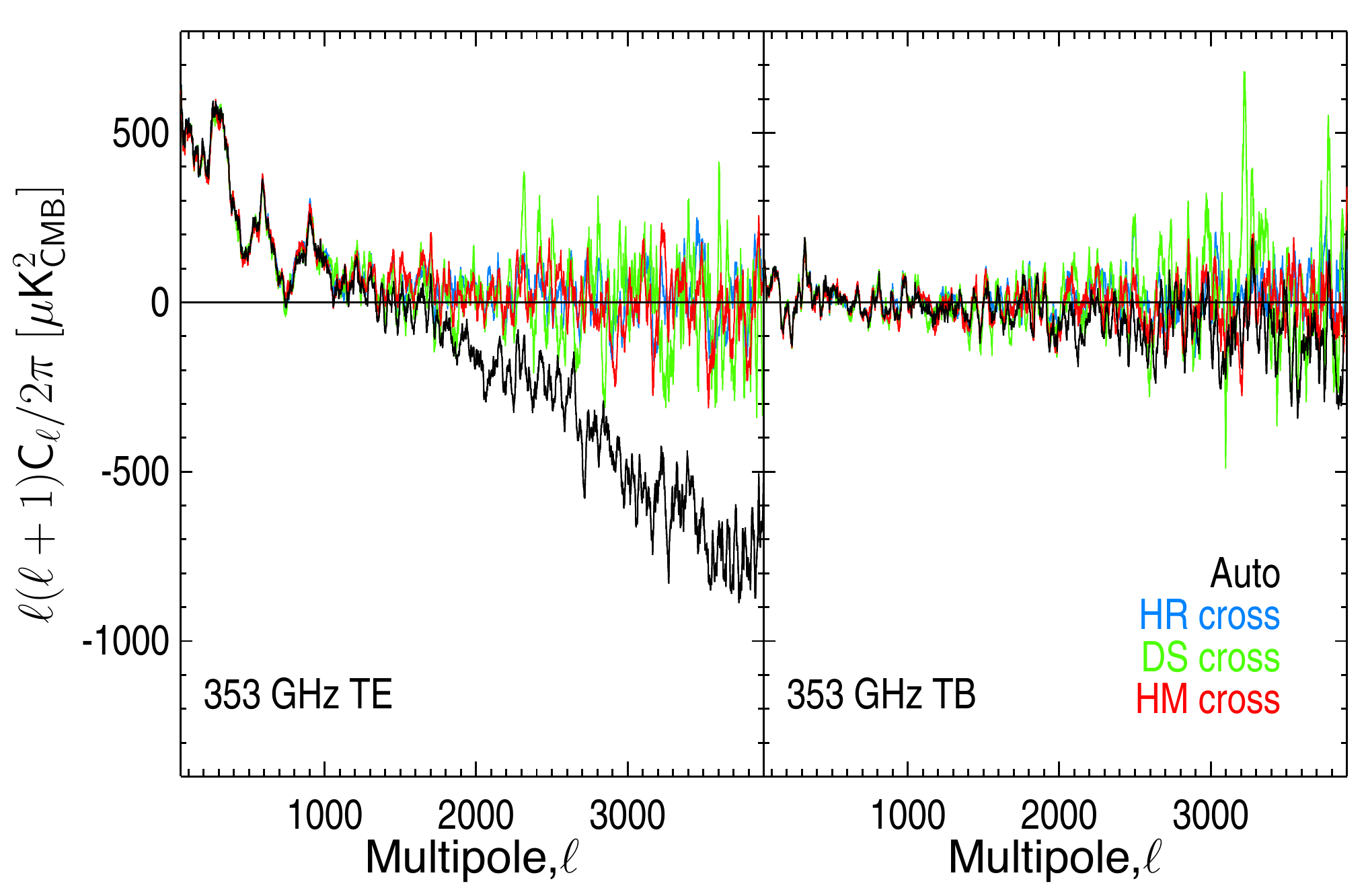}
	\caption{Pseudo-power spectra for $TE$ ({\emph left}) and $TB$ ({\emph right}) for each frequency (\emph{from top to bottom}: 100, 143, 217, and 353\GHz). The auto-spectra are shown in black. Cross-spectra of half-ring (HR), half-mission (HM), and detector-set (DS) half-differences are shown in blue, red, and green, respectively. A Galaxy and point source mask, leaving 40\,\% of the sky, was used in all cases.}
	\label{fig:T-P_correlation}
\end{figure}

In absence of spatial correlations, noise correlations between temperature $I$ and polarized $Q$ and $U$ modes vanish in the harmonic domain, thanks to the orthogonality of the spherical harmonic decomposition. Consequently the $TE$ and $TB$ auto-spectra are not biased by noise in the way that the $TT$, $EE$, and $BB$ spectra are. In practice, transfer function deconvolution, filtering, and pixelization effects can produce spatial correlations at high multipoles, resulting in a noise bias that is observed in the $TE$ and $TB$ angular power spectra. In Fig.~\ref{fig:T-P_correlation} we compare the auto and cross-spectra for the half-ring, half-mission, and detector set splits. These pseudo-spectra have been built by masking Galactic emission and point sources (approximately 40\,\% of the sky). The auto-spectra are biased at high multipoles (starting at $\ell\approx 1500$). The amplitude of this bias and its sign depend on the frequency and on the mode considered. Nevertheless, none of the cross-spectra show significant departures from the null expectation.

We observe that the amplitude of the noise bias in auto-spectra is mitigated when adding more independent data sets, such as detectors or surveys (survey maps show larger amplitude than half-mission and full-mission). These results are fully reproduced in the FFP8 simulations.

\subsection{Subset map cross-covariance}
\label{sec:detset-correlation}
When mapping subsets of the available data (selecting detectors and/or time spans) we have a choice between solving for independent baseline offsets for the subset in question or reusing full-mission, full-frequency baselines (as in the 2015 HFI map release). Full-mission baselines are more accurate, leaving less large-scale noise in the maps, but introduce noise correlation between detector-set maps.

\begin{figure}[htbp]
	\centering
	\includegraphics[width=\columnwidth]{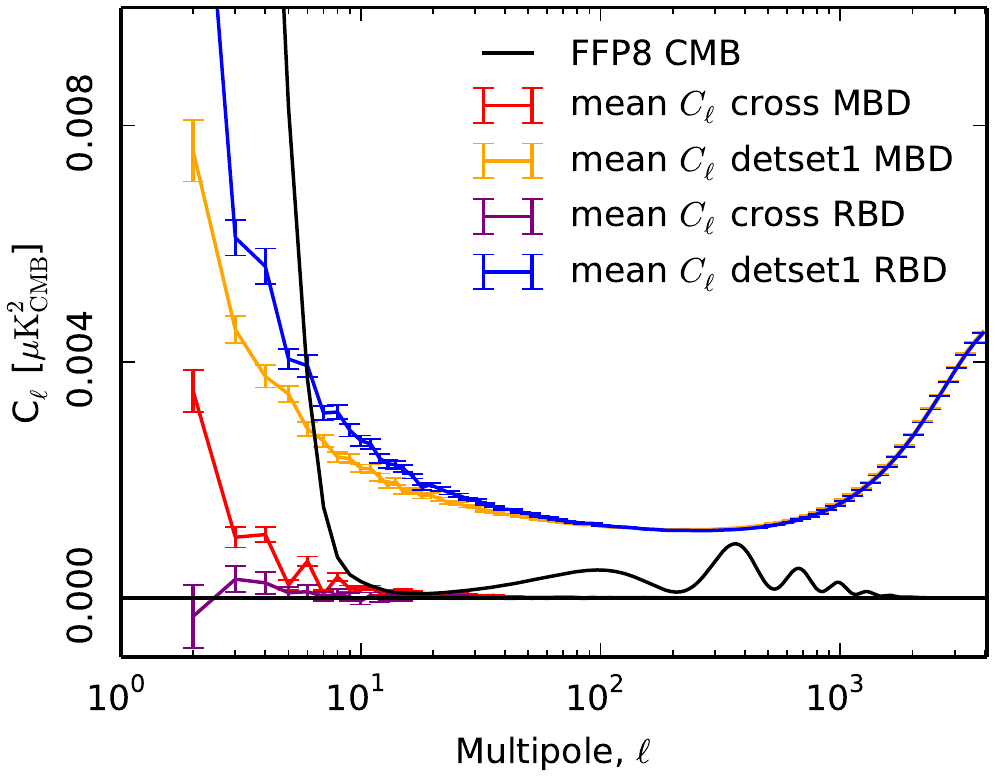}
	\caption{Average $EE$ spectra obtained from 100 simulations of detector-set noise maps. The latter have been produced using either the full-mission baseline destriping (MBD) or the destriping run on each subset independently (run baseline destriping or RBD). Both the increased low-$\ell$ noise in the RBD case and the increased cross-spectrum noise bias in the MBD case are apparent.} 
	\label{fig:subsetbias}
\end{figure}

We can measure the resulting bias in cross-spectra through noise simulations. Comparison between noise spectra from a Monte Carlo analysis at 100\GHz\ using both methods of destriping is shown in Fig.~\ref{fig:subsetbias} for the case of detector sets. Noise spectra for each detector set show more large-scale power when using independent baselines than in the full mission case. On the other hand, the cross-power spectrum is biased by up to a few times $10^{-3}\mu$K$^2$ below $\ell=10$; the same is true for the half-mission subset.

\section{Systematic effects}
\label{sec:systematics}

We now describe the major systematic effects that could potentially affect the maps: the gain variations; errors in the absolute gain determination; errors in the polarization efficiency and orientation; and, most of all, the detector-to-detector gain mismatch. The latter includes bandpass mismatch (which affects the response to foregrounds of the detectors at the same frequency) and relative gain uncertainties, both of which create intensity-to-polarization leakage. All these effects are constrained using tests involving the combination of maps, residuals in maps, cross-power spectra, and dedicated simulations.

\subsection{Gain stability}
\label{sec:bogopix}

Gain stability has been significantly improved with respect to the \Planck\ 2013 release. This is mainly due to the ADC correction, combined with the new determination of the time transfer function (see \citetalias{planck2014-a08}). The amplitude of the apparent gain variation has been improved from 1--2\,\% to less than 0.5\,\% for all cases. Residual gain variations are compatible with zero when including the correction for the long-time-constant residuals, as discussed in Sect.~\ref{sec:vltc}.

We check the stability of the gain over time using the same tool as in \cite{planck2013-p03f}, called \bogopix. For each bolometer, the code fits simultaneously for the gain $g_r$ and the offsets $o_r$ for each ring, marginalizing over the sky signal $T$:
\begin{equation}
	P_t = g_r \times(A_{tp} \cdot T_p + t_{\mathrm{orb}}) + o_r + n_t \,.
\label{eq:bogopix}
\end{equation}
Given the low amplitude of the observed gain variations (less than 0.5\,\%), we linearize Eq.~\eqref{eq:bogopix} and solve by iteration \citep[see][]{planck2013-p03f}; one or two iterations are sufficient to
ensure convergence. To initialize the iterations, we start from the constant gain solution $G$ (see Sect.~\ref{sec:orbdipcal}).

We compute the gain variations from single-bolometer data (neglecting polarization). Polarized signals will affect the gain determination. To reduce this bias, we ignore sky regions where the polarized emission is the strongest, which lie mostly in the Galactic plane.

\begin{figure}[!htbp]
	\centering
	\includegraphics[width=\columnwidth]{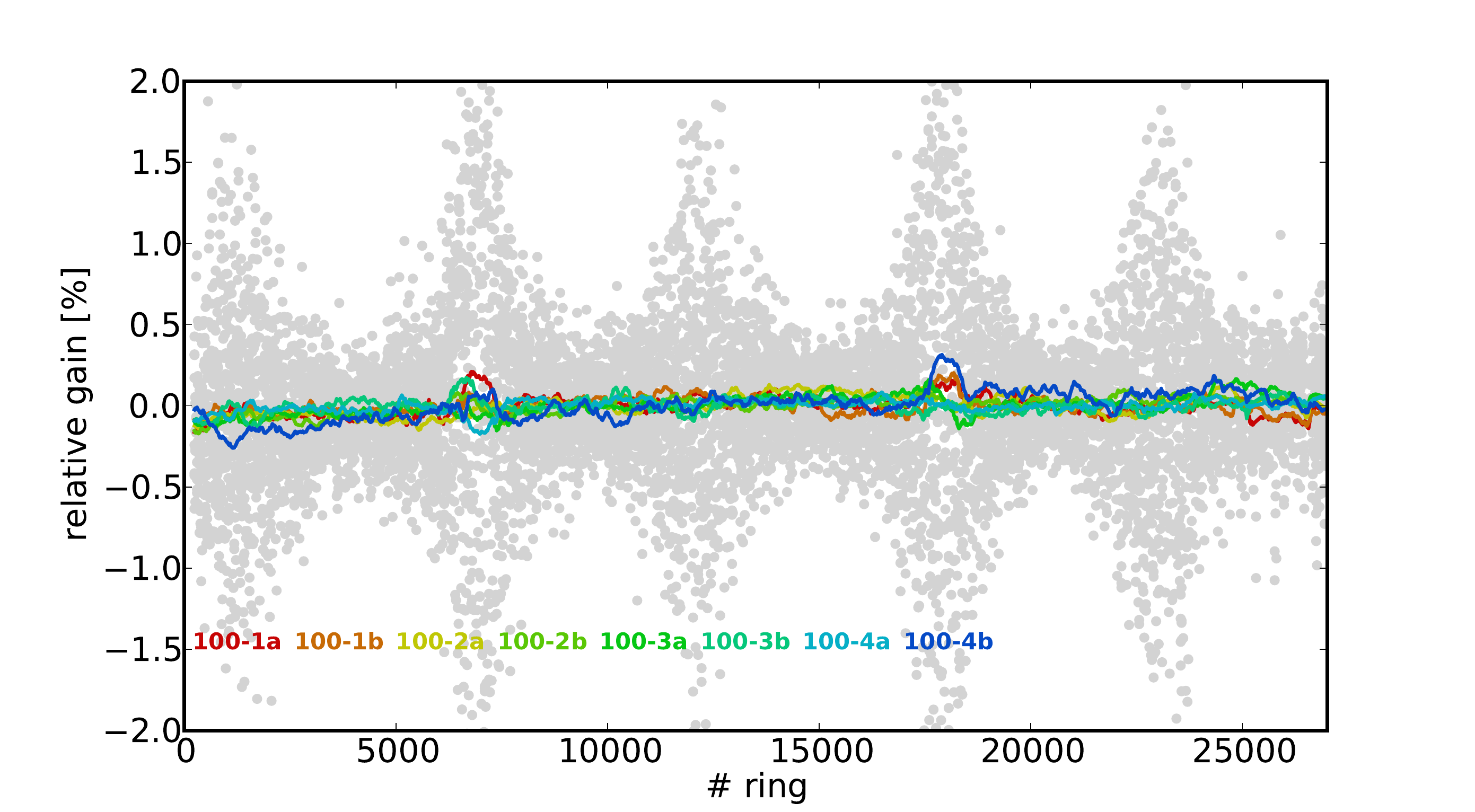}\\
	\includegraphics[width=\columnwidth]{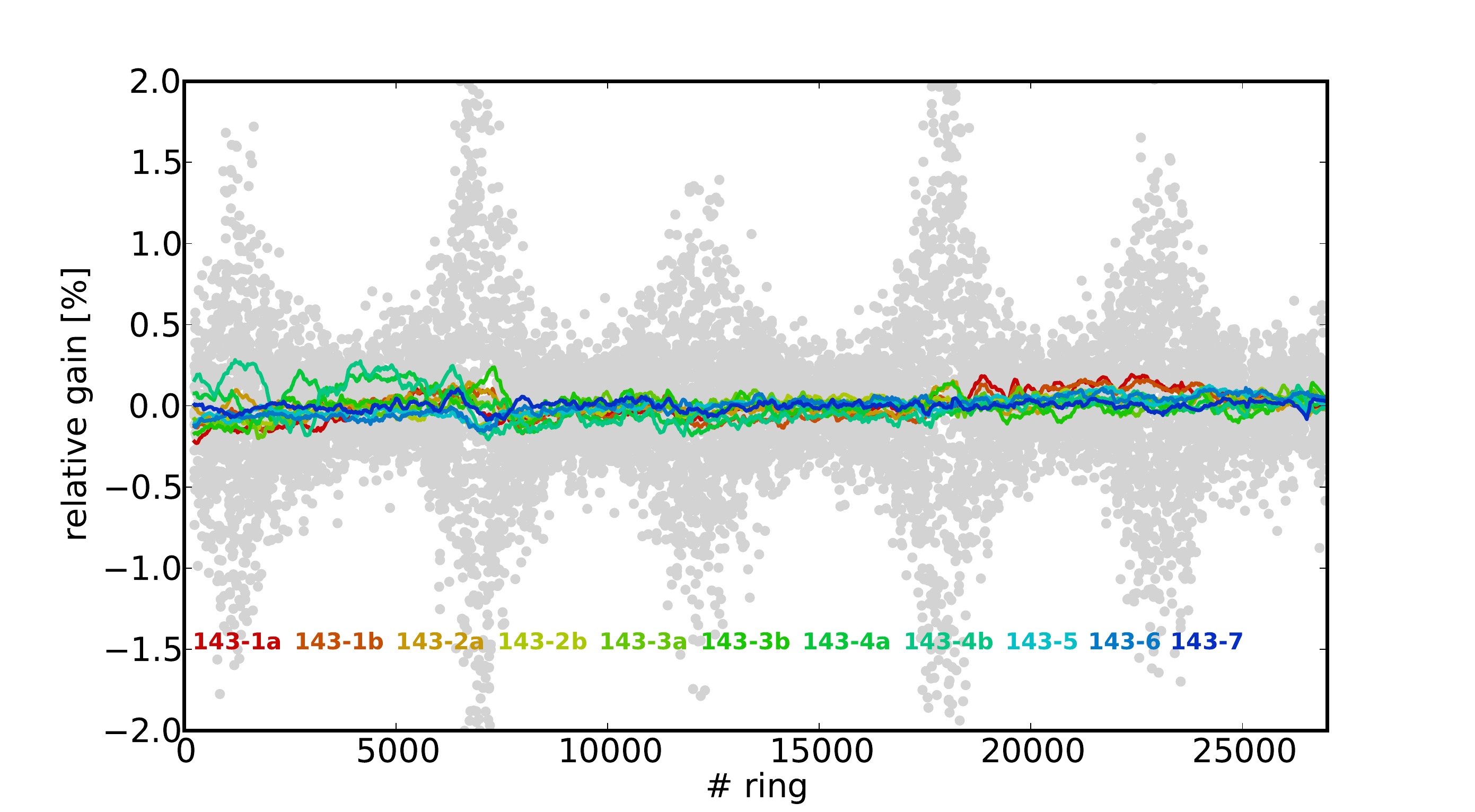}\\
	\includegraphics[width=\columnwidth]{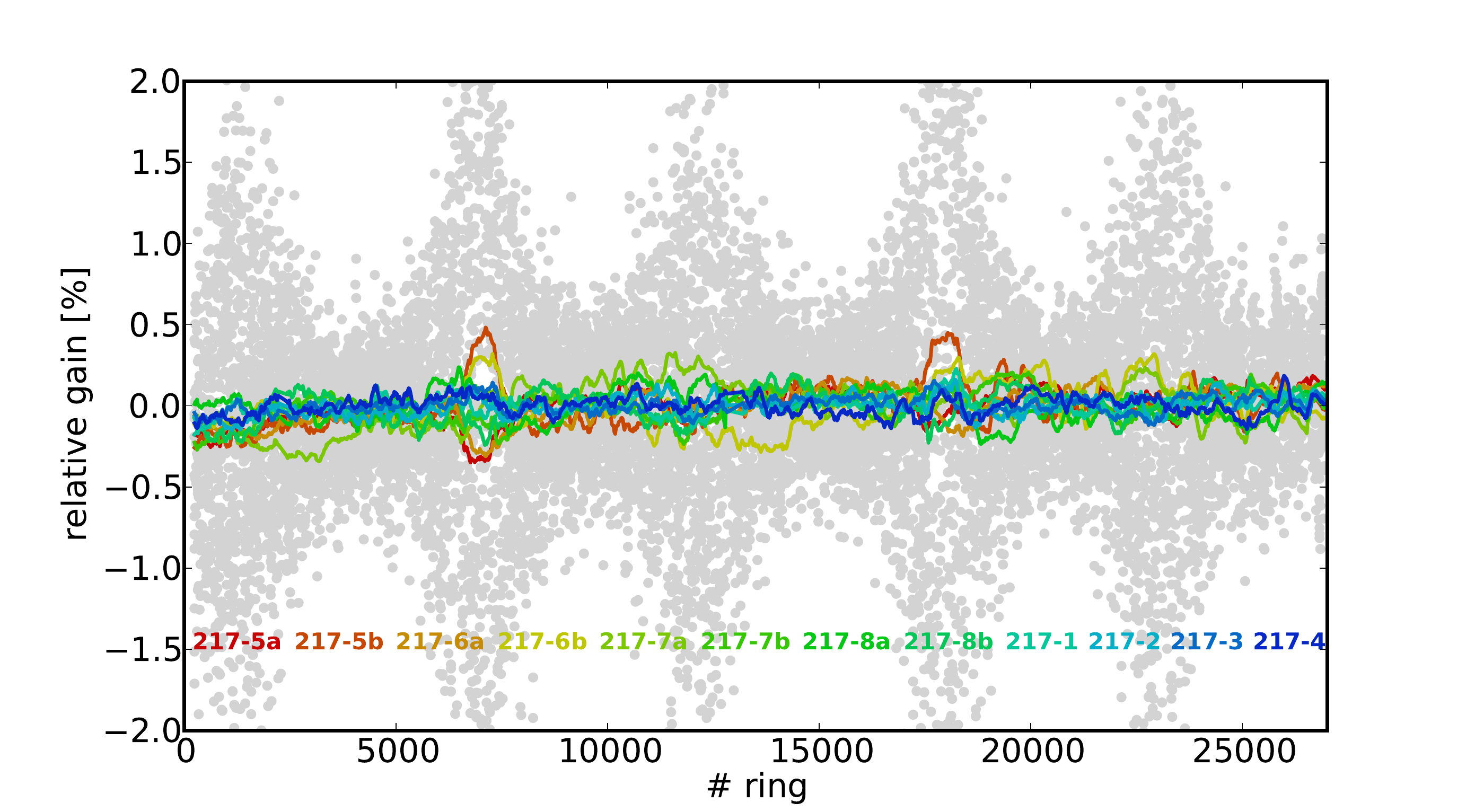}
	\caption{Gain variation with ring number for each bolometer estimated using \bogopix. {\it From top to bottom:}~100, 143, and 217\GHz. Gain values for individual rings (gray dots) have been smoothed with a 4-day width. The gain variations are lower than 0.3, 0.4, and 0.5\,\% at 100, 143, and 217\GHz\ respectively.}
	\label{fig:bogo_results}
\end{figure}

Figure~\ref{fig:bogo_results} shows the results of \bogopix\ for bolometers at 100, 143, and 217\GHz, smoothed over a 4-day period. At higher frequencies (353\GHz\ and above), the gain variations are much lower than the gain uncertainty. Owing to the \Planck\ scanning strategy, the Galactic foreground is larger for some rings, whilst the orbital dipole amplitude is almost constant. This increases the dispersion around those regions and potentially induces some bias in the gain determination. In the end, we find gain variations with amplitudes lower than 0.3, 0.4, and 0.5\,\% at 100, 143, and 217\GHz, respectively.

The residual apparent gain variations are essentially coming from: the uncertainty in the current ADC correction; the uncertainty in the long-time-constant estimation; and the effect of long-term thermal variations on the bolometer and electronics response.

\subsection{Calibration accuracy}
\label{sec:calib_accuracy}

\subsubsection{Inter-frequency accuracy}
\label{sec:intercal}
The precision of the calibration can be assessed by looking for residual dipoles in the maps. If the calibration of a map is slightly incorrect, the removal of the solar dipole in the mapmaking process leaves a residual dipole. However, identifying such a residual dipole is difficult because of the presence of other sources of power at $\ell=1$, mostly due to Galactic emission and zodiacal light, but also to imperfect correction of systematic effects such as far sidelobes.

Following the method presented in Sect.~\ref{sec:soldip} used to estimate the solar dipole direction and amplitude, we cleaned the Galactic emission from the HFI maps using a local correlation with the 857\GHz\ map.
Adding the solar dipole that was removed in the mapmaking process (Sect.~\ref{sec:soldip}) to $D_{\mathrm{res}}$ produces a map that contains the true solar dipole. We then fit for its amplitude, fixing its direction to the official \Planck\ value (lon, lat $= 264\pdeg00,48\pdeg24$) to limit the effect of other residuals still present in $D_{\mathrm{res}}$ (Galactic, systematic effects). The fit is done using sky pixels where $I_{857}<2$ (faintest 37\,\% of the sky) to limit the effect of Galactic emission residuals (CO, free-free emission, and small scale dust SED variations). Table~\ref{tab:amplitude} gives the ratio of fitted amplitude to the removed dipole found for the HFI maps at frequencies from 100 to 545\GHz. Because effects other than a miscalibration can contribute to a residual dipole in the maps, these ratios provide upper limits on the calibration accuracy at each frequency. The results indicate that the calibration at 100 and 143\GHz\ is precise at a level of few $10^{-4}$. At 217\GHz, the fit is compatible with a residual dipole at the 0.2\,\% level. At higher frequencies, the fits indicates residuals at 0.52\,\% and 1.23\,\% at 353 and 545\GHz, respectively.

\begin{table}
\begingroup
\newdimen\tblskip \tblskip=5pt
\caption{Ratio of amplitudes of the fitted dipole ($A_\mathrm{fit}$) and the removed dipole ($A_\mathrm{rm}=3364.5$\muK). The direction of the dipole removed from the data (lon=$264\pdeg00$, lat=$48\pdeg24$) was constrained here. The fit was computed for different sky fractions from 30 to 70\,\%. Statistical and systematic uncertainties are also indicated.}
\label{tab:amplitude}
\nointerlineskip
\vskip -3mm
\footnotesize
\setbox\tablebox=\vbox{
   \newdimen\digitwidth 
   \setbox0=\hbox{\rm 0} 
   \digitwidth=\wd0 
   \catcode`*=\active 
   \def*{\kern\digitwidth}
   \newdimen\signwidth 
   \setbox0=\hbox{-} 
   \signwidth=\wd0 
   \catcode`!=\active 
   \def!{\kern\signwidth}
\halign{\hbox to 2cm{#\leaderfil}\tabskip 1em&\hfil# \tabskip 1em&\hfil#\hfil\tabskip 1em&\hfil#\hfil\tabskip 0pt&\hfil#\hfil\tabskip  0pt\cr                            
\noalign{\doubleline}
	\omit\hfil Frequency \hfil & \omit\hfil $A_\mathrm{fit}/A_\mathrm{rm}$ \hfil & \omit\hfil Statitiscal \hfil & \omit\hfil Systematic \hfil\cr 
	\omit\hfil $[$GHz$]$ \hfil & & uncertainties & uncertainties \cr 
	\noalign{\vskip 3pt\hrule\vskip 5pt}
	100 & 1.00010 & $\pm 0.00006$ & $\pm 0.0001$\cr
	143 & 0.99988 & $\pm 0.00012$ & $\pm 0.0001$\cr
	217 & 1.00184 & $\pm 0.00027$ & $\pm 0.0003$\cr
	353 & 1.00568 & $\pm 0.00185$ & $\pm 0.0020$\cr
	545 & 1.02515 & $\pm 0.01627$ & $\pm 0.0190$\cr
\noalign{\vskip 5pt\hrule\vskip 3pt}
}
}
\endPlancktablewide                 
\endgroup
\end{table}

These results are in agreement with those obtained while performing component separation, as shown in Table~4 from \cite{planck2014-a11} and in \cite{planck2014-a12}. They are also in agreement with the results from the cosmological parameter determination, where intercalibration coefficients are also fitted for~\citep[see][]{planck2014-a13}.
The agreement between these measurements computed over different multipole ranges highlights the quality of the \Planck-HFI calibration, together with the accuracy of the transfer function reconstruction.

\subsubsection{Intra-frequency accuracy}
\label{sec:relative_calib}

For polarization reconstruction with \Planck-HFI data, we have to combine data from several detectors. Any relative calibration error will induce an intensity-to-polarization leakage (see Sect.~\ref{sec:leakage}). For the CMB channels, we have assessed the relative calibration accuracy for each detector at a given frequency using two complementary methods.

As in~section 6.2 (see figure~14) of \citet{planck2013-p03f}, we derive relative inter-calibration factors for each detector (for 100 to 353\GHz), rescaling their cross-pseudo-power spectra, estimated over 40\,\% of the sky (30\,\% for 353\GHz) in the $\ell$ range $25$--$300$, which encompasses the first acoustic peak. We used colour-correction factors at 353\GHz, because even at high latitude the dust emission is large. As in 2013, we keep the maximum of these factors as a conservative estimate of the relative calibration accuracy. In 2015, we find 0.09, 0.07, 0.16, and 0.78\,\% for 100, 143, 217, and 353\GHz, respectively (compared to 0.39, 0.28, 0.21, and 1.35\,\% in 2013). Since single-detector maps are built ignoring polarization, these values should be considered as conservative upper limits on the relative detector-to-detector calibration accuracy.

We complemented these estimations by analysing the solar dipole residual on the differences of single detector maps. We fit the dipole amplitude fixing its direction while masking 30\,\% of the sky in the Galactic plane to avoid regions affected by band-pass differences. We find maximal amplitudes  0.5, 0.6, and 3.0\,$\muK$ for 100, 143, and 217\GHz, respectively, which -- relative to the solar dipole amplitude ($3364.5 \muK$, see Sect.~\ref{sec:soldip}) -- gives accuracies of the same order as the aforementioned spectra analysis. As in this previous method, the main limitation comes from polarization which is ignored in the single detector maps.

While significantly better than for the 2013 release, calibration mismatch between bolometers at a given frequency is one of the main systematic residuals contaminating the HFI large angular scales in polarization, as explained in the next section.

\subsection{Intensity-to-polarization leakage}
\label{sec:leakage}
Any gain mismatch between the measurements of detectors belonging to the same frequency channel will result in intensity-to-polarization and cross-polarization leakage in the channel maps. In \Planck, the dominant leakage effect has three main origins:
\begin{itemize}
	\item monopole mismatch from the uncertainty in the mean offset determination;
	\item gain mismatch that produces leakage from the whole intensity signal into polarization;
	\item bandpass mismatch (hereafter BPM) that mainly generates intensity-to-polarization leakage from foreground emission (with a non-CMB spectrum). In the case of HFI, the leakage effect is dominated by CO and thermal dust emission.
\end{itemize}
All these leakage sources are especially important for the large angular scales. Beam-mismatch polarization leakage occurs at small angular scales and is discussed in \citetalias{planck2014-a08}. Although the first two mismatches can be minimized by obtaining more accurate measurements of offsets and gains respectively, the BPM cannot be removed in the mapmaking process if we want to project CMB and foregrounds at the same time.

The power absorbed by a given bolometer $b$ at time $t$ is expressed using the Stokes parameters ($I_p \, , Q_p\, , U_p$) which characterize the emission in intensity and polarization in the corresponding sky pixel $p$. The polarized HFI channels are calibrated using the CMB orbital dipole and the total \emph{calibrated} power absorbed by the bolometer $b$ can be written as
\begin{eqnarray}
	\nonumber
	m_{t}^{b} & = & (1+\epsilon^b_{\mathrm{gain}}) \\\nonumber
	&\times& \left\{\sum_{k} C_k(1+\epsilon^b_{\mathrm{BP},k}) \left[I_p^k+\rho^b\left( Q_p^k\cos \phi^b_t+U_p^k\sin \phi^b_t\right)\right]\right\}\\
	&+& \epsilon^b_{\mathrm{offset}}+n_t\,, 
	\label{eq:measure_bolo}
\end{eqnarray}
where the polarization efficiency $\rho^b$ and the polarization angle $\phi^b_t=2(\psi_t + \alpha^b)$ are explicitly dependent on the bolometer $b$, with the index $k$ ranging over the different sky components. Additionnaly we have the following definitions:
\begin{itemize}
	\item $\epsilon_{\mathrm{gain}}^b$ encodes the gain mismatch of bolometer $b$ with respect to the mean calibration of the channel;
	\item $\epsilon_{\mathrm{offset}}^b$ corresponds to the overall offset of bolometer $b$, which is small but not vanishing;
	\item $C_k$ is the average transmission of sky component $k$ in a given channel and $\epsilon^b_{\mathrm{BP},k}$ is the bandpass mismatch specific to bolometer $b$, and affecting all sky components except the CMB (see description below).
\end{itemize}
Each of these $\epsilon$ terms is responsible for leakage from intensity to polarization in a manner that can in principle be quantified and corrected for, as described hereafter. Note that we only consider first-order terms in $\epsilon$, as any higher-order contribution is negligible.

\subsubsection{Bandpass mismatch (BPM)}
\label{sec:bpm}
Each emission component $k$ (where $k$ = CMB, dust, synchrotron, etc.) is integrated over the bandpass of the detector according to a given spectrum $f_k(\nu)$. Since the polarized HFI channels are calibrated using the CMB orbital dipole, we define the transmission coefficients
\begin{eqnarray}
	\label{eq:leak_coeff}
	C_k^b&=&\frac{\int f_k(\nu) H^b_{\nu} d\nu}{\int f_{\mathrm{CMB}}(\nu) H^b_{\nu} d\nu}\\\nonumber
	&\equiv&C_k(1+\epsilon^b_{\mathrm{BP},k})\;, 
\end{eqnarray}
where $H^b_{\nu}$ is the spectral response of bolometer $b$ and $C_k=\sum_b C_k^b/N_{\mathrm{bolo}}$ is the average value of the $C_k^b$ in a given channel. These transmission coefficients express the $k$-component emission in CMB units. If all bolometers had the same spectral responses then $\epsilon^b_{\mathrm{BP},k}$ would be equal to zero, i.e. $C_k^b=C_k$, in which case no BPM-related leakage would be produced.

Considering only bandpass mismatch corresponds to setting $\epsilon^b_{\mathrm{gain}}=0$ and $\epsilon_{\mathrm{offset}}^b=0$ in Eq.~\eqref{eq:measure_bolo}. Then, ordering all the data samples, $m_t^b$, for a bolometer observing a position $p$ on the sky into a single vector $\vec{D}^b$, defining $\tens{A}$ to be the pointing matrix in temperature and polarization, and $\tens{n}$ the noise vector, Eq.~\eqref{eq:measure_bolo} reads
\begin{equation}
	\vec{D}^b = 
	\sum_{k} C_k \tens{A} \left(\begin{array}{c}I_p^k\\Q_p^k\\U_p^k\end{array}\right)
	+\sum_{k} C_k \epsilon_{\mathrm{BP},k}^b \tens{A}\left(\begin{array}{c}I_p^k\\Q_p^k\\U_p^k\end{array}\right)
	+ \vec{n}\;.
\end{equation}
Using all bolometers $b$ within a channel, the mapmaking procedure solves for the total signal Stokes parameters $(I^{\mathrm{tot}}_p\,, Q^{\mathrm{tot}}_p\,,U^{\mathrm{tot}}_p)$ in pixel $p$, formally
computing,
\begin{equation}
	\left(\begin{array}{c}I_p^{\mathrm{tot}}\\Q_p^{\mathrm{tot}}\\U_p^{\mathrm{tot}}\end{array}\right) = \left(\tens{A}^\mathrm{T}\tens{N}^{-1}\tens{A}\right)^{-1}\tens{A}^T\tens{N}^{-1}\vec{D}\;,
\end{equation}
which becomes
\begin{equation}
	\left(\begin{array}{c}I_p^{\mathrm{tot}}\\Q_p^{\mathrm{tot}}\\U_p^{\mathrm{tot}}\end{array}\right)
	=
	\sum_{k} C_k \left(\begin{array}{c}I_p^k\\Q_p^k\\U_p^k\end{array}\right)
	+ \sum_{k} C_k \sum_{b=0}^{N_{\mathrm{b}}-1} \epsilon_{\mathrm{BP},k}^b \Gamma_p^b\left(\begin{array}{c}I_p^k\\Q_p^k\\U_p^k\end{array}\right)\;,
\label{eq:leak}
\end{equation}
where $\Gamma^b_p\equiv\left(\tens{A}^\mathrm{T}\tens{N}^{-1}\tens{A}\right)^{-1}\tens{A}^\mathrm{T}\tens{N}^{-1}\Delta^b \tens{A}$. We have introduced the matrix $\Delta^b$, the elements of which are equal to zero except for the diagonal elements relevant to bolometer $b$, which are set to 1. The last term of Eq.~\eqref{eq:leak} is the leakage term in pixel $p$, where intensity will leak into $Q$ and $U$, $Q$ into $I$ and $U$, and $U$ into $I$ and $Q$, according to the mismatch coefficients $\epsilon_{\mathrm{BP},k}^b$ and the values of the $3\times 3$ matrix
\begin{equation}
	\Gamma^b_p=\left(
		\begin{array}{ccc}
			\Gamma_{II} &\Gamma_{QI} &\Gamma_{UI} \\
			\Gamma_{IQ} & \Gamma_{QQ}&\Gamma_{UQ}\\
			\Gamma_{IU} & \Gamma_{QU}&
			\Gamma_{UU}
		\end{array} \right)_p^b\;.
\end{equation}

Considering all pixels, the quantities $\Gamma^b_{XX}$ correspond to nine sky maps for bolometer $b$. These maps can be fully determined from the mapmaking solution and may be understood as patterns of the mismatch leakage. In practice, cross-polarization leakage and polarization-to-intensity leakage are negligible compared to the intensity-to-polarization contribution and we therefore consider the latter only. The $\Gamma_{IQ}$ and $\Gamma_{IU}$ maps have been systematically produced by the mapmaking pipeline.\footnote{The $\Gamma_{II}$ pattern map quantifies the correction that should, in principle, be brought to the $I$ channel map, given that some intensity has leaked into polarization. The correction is, however, negligible and not taken into account here. The same is true for $\Gamma_{QQ}$ and $\Gamma_{UU}$.} With these assumptions, the BPM-induced leakage in $Q$ and $U$ for the sky component $k$ reads
\begin{equation}
	L^{\mathrm{BP},k}_{IQ,IU} = C_k I^k \sum_{b=0}^{N_{\mathrm{bolo}}-1} \epsilon_{\mathrm{BP},k}^b\, \Gamma^b_{IQ,IU}\;.
	\label{eq:leak_map_bp}
\end{equation}

In consequence, for a given calibrated intensity template of the sky component $k$ (i.e., $I^k_{\mathrm{template}}=C_k I^k$) we may compute leakage correction maps as 
\begin{equation}
	L^{\mathrm{corr},k}_{IQ,IU} = I^k_{\mathrm{template}}
	\sum_{b=0}^{N_{\mathrm{bolo}}-1} \epsilon_{\mathrm{BP},k}^b\, \Gamma^b_{IQ,IU}= I^k_{\mathrm{template}}\sum_{b=0}^{N_{\mathrm{bolo}}-1} \frac{C_k^b}{C_k}\, \Gamma^b_{IQ,IU}\;,
	\label{eq:leak_corr}
\end{equation}
where the last equality uses the fact that $\sum_{b=0}^{N_{\mathrm{bolo}}-1}\Gamma^b_{IQ,IU}=0$ by construction. 

\begin{figure}[htbp]
	\centering
	\includegraphics[width=.48\columnwidth]{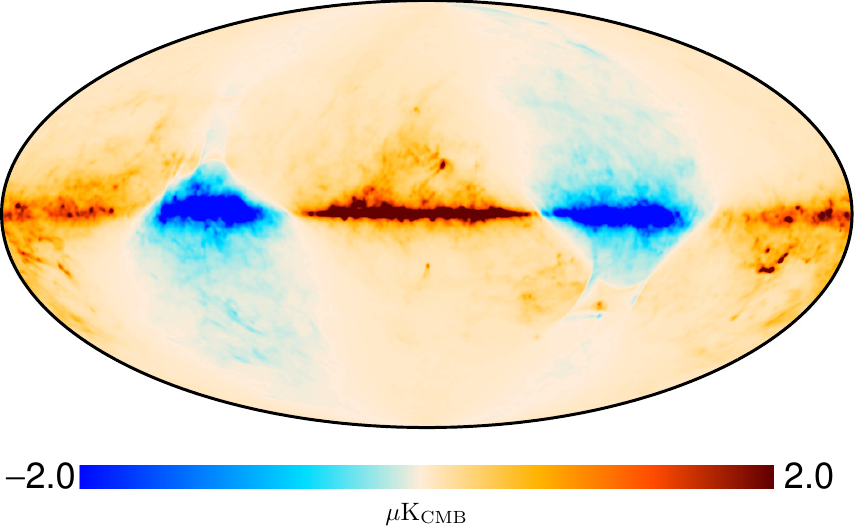} 
	\includegraphics[width=.48\columnwidth]{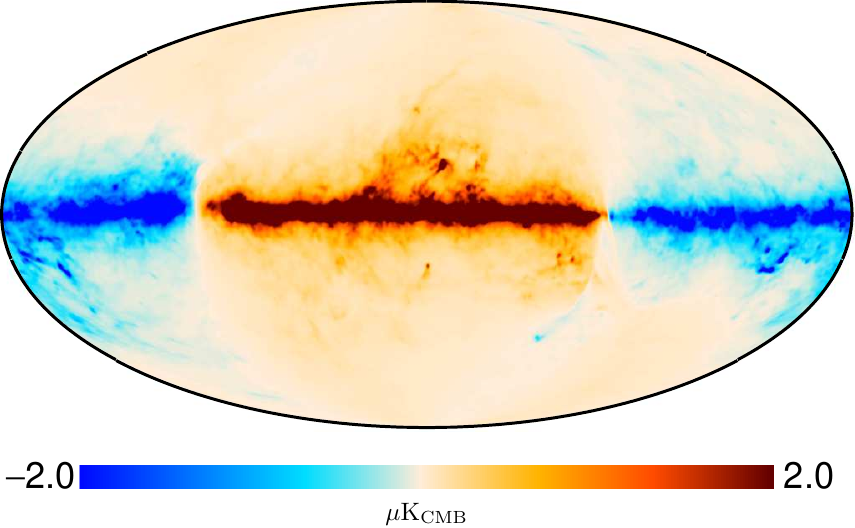}\\
	\includegraphics[width=.48\columnwidth]{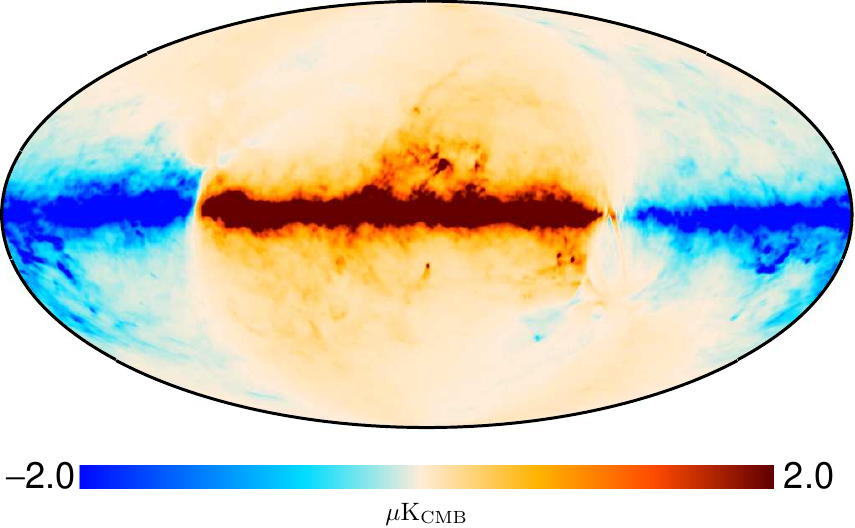} 
	\includegraphics[width=.48\columnwidth]{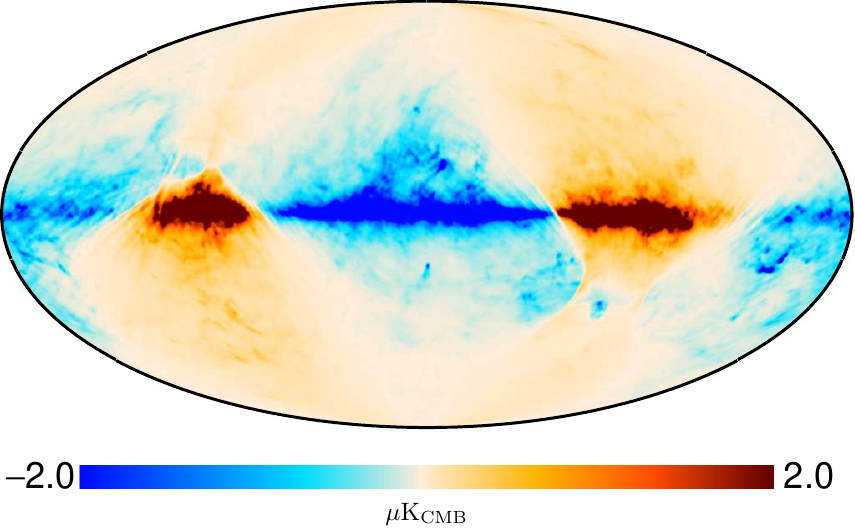}\\
	\includegraphics[width=.48\columnwidth]{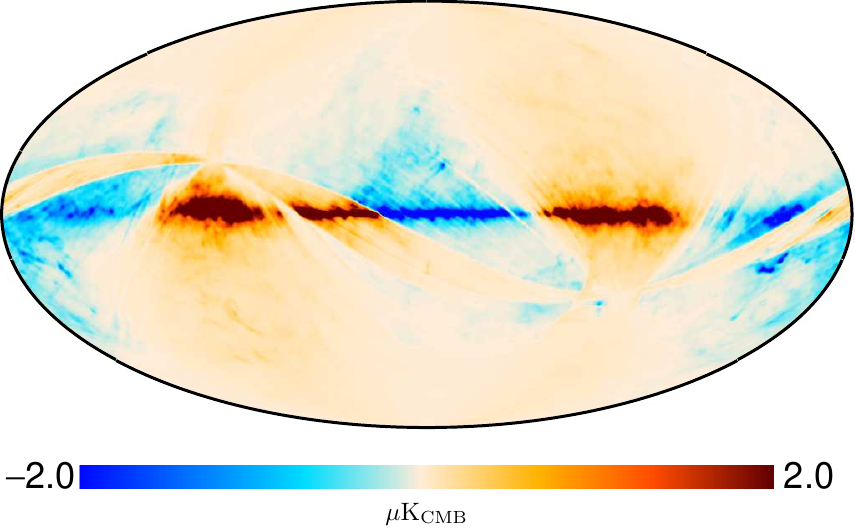} 
	\includegraphics[width=.48\columnwidth]{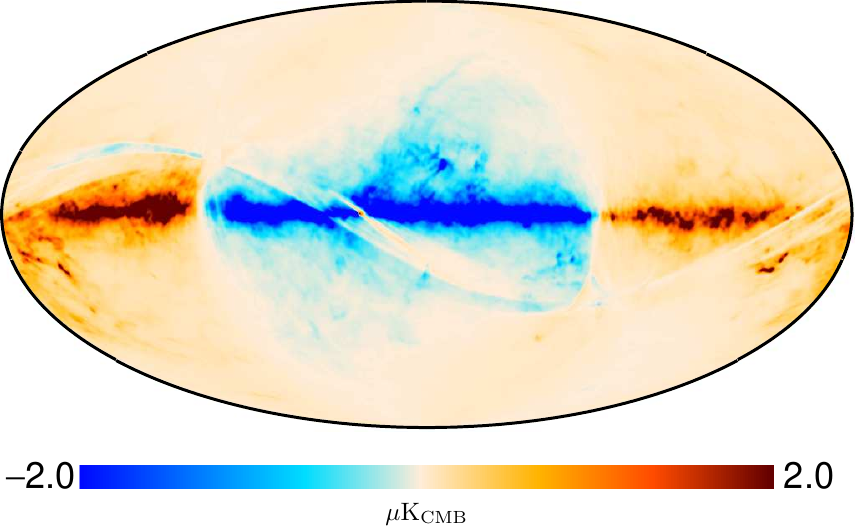}\\
	\includegraphics[width=.48\columnwidth]{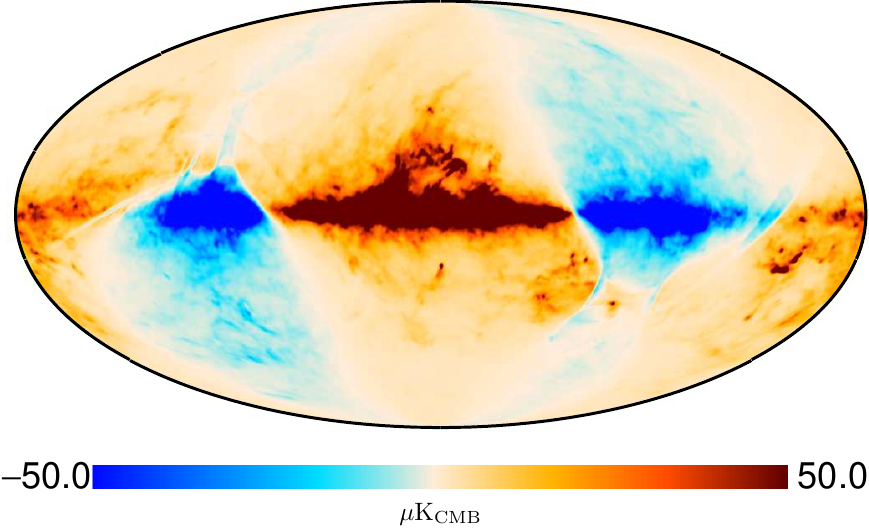} 
	\includegraphics[width=.48\columnwidth]{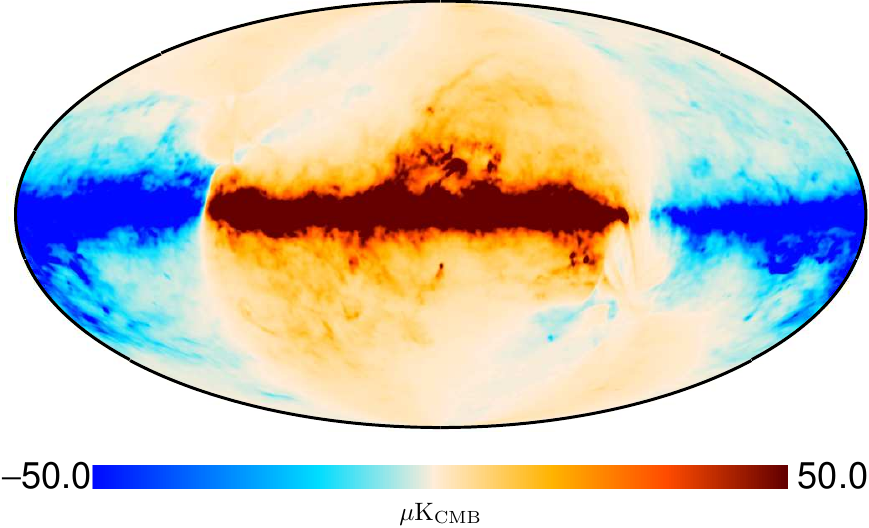}
	\caption{Dust leakage correction maps from ground-based measurements of the bandpass in $Q$ (left) and $U$ (right) at all HFI channels: 100, 143, 217, and 353\GHz\ (from top to bottom).}
	\label{fig:leak_corr_maps_dust}
\end{figure}

Leakage correction maps have been produced for all polarized HFI channels. The relevant foregrounds at these frequencies are dust (all channels) and CO (all channels except 143\GHz). 
To do so, the coefficients $C_{\mathrm{dust}}^b$ have been computed from Eq.~\eqref{eq:leak_coeff}, where the spectral responses of the bolometers $H^b(\nu)$ are those obtained from pre-launch ground-based measurements of the bandpasses~\citep{planck2013-p03d}. The dust spectrum is taken as a greybody with spectral index $\beta=1.62$ and temperature $T=19.7$\,K, which are the all-sky average values found in~\citet{planck2013-p06b}. For the intensity template required in Eq.~\eqref{eq:leak_corr}, we use the thermal dust intensity maps at 353\GHz\ obtained from the \Planck\ thermal dust model \citep{planck2013-p06b}.
Combining all these ingredients, the dust correction maps $L^{\mathrm{corr\;(dust)}}_{IQ,IU}$ are produced according to Eq.~\eqref{eq:leak_corr} and delivered in the 2015 HFI data release (Fig.~\ref{fig:leak_corr_maps_dust}).

Note, however, that the reliability of these corrections is limited by uncertainties both in the physical nature of the foreground components and in the determination of the bolometer spectral responses. For the sake of simplicity, a constant spectral index and a constant temperature across the sky have been assumed for the thermal dust emission. Furthermore, the calibrated thermal dust intensity templates are those derived from the 2013 \Planck\ thermal dust model which, while close, does not strictly correspond to the calibration of the 2015 maps. Also, the leakage corrections are particularly sensitive to the differences in transmission between bolometers (i.e., the $C_k^b$ coefficients); small uncertainties on those will yield large uncertainties in the final correction maps. In conclusion, the bandpass leakage corrections should not be taken at face value, but should be thought of as order-of-magnitude estimates only. We only advocate the use of these correction maps to test the stability and estimate uncertainties of any further results using the HFI polarization maps. A result solely obtained by applying the corrections will not be reliable.

\subsubsection{Calibration and monopole mismatches}
\label{sec:calib_mismatch}

Using the same formalism as above, calibration mismatch is computed by setting $\epsilon_{\mathrm{BP},k}^b=0$ and $\epsilon_{\mathrm{offset}}^b=0$ in Eq.~\eqref{eq:measure_bolo}. Following closely Sect.~\ref{sec:bpm}, one finds that the total intensity-to-polarization leakage due to calibration mismatch is
\begin{eqnarray}
	\nonumber
	L^{\mathrm{gain}}_{IQ,IU} &=& \sum_k C_k I^k\times \sum_{b=0}^{N_{\mathrm{bolo}}-1} \epsilon_{\mathrm{gain}}^b\, \Gamma^b_{IQ,IU}\\
	&\approx& I^{\mathrm{dipole}}\times \sum_{b=0}^{N_{\mathrm{bolo}}-1}\; \epsilon_{\mathrm{gain}}^b\, \Gamma^b_{IQ,IU}\;,
	\label{eq:leak_map_calib}
\end{eqnarray}
where at first order, for low-frequency maps, the solar dipole signal ($k=\mathrm{dipole}$, $C_{\mathrm{dipole}}=1$ by construction) provides the dominant contribution to the calibration mismatch leakage effect.

Setting $\epsilon_{\mathrm{BP},k}^b=0$ and $\epsilon_{\mathrm{gain}}^b=0$ in Eq.~\eqref{eq:measure_bolo}, one shows in a similar fashion that the monopole intensity-to-polarization leakage is simply 
\begin{equation}
	L^{\mathrm{mono}}_{IQ,IU} = \sum_{b=0}^{N_{\mathrm{bolo}}-1}\; \epsilon_{\mathrm{offset}}^b\, \Gamma^b_{IQ,IU}\;,
	\label{eq:leak_map_mono}
\end{equation}
where the monopole mismatch is modelled using a constant sky template $I^{\mathrm{monopole}}=1$, while the amplitude of the mismatch is encoded in $\epsilon^b_{\mathrm{offset}}$.

Although the BPM coefficients $\epsilon_{\mathrm{\mathrm{BP},dust}}^b$ can be evaluated directly from foreground modelling (assuming a given spectrum of the dust and using the spectral responses of the detectors), this is not the case for $\epsilon_{\mathrm{gain}}^b$ and $\epsilon_{\mathrm{offset}}^b$. It is therefore not possible to provide correction maps for these leakage effects by computing Eq.~\eqref{eq:leak_map_calib} and \eqref{eq:leak_map_mono} directly. However, one may consider the possibility of fitting these quantities from the maps themselves, by using the $I^{\mathrm{dipole}}\times\Gamma^b_{IQ,IU}$ and $\Gamma^b_{IQ,IU}$ as templates of the gain and monopole leakages respectively. Such a method, dubbed the ``generalized global fit'' (GGF), has been implemented and is further described in Appendix~\ref{annex:ggf}. 

The leakage maps produced with the GGF method for the 353\GHz\ channel are delivered in the 2015 release and corrected for BPM, calibration, and monopole leakage simultaneously.

\subsection{In-flight validation of the polarimeter efficiency and orientation}
As discussed in Sect.~\ref{sec:pol_eff}, the polarimeter efficiency and orientation used in this release are taken from ground measurements~\citep{rosset2010}. In order to validate these numbers in flight, we used the Crab nebula maps obtained with the IRAM~30\,m telescope and the 90\GHz\ XPOL polarimeter \citep{aumont2010}. These maps consist of $I$, $Q$, and $U$ measurements with an angular resolution of 27\arcs\ of a 10\arcm-wide region around the Crab nebula (Tau A, M1, or NGC\,1952, at J2000 coordinates RA = $5^{\mathrm h} 34^{\mathrm m} 32^{\mathrm s}$ and Dec = $22\deg 00\arcm 52\arcs$). The same region was observed by \Planck\ once per survey, with different scan directions for odd and even surveys. We compared single survey, single bolometer maps of the Crab region with a model obtained from the IRAM maps, and solved for the best values of polarimeter angle and efficiency.

From single survey, single bolometer data, we can only solve for an intensity map, which projects on the sky the total power $P_t$ described in Eq.~\eqref{eq:pabs}. This power depends on the true value of the polarization angle $\alpha$ specific to the detector.

We compared the single bolometer, single survey maps with a model obtained with the following procedure:
\begin{itemize} 
	\item we pixelized the IRAM observations on a \healpix\ grid with $\Nside=2048$, rotating to Galactic coordinates;
	\item we convolved these maps with the single bolometer, single survey effective beams using FEBeCoP~\citep{mitra2010};
	\item using the Crab IRAM map as a template, and the polarization angles $\alpha$, we modelled the intensity map described above in the Crab region, as a function of an angular offset $\Delta \alpha$.
\end{itemize}

\begin{figure}[htbp]
	\centering
	\includegraphics[width=0.5\textwidth]{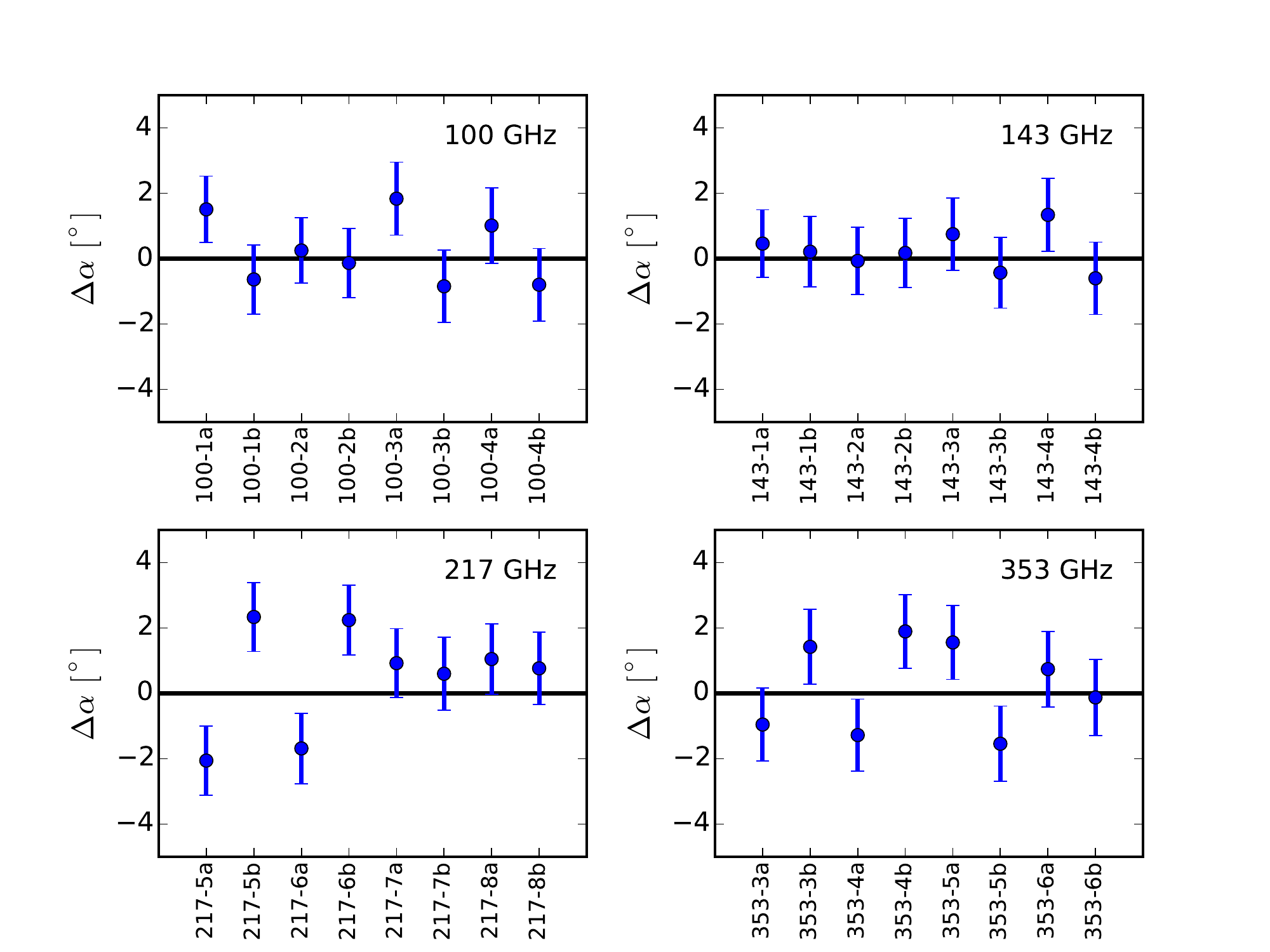}
	\caption{Estimated angular offset from comparison of the single bolometer, single detector maps with the IRAM Crab nebula maps, combining the first four surveys.}
	\label{fig:polar_angles}
\end{figure}

We then fitted for the values of the angular offset $\Delta \alpha$. To do that, we first removed the background from the single bolometer maps. We built a noise model combining the single detector pixel variance with the noise of the IRAM observation, taking into account the smoothing applied. We used the \cite{rosset2010} values as a prior.
The resulting angular offsets are presented in Fig.~\ref{fig:polar_angles}. Corrections are compatible with zero, and this analysis does not favour an update of the ground-based parameters.

We used the same procedure also to fit the polarization efficiency $\rho$, but the result is completely dominated by the ground-based calibration prior.

If we assume that the CMB anisotropies have vanishing $TB$ and $EB$ power spectra, i.e., that there are no parity violating physical mechanisms in the early Universe, we can also check whether the overall polarizer angle of \Planck-HFI is compatible with zero. \cite{planck2014-a23} show that the CMB $TB$ and $EB$ spectra measured by HFI are consistent with zero. Their analysis gives a polarizer angle within $0.3^\circ$ of zero, which is identical to the systematic error of the ground-based measurements. This is a factor of five improvement over the \textit{WMAP} final results~\citep{hinshaw2012} and comparable with ACT~(\citealt{naess2014}; see also the review by \citealt{gubitosi2013}).

\section{Conclusions}
\label{sec:conclusion}

\begin{table*}[t]
\centering{}\caption{Main characteristics of HFI Full Mission Maps. \label{tab:summary}}
\begingroup
\newdimen\tblskip \tblskip=5pt
\nointerlineskip
\vskip -3mm
\footnotesize
\setbox\tablebox=\vbox{
   \newdimen\digitwidth 
   \setbox0=\hbox{\rm 0} 
   \digitwidth=\wd0 
   \catcode`*=\active 
   \def*{\kern\digitwidth}
   \newdimen\signwidth 
   \setbox0=\hbox{+} 
   \signwidth=\wd0 
   \catcode`!=\active 
   \def!{\kern\signwidth}
\halign{\hbox to 2.5in{#\leaderfil}\tabskip 2.2em&
\hfil#\hfil& 
\hfil#\hfil& 
\hfil#\hfil& 
\hfil#\hfil& 
\hfil#\hfil& 
\hfil#\hfil& 
\hfil#\hfil\/\tabskip=0pt\cr 
\noalign{\doubleline}
\omit\hfil Quantity \hfil&                                                    &           &           &           &           &           & Notes\cr
\noalign{\vskip 3pt\hrule\vskip 5pt}                                      
Reference frequency $\nu$ [\GHz]           &      100&        143&        217&        353&        545&        857& \tablefootmark{a1}\cr
Number of bolometers                      &        8&         11&         12&         12&          3&          4& \tablefootmark{a2}\cr
\noalign{\vskip 3pt\hrule\vskip 3pt}  
Effective beam solid angle $\Omega$ [arcmin$^2$]      &   106.22&      60.44&      28.57&      27.69&      26.44&      24.37& \tablefootmark{b1}\cr
Error in solid angle $\sigma_\Omega$ [arcmin$^2$]     &   **0.14&      *0.04&      *0.04&      *0.02&      *0.02&      *0.02& \tablefootmark{b2}\cr
Spatial variation (\rms) $\Delta\Omega$ [arcmin$^2$] & **0.20&      *0.20&      *0.19&      *0.20&      *0.21&      *0.12& \tablefootmark{b3}\cr
Effective beam FWHM$_1$ [arcmin]                      &   **9.68&      *7.30&      *5.02&      *4.94&      *4.83&      *4.64& \tablefootmark{b4}\cr
Effective beam FWHM$_2$ [arcmin]                      &   **9.66&      *7.22&      *4.90&      *4.92&      *4.67&      *4.22& \tablefootmark{b5}\cr
Effective beam ellipticity $\epsilon$                 &   *1.186&      1.040&      1.169&      1.166&      1.137&      1.336& \tablefootmark{b6}\cr
Variation (\rms) of the ellipticity $\Delta\epsilon$ &*0.024&      0.009&      0.029&      0.039&      0.061&      0.125& \tablefootmark{b7}\cr
\noalign{\vskip 3pt\hrule\vskip 5pt}                                      
Sensitivity per beam solid angle [$\mu\mathrm{K_{CMB}}$]       &       7.5&       4.3&       8.7&       29.7*&           &           & \tablefootmark{c1}\cr
\phantom{Sensitivity per beam solid angle}                        [kJy sr$^{-1}$]               &          &          &          &           &        9.1&        8.8& \tablefootmark{c1}\cr
 Temperature Sensitivity           [$\mu\mathrm{K_{CMB}}$\,deg]&      1.29&      0.55&      0.78&       *2.56&           &           & \tablefootmark{c2}\cr
 \phantom{Temperature Sensitivity}         [kJy sr$^{-1}$\,deg]&          &          &          &           &       0.78&       0.72& \tablefootmark{c2}\cr
 Polarization Sensitivity          [$\mu\mathrm{K_{CMB}}$\,deg]&      1.96&      1.17&      1.75&       *7.31&           &           & \tablefootmark{c3}\cr
\noalign{\vskip 3pt\hrule\vskip 3pt}
Calibration accuracy [\%]& $0.09$& $0.07$& $0.16$& $0.78$& $1.1(+5)$& $1.4(+5)$ & \tablefootmark{d}\cr
\noalign{\vskip 3pt\hrule\vskip 3pt}  
CIB monopole prediction           [\MJysr]&   0.0030&     0.0079&     0.033*&     0.13**&     0.35**&     0.64**& \tablefootmark{e1}\cr
Zodiacal light level correction [K$_{\mathrm{CMB}}$] & $4.3 \times 10^{-7}$ & $9.4 \times 10^{-7}$ & $3.8\times 10^{-6}$ & $3.4 \times 10^{-5}$ &  &  & \tablefootmark{e2}\cr
\phantom{Zodiacal light level correction} [\MJysr] &  &  &  &  & 0.04** & 0.12** & \tablefootmark{e2}\cr
\noalign{\vskip 5pt\hrule\vskip 3pt}}}
\endPlancktablewide
\raggedright
\tablenote \textit{a1} {~Channel map reference frequency, and channel identifier.}\par
\tablenote \textit{a2} {~Number of bolometers whose data were used in producing the channel map.}\par
\tablenote \textit{b1} {~Mean value over bolometers at the same frequency. See Sect.~4.2 in paper 1.}\par
\tablenote \textit{b2} {~As given by simulations. }\par
\tablenote \textit{b3} {~Variation (\rms) of the solid angle across the sky. }\par
\tablenote \textit{b4} {~FWHM of the Gaussian whose solid angle is equivalent to that of the effective beams.}\par
\tablenote \textit{b5} {~Mean FWHM of the elliptical Gaussian fit.}\par
\tablenote \textit{b6} {~Ratio of the major to minor axis of the best-fit Gaussian averaged over the full sky. }\par
\tablenote \textit{b7} {~Variability (\rms) on the sky. }\par
\tablenote \textit{c1} {~Estimate of the noise per beam solid angle as given in \textit{b1}.}\par
\tablenote \textit{c2} {~Estimate of the noise in intensity scaled to $1\deg$ assuming that the noise is white.}\par
\tablenote \textit{c3} {~Estimate of the noise in polarization scaled to $1\deg$ assuming that the noise is white.}\par
\tablenote \textit{d\phantom{1}} {~Calibration accuracy (at 545 and 857\GHz: the 5\% accounts for the model uncertainty).}\par
\tablenote \textit{e1} {~According to the~\cite{bethermin2012} model, whose uncertainty is estimated to be at the 20\,\% level (also for constant $\nu I_\nu$).}\par
\tablenote \textit{e2} {~Zero-level correction to be applied on Zodical light corrected maps.}\par
\endgroup
\end{table*}

This paper has described the processing applied to construct the \Planck-HFI maps delivered in the 2015 release. It has also assessed the main characteristics of the maps in terms of noise and systematics, in particular resulting from ADC corrections and bolometer long time constants. Since the last release, the calibration has been upgraded and is now significantly more accurate. At low frequency, it is now independent and based on the orbital dipole signal, while the planets Uranus and Neptune are used to calibrate the high end of HFI, achieving 6.1 and 6.4\,\% absolute photometric calibration at 545 and 857\GHz, respectively. This has allowed us to measure a consistent CMB solar dipole with an unprecedented accuracy better than $10^{-3}$ and in agreement with the independent determination by LFI.

Table~\ref{tab:summary} gives a quantitative assessment of the main characteristics of the \Planck\ HFI maps from the 2015 release. They now cover the entire \Planck\ HFI cold mission (885 days). The HFI aggregated sensitivity (referring to a weighted average of the 100, 143, and 217\GHz\ channel maps) is $26\,\mathrm{\mu K_{CMB}.arcmin}$ in temperature and $52\,\mathrm{\mu K_{CMB}.arcmin}$ in polarization.

The noise in the maps shows some small low-frequency excess on top of white noise prior to time constant deconvolution. The latter then naturally raises the higher part of the noise spectra in the multipole domain. We have identified a low-level noise correlation in particular between half-ring and detector subsets that is not directly reproduced by simulations, although the level is small compared to the CMB signal.

The raw sensitivity must be matched by a long list of constraints on any possible systematic effects. This list includes: an absolute calibration at a level of 0.1\,\% to 1.4\,\% depending on the frequency; a resulting apparent gain variation of less than 0.5\,\%; and a knowledge of the polarization angle and polarization absolute value respectively, at the degree level and the 1\,\% level. The instrumental beam has been measured at the percent level by using multiple planet crossings.

Despite the huge progress made in the understanding of all the aforementioned systematic effects, \Planck-HFI polarization maps are still dominated by systematic residuals at large scales. These are essentially coming from the temperature-to-polarization leakage resulting from the mismatch between the bolometers that are combined to reconstruct linear polarization maps.
The origins of the leakage effects include: mismatch of the zero level from uncertainty in the offset determination; mismatch from gain uncertainty (even at the $10^{-3}$ level); and bandpass mismatch. Corresponding first order corrections for monopole, dipole, and bandpass mismatch are provided (as described in Appendix~\ref{annex:ggf}) but residuals are still found to be larger than noise at very large scales. As a consequence, the \Planck-HFI polarization maps at large scales cannot yet be directly used for cosmological studies.

\begin{acknowledgements}
The Planck Collaboration acknowledges the support of: ESA; CNES and CNRS/INSU-IN2P3-INP (France); ASI, CNR, and INAF (Italy); NASA and DoE (USA); STFC and UKSA (UK); CSIC, MINECO, JA, and RES (Spain); Tekes, AoF, and CSC (Finland); DLR and MPG (Germany); CSA (Canada); DTU Space (Denmark); SER/SSO (Switzerland); RCN (Norway); SFI (Ireland); FCT/MCTES (Portugal); ERC and PRACE (EU). A description of the Planck Collaboration and a list of its members, indicating which technical or scientific activities they have been involved in, can be found at \href{http://www.cosmos.esa.int/web/planck/planck-collaboration}{http://www.cosmos.esa.int/web/planck/planck-collaboration}.
\end{acknowledgements}

\bibliographystyle{aat}
\bibliography{Planck_bib,mapmaking_bib}


\appendix
\section{HFI map product description}
\label{annex:official}

Here we summarize the  HFI map products that are part of  the Planck 2015 data release.

\subsection{Map products}
\label{sec:mapproducts}
The 2015 release contains many different maps whose details are described in the subsections below. All maps are given as \healpix\ vectors with NESTED ordering, in Galactic coordinates, at the resolution corresponding to $\Nside=2048$ (for high resolution maps). Depending on the type of product, these vectors are packaged into a binary table and written into a FITS file. The table contains in most cases 50\,331\,648 rows (the length of the \healpix\ vector for $\Nside=2048$) and either three columns ($I$, $II$, $H$, respectively for intensity, noise (co)variance, and hits) for the temperature-only cases or 10~columns ($I$, $Q$ and $U$ signals plus the six various $I$, $Q$, and $U$ noise (co)variances and hits) for the polarized cases. Pixels with a condition number larger than $10^{3}$ see their hit number brought down to zero. A pixel with zero hit has an intensity value of $-1.6375\times 10^{30}$ (the \healpix\ conventional null value).

The general matrix of products is pictured in Fig.~\ref{fig:MapMatrix}. The main products are the maps of the full channels, covering the full mission. For characterization and analysis purposes,  the channels have been split into independent detector sets. The detector sets for the polarized channels are  groups of four PSBs that can be used to build a sky map in temperature and polarization. 
Bolometers insensitive to polarization (SWB) of the  same frequency channels are not used in the detector sets. The detector set names and the bolometers they use are listed in Table~\ref{tab:detset}. For completeness, full mission maps built using each of the unpolarized bolometers are also provided.

\begin{table*}[t]
\centering{}  \caption{ Detector set definitions \label{tab:detset}}
\begingroup
\newdimen\tblskip \tblskip=5pt
\nointerlineskip
\vskip -3mm
\footnotesize
\setbox\tablebox=\vbox{
   \newdimen\digitwidth 
   \setbox0=\hbox{\rm 0} 
   \digitwidth=\wd0 
   \catcode`*=\active 
   \def*{\kern\digitwidth}
   \newdimen\signwidth 
   \setbox0=\hbox{+} 
   \signwidth=\wd0 
   \catcode`!=\active 
   \def!{\kern\signwidth}
\halign{\hbox to 2.5in{#\leaderfil}\tabskip 2.2em&
\hfil#\hfil& 
\hfil#\hfil\/\tabskip=0pt\cr 
\noalign{\doubleline}
\omit\hfil Frequency \hfil&  DetSet1 &    DetSet2\cr
\noalign{\vskip 3pt\hrule\vskip 5pt}                                      
100\GHz   & 100-1a/b 100-4a/b  &   100-2a/b 100-3a/b\cr
143\GHz   & 143-1a/b 143-3a/b  &   143-2a/b 143-4a/b\cr
217\GHz   & 217-5a/b 217-7a/b  &   217-6a/b 217-8a/b\cr
353\GHz   & 353-3a/b 353-5a/b  &   353-4a/b 353-6a/b\cr
\noalign{\vskip 5pt\hrule\vskip 3pt}}}
\endPlancktablewide
\raggedright
\endgroup
\end{table*}

The mission duration has been split into single surveys,  ``years''  and the two halves of the  mission duration. The surveys are defined as the observations within a contiguous rotation range of $180\deg$ for the spin axis, and as a consequence each survey does not cover the full sky. Note that the fifth survey
had not been completed when HFI stopped observing, and was also interrupted by various end-of-mission tests. That last survey was also performed with a different scanning strategy than the first four surveys~\citep{planck2014-a01}. The date and ring number corresponding to the beginning and end of each of the time split are summarized in Table~\ref{tab:timesplit}.
Single survey maps are provided for the full channels only. Yearly maps are provided for Year~1 and Year~2 for the full channels, the detector sets and the SWBs, where the years span Surveys 1--2 and 3--4. Half-mission maps are also provided, where each half contains one half of the valid rings (or stable pointing periods). There are 347 discarded rings (and 26\,419 valid ones), most of which occurred during the (partial) last survey, when various end-of-life tests were performed.
\begin{table*}[t]
\centering{}  \caption{Date and ring numbers for the beginning and end of each time split.\label{tab:timesplit}}
\begingroup
\newdimen\tblskip \tblskip=5pt
\nointerlineskip
\vskip -3mm
\footnotesize
\setbox\tablebox=\vbox{
   \newdimen\digitwidth 
   \setbox0=\hbox{\rm 0} 
   \digitwidth=\wd0 
   \catcode`*=\active 
   \def*{\kern\digitwidth}
   \newdimen\signwidth 
   \setbox0=\hbox{+} 
   \signwidth=\wd0 
   \catcode`!=\active 
   \def!{\kern\signwidth}
\halign{\hbox to 2.5in{#\leaderfil}\tabskip 2.2em&
\hfil#\hfil& 
\hfil#\hfil& 
\hfil#\hfil& 
\hfil#\hfil\/\tabskip=0pt\cr 
\noalign{\doubleline}
\omit\hfil Time split \hfil&        start date & first ring & end date & end ring \cr
\noalign{\vskip 3pt\hrule\vskip 5pt}                                      
 Full mission  & 12/08/2009   & **240 & 13/01/2012   & 27005 \cr
 Nominal mission  & 12/08/2009  & **240 &   28/11/2010   & 14723  \cr
 Half mission 1  & 12/08/2009  & **240 & 15/10/2010  & 13471  \cr 
 Half mission 2  &15/10/2010  & 13472 & 13/01/2012  & 27005 \cr
 Year 1  & 12/08/2009  & **240 &   12/08/2010  & 11194 \cr
 Year 2  &  12/08/2010  & 11195 & 29/07/2011 & 21720 \cr
 Survey 1 &  12/08/2009  & **240  &08/02/2010  & *5720   \cr
 Survey 2 & 08/02/2010 & *5721 &  12/08/2010 & 11194 \cr
 Survey 3 & 12/08/2010 & 11195  & 08/02/2011  & 16691 \cr
 Survey 4 &08/02/2011  & 16692 & 29/07/2011 & 21720 \cr
 Survey 5 & 29/07/2011 & 21721 & 13/01/2012  & 27005 \cr
\noalign{\vskip 5pt\hrule\vskip 3pt}}}
\endPlancktablewide
\raggedright
\endgroup
\end{table*}

Half-ring maps are produced by splitting each ring into two equal duration parts. The difference of the two half-ring maps provides a useful estimate of the high frequency noise and possibly other systematics. Note that this is the only case where the destriping offsets are different from the offsets in the standard case. Half-ring maps are provided for the full mission only. They are given for the full channels, the detector sets, and the SWB maps.

Maps are corrected for zodiacal light emission. Correction maps are also provided for the frequency maps and the various time splits.


Units are K$_\mathrm{CMB}$ for frequencies up to 353\GHz\ and \MJysr assuming a constant $\nu I_{\nu}$ law above.

\subsection{Stokes parameter correlations}
\label{annex:IQUcorrel}

The \Planck\ HFI delivery includes pixelized maps of Stokes covariances ($II$,$IQ$,$IU$,$QQ$,$QU$,$UU$) solved during the mapmaking process for each pixel independently.

\begin{figure*}[htbp]
	\centering
        \begin{tabular}{ccc}
          \includegraphics[width=.32\textwidth,viewport=0 125 790 540,clip]{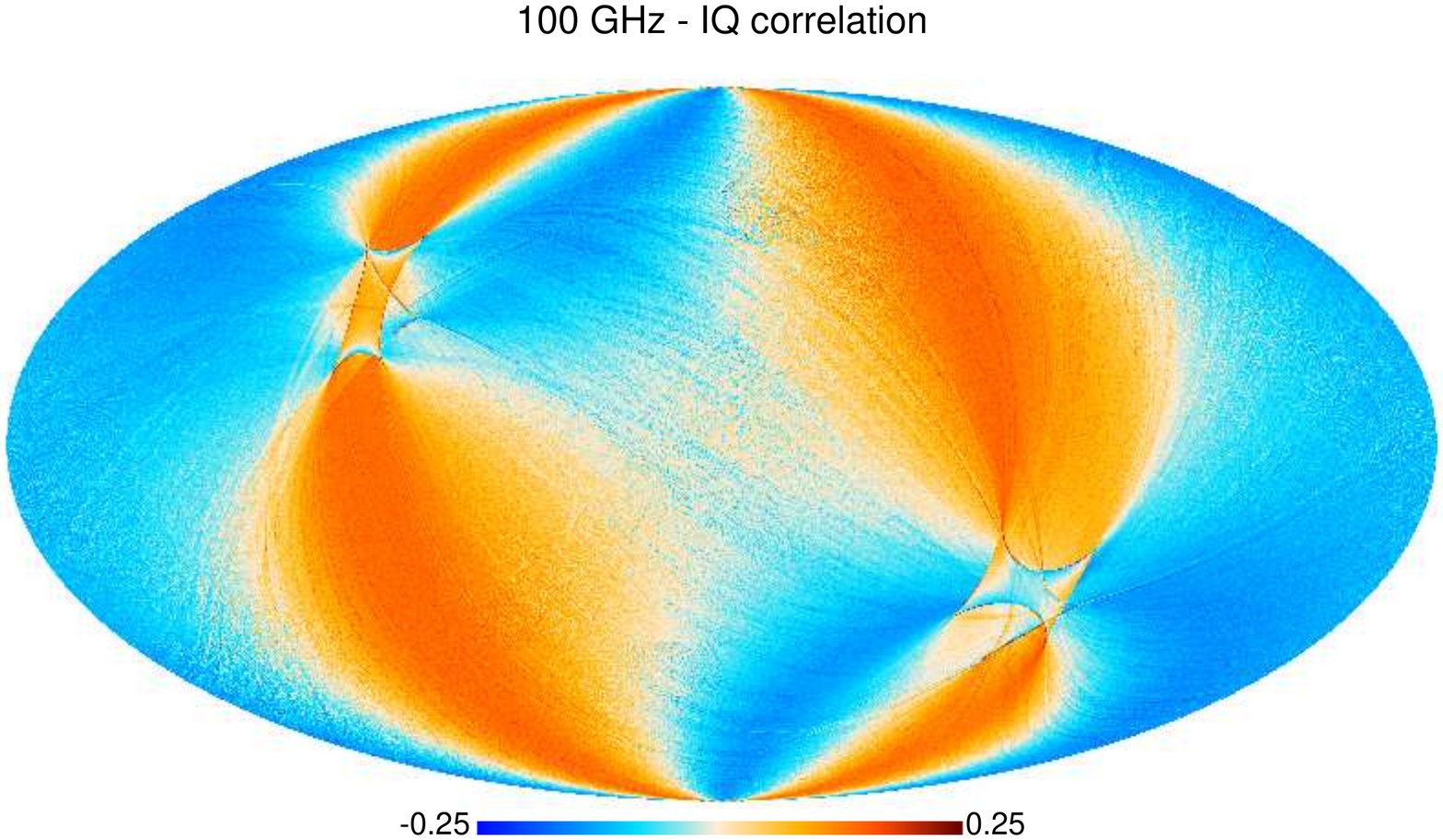} & \includegraphics[width=.32\textwidth,viewport=0 125 790 540,clip]{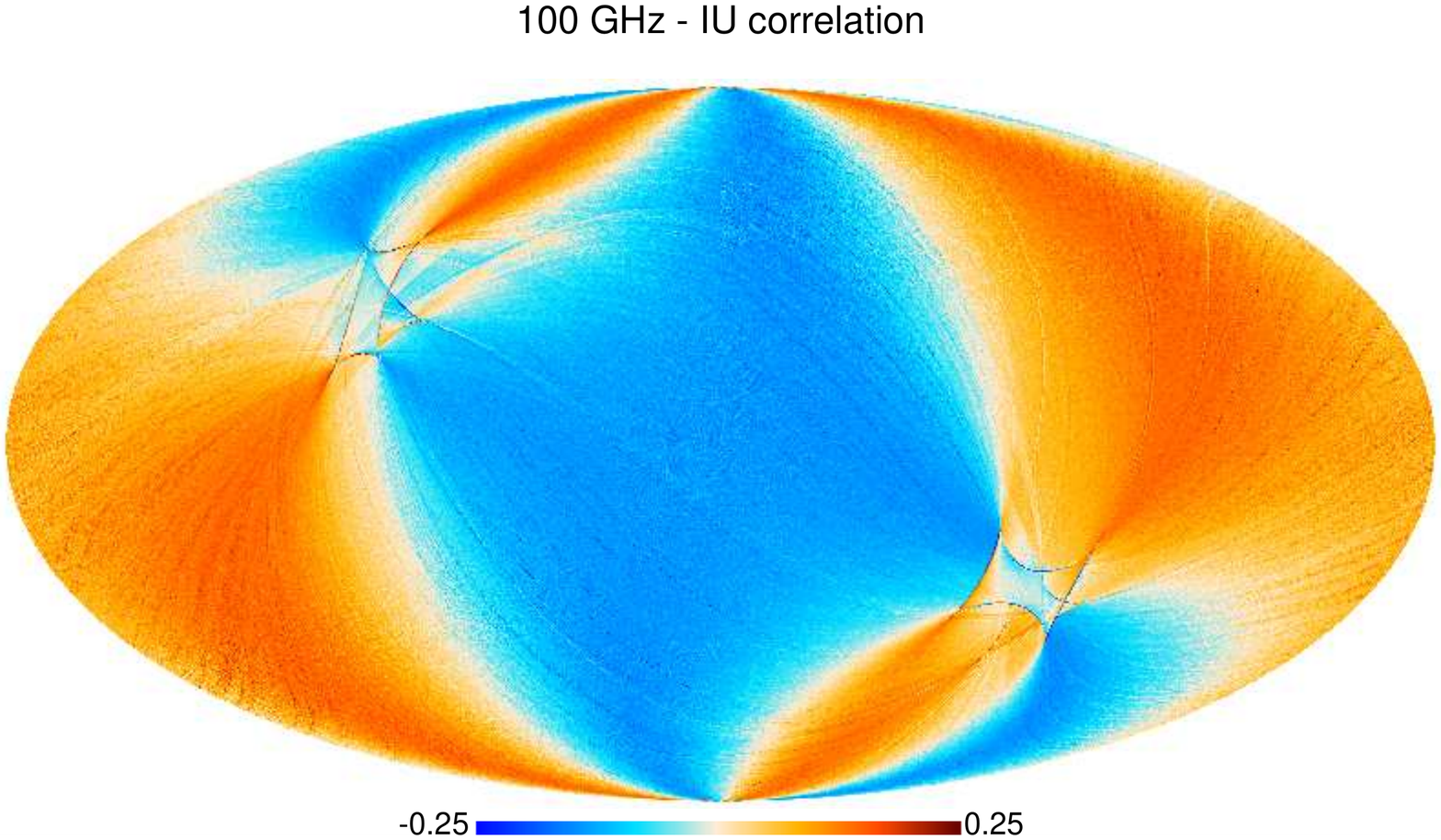} & \includegraphics[width=.32\textwidth,viewport=0 125 790 540,clip]{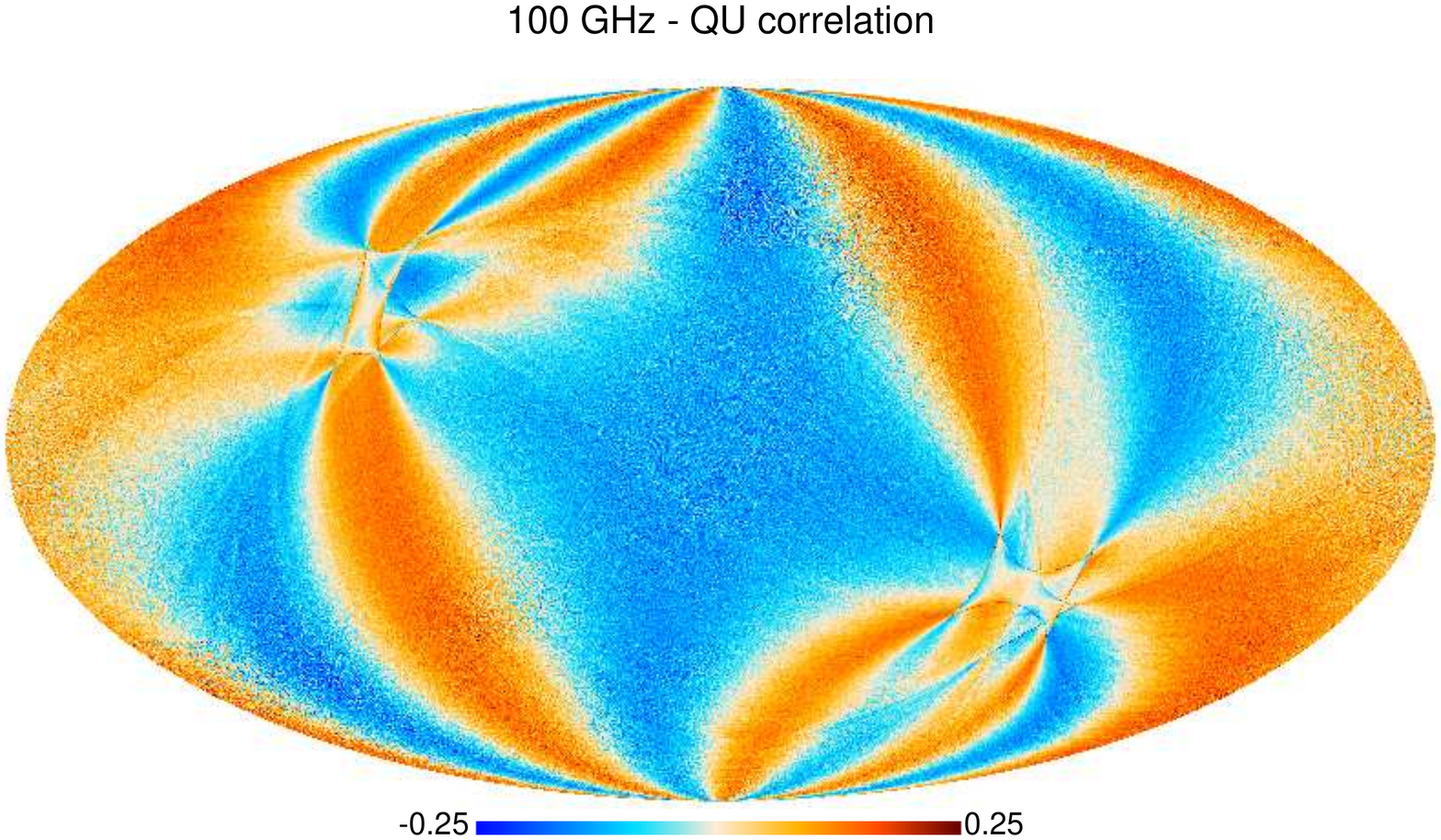} \\
          \includegraphics[width=.32\textwidth,viewport=0 125 790 540,clip]{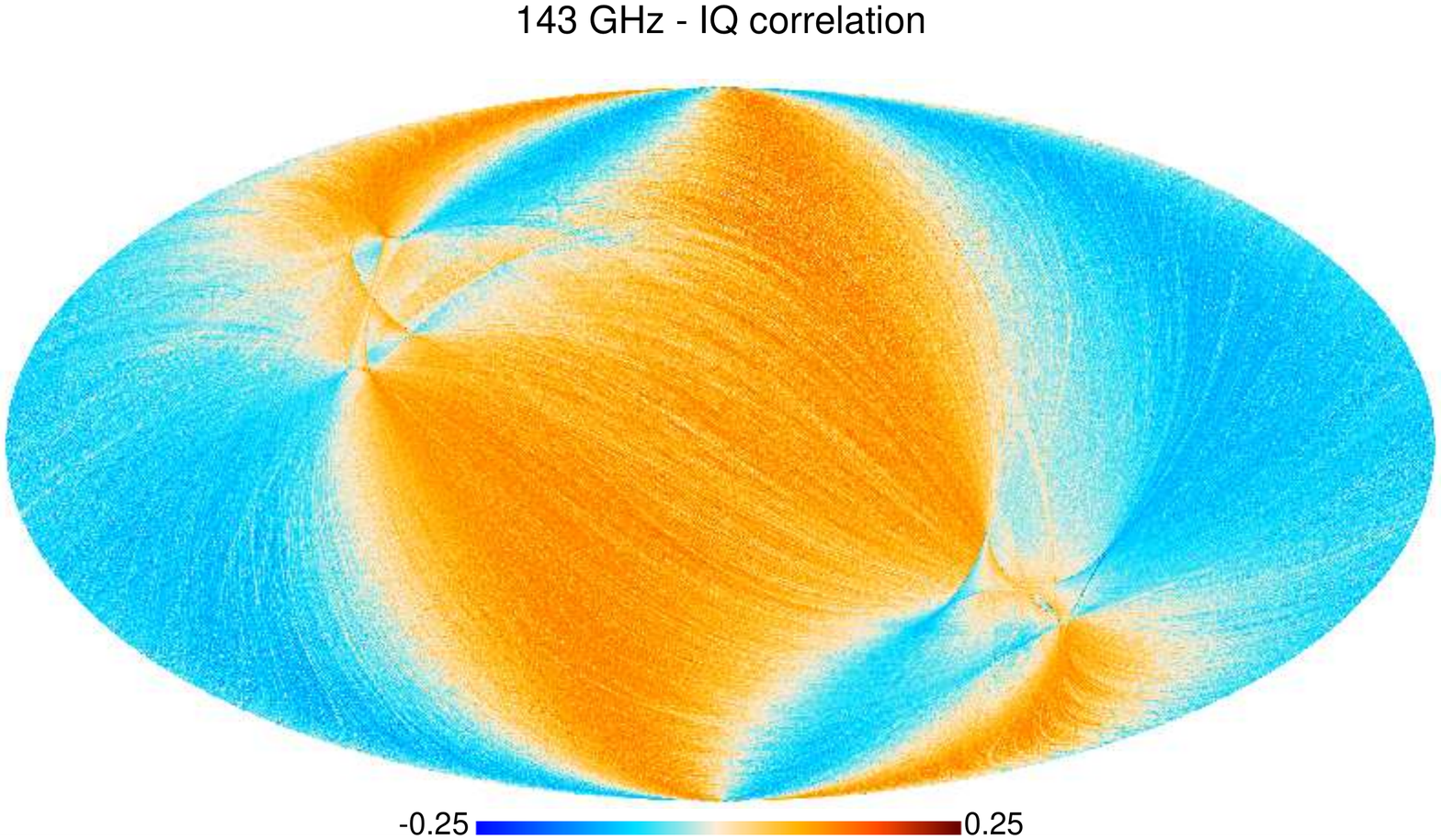} & \includegraphics[width=.32\textwidth,viewport=0 125 790 540,clip]{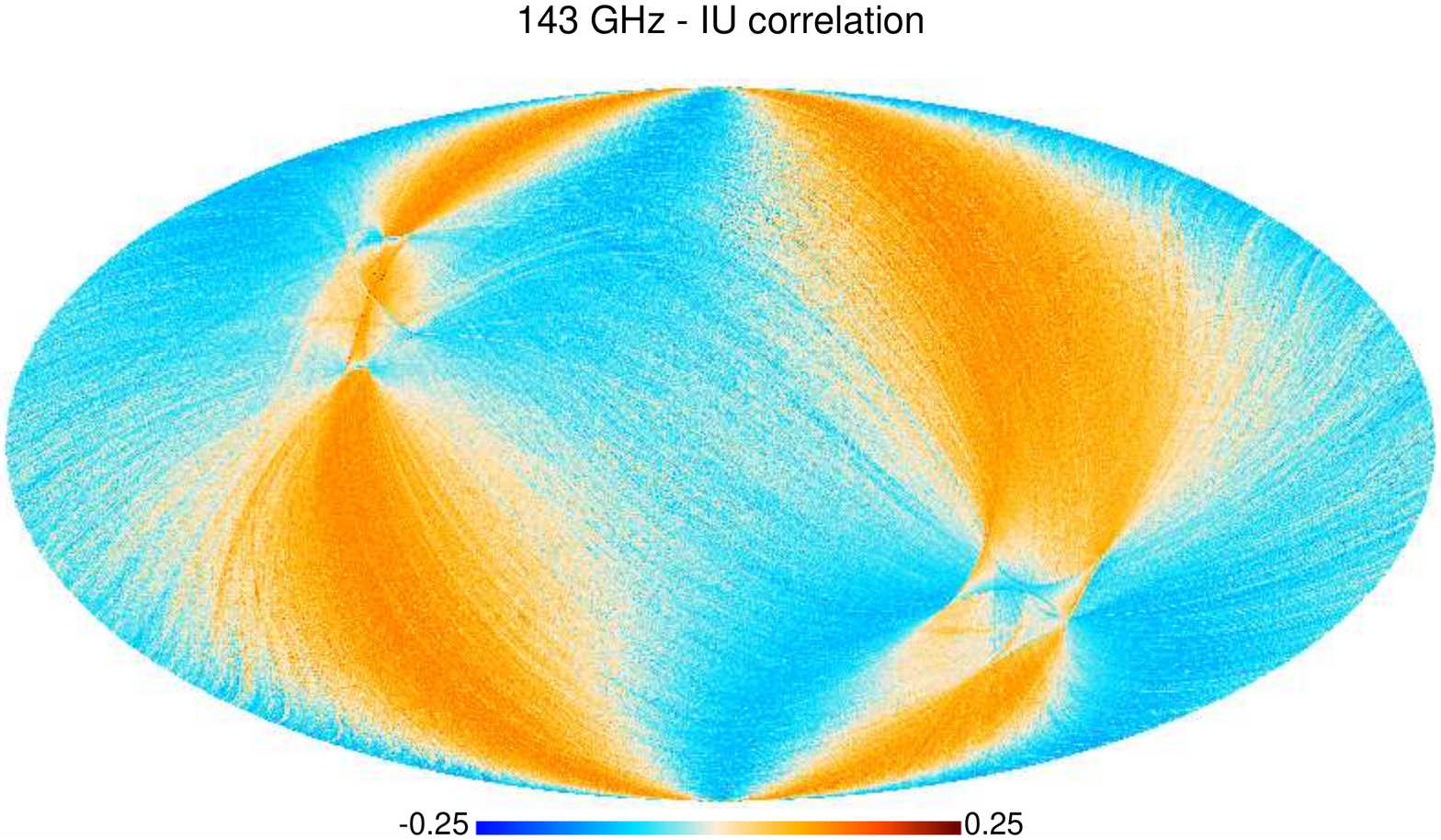} & \includegraphics[width=.32\textwidth,viewport=0 125 790 540,clip]{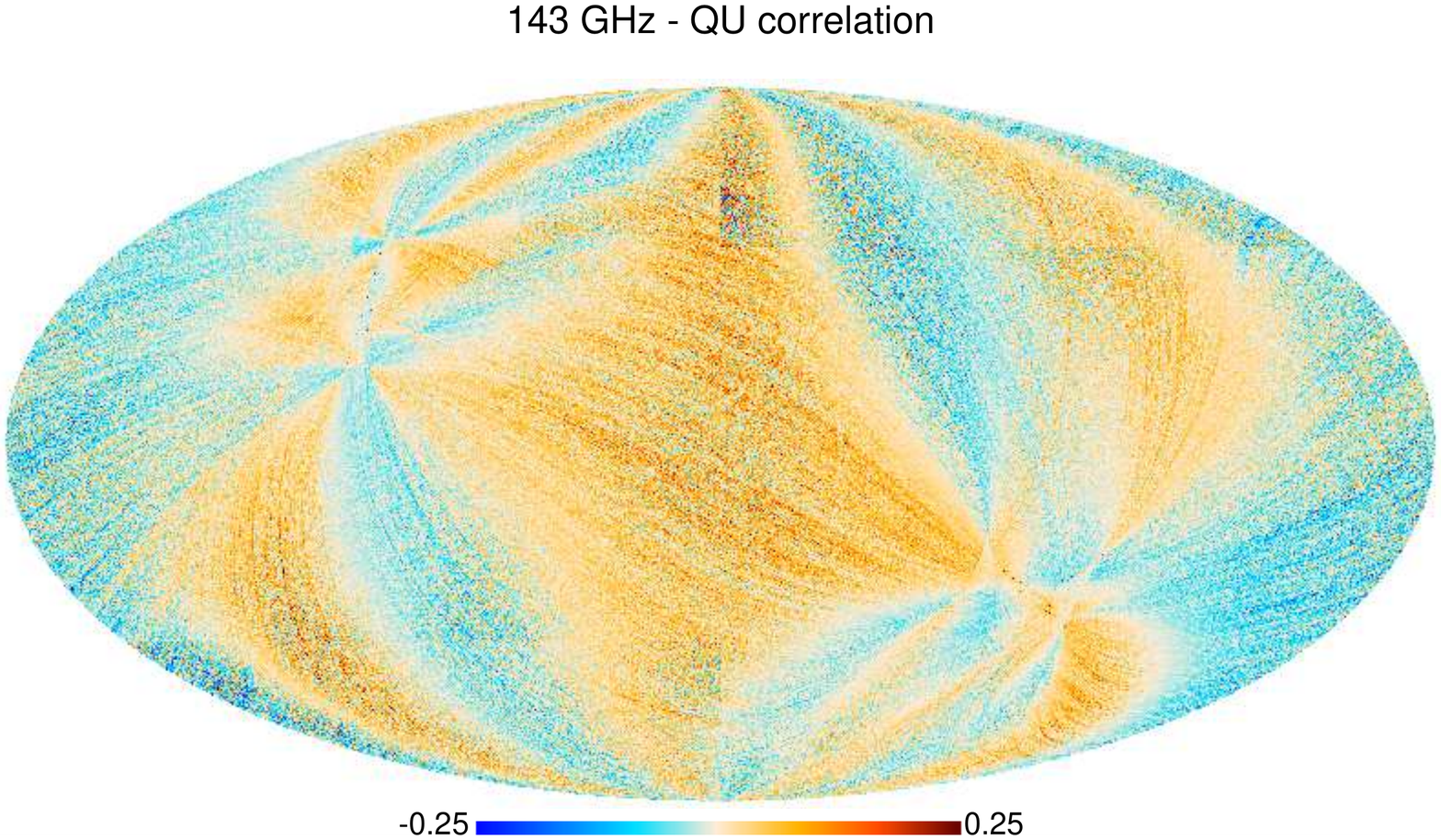} \\
          \includegraphics[width=.32\textwidth,viewport=0 125 790 540,clip]{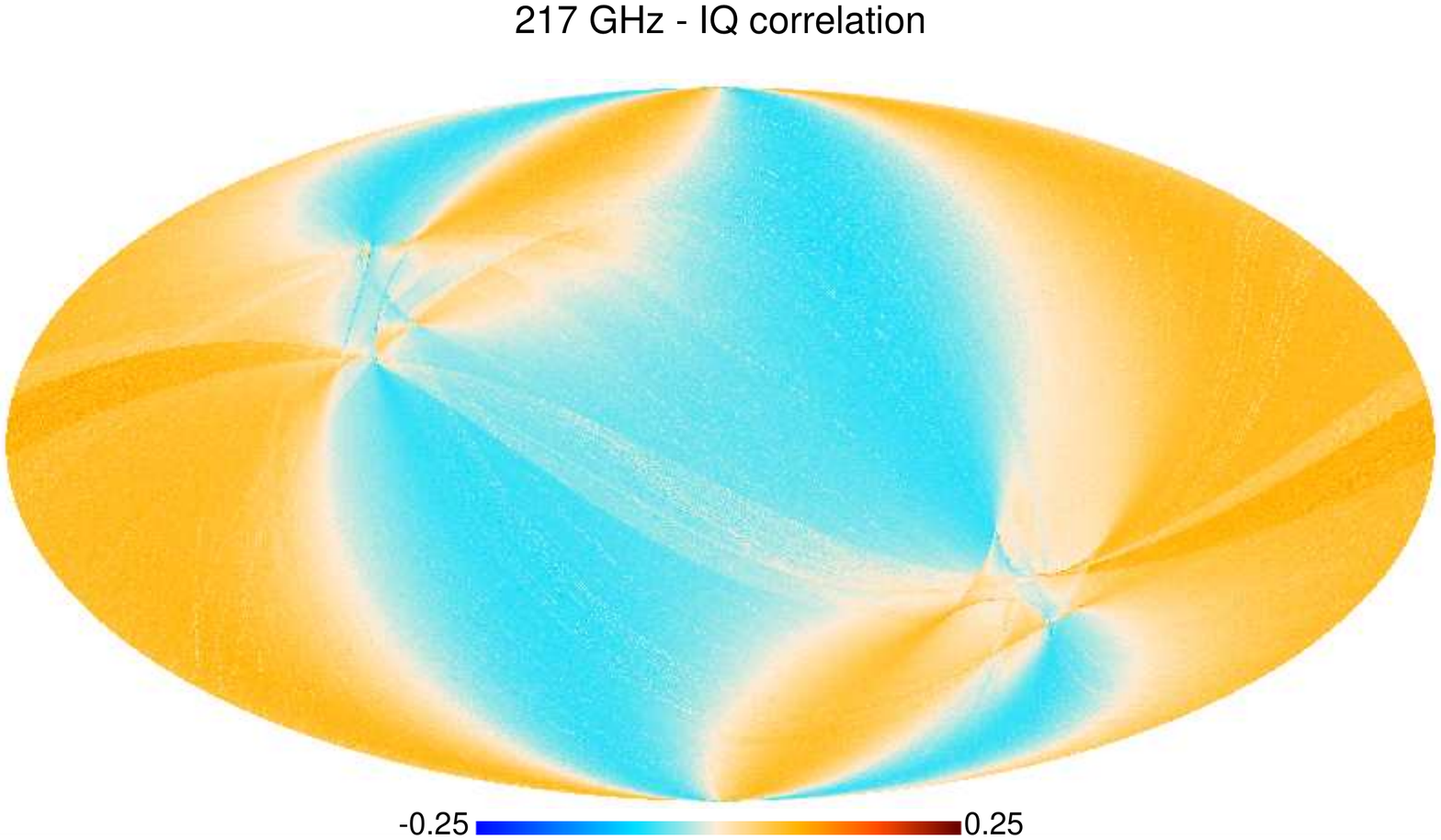} & \includegraphics[width=.32\textwidth,viewport=0 125 790 540,clip]{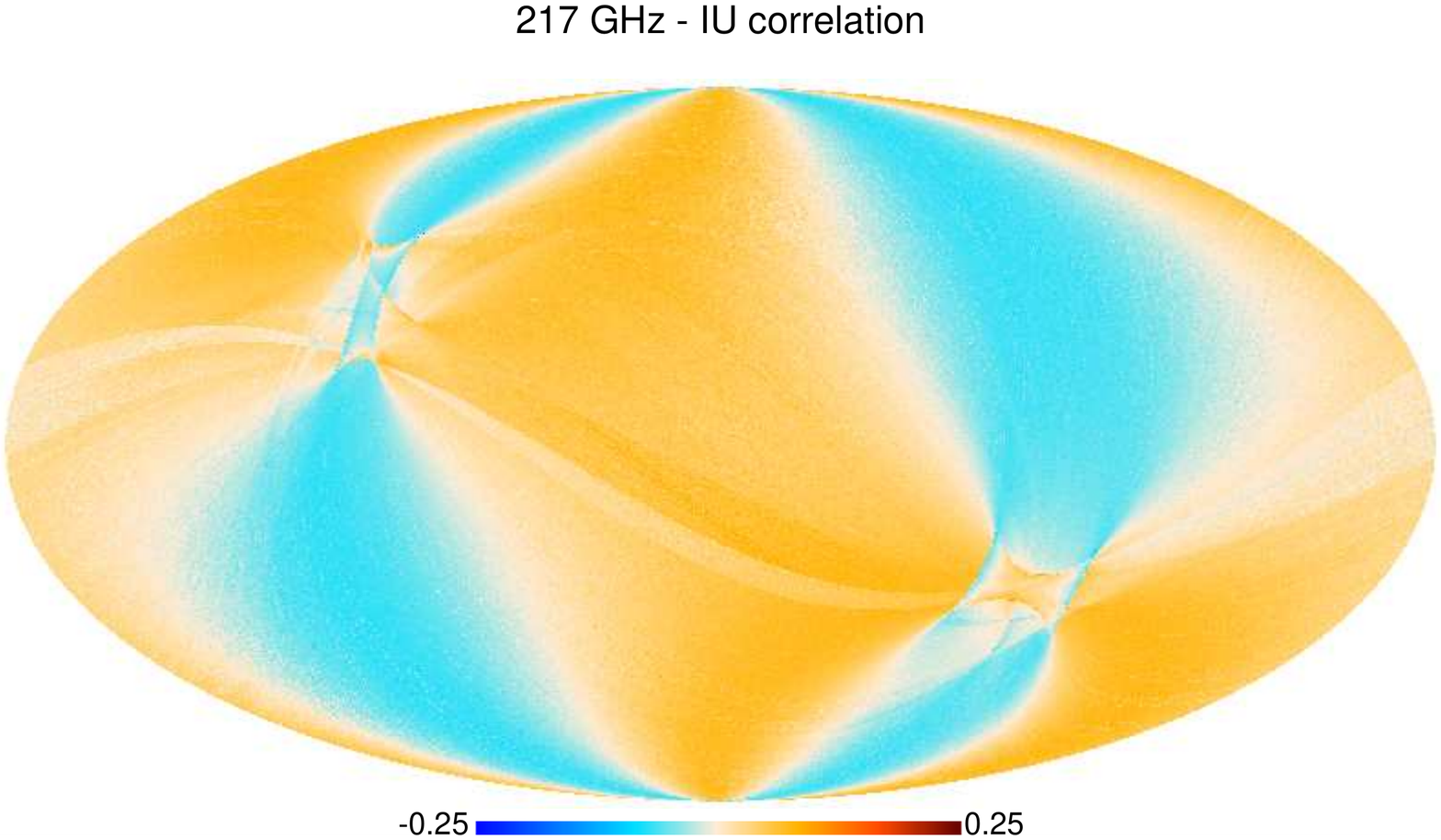} & \includegraphics[width=.32\textwidth,viewport=0 125 790 540,clip]{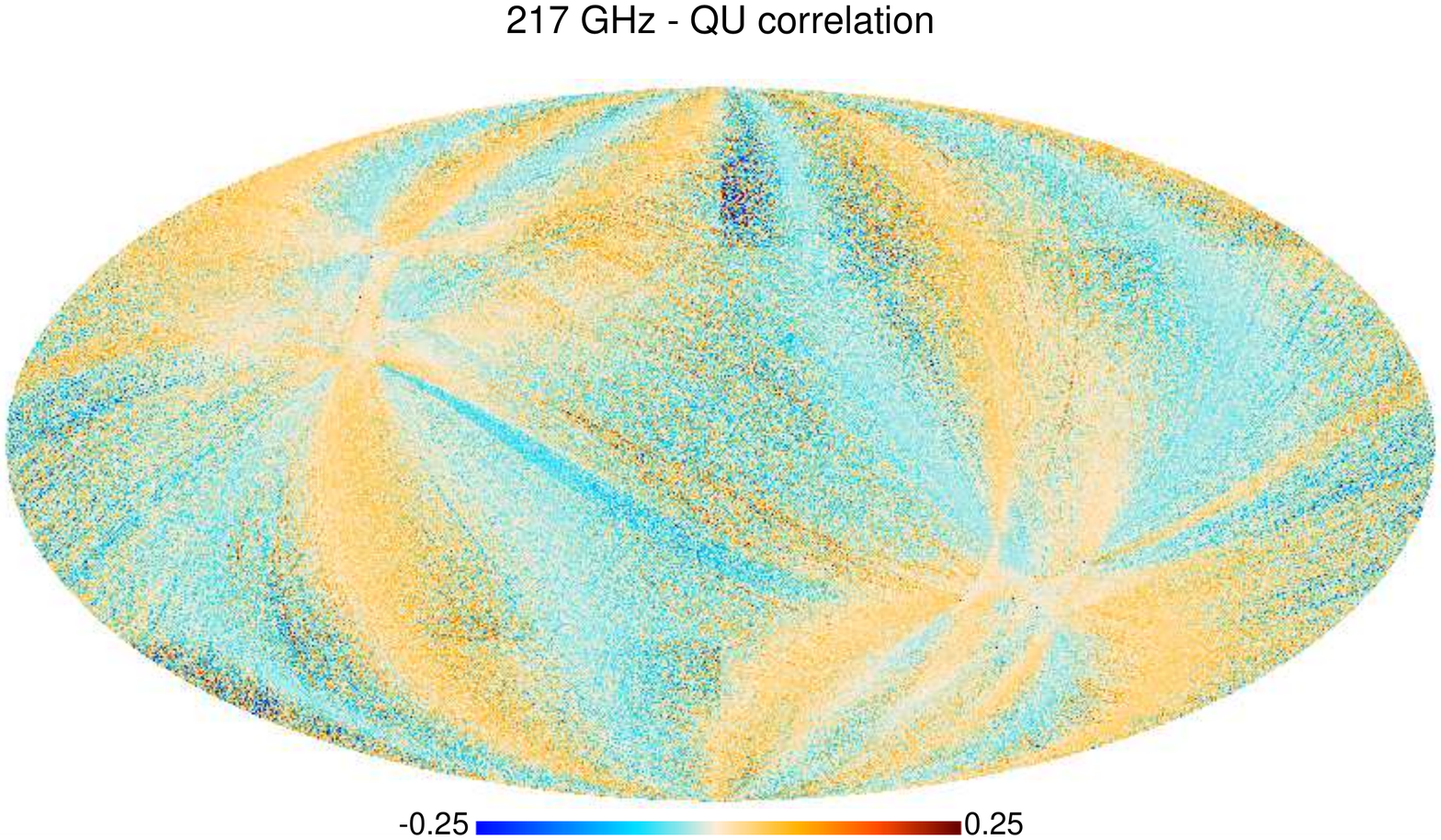} \\
          \includegraphics[width=.32\textwidth,viewport=0 125 790 540,clip]{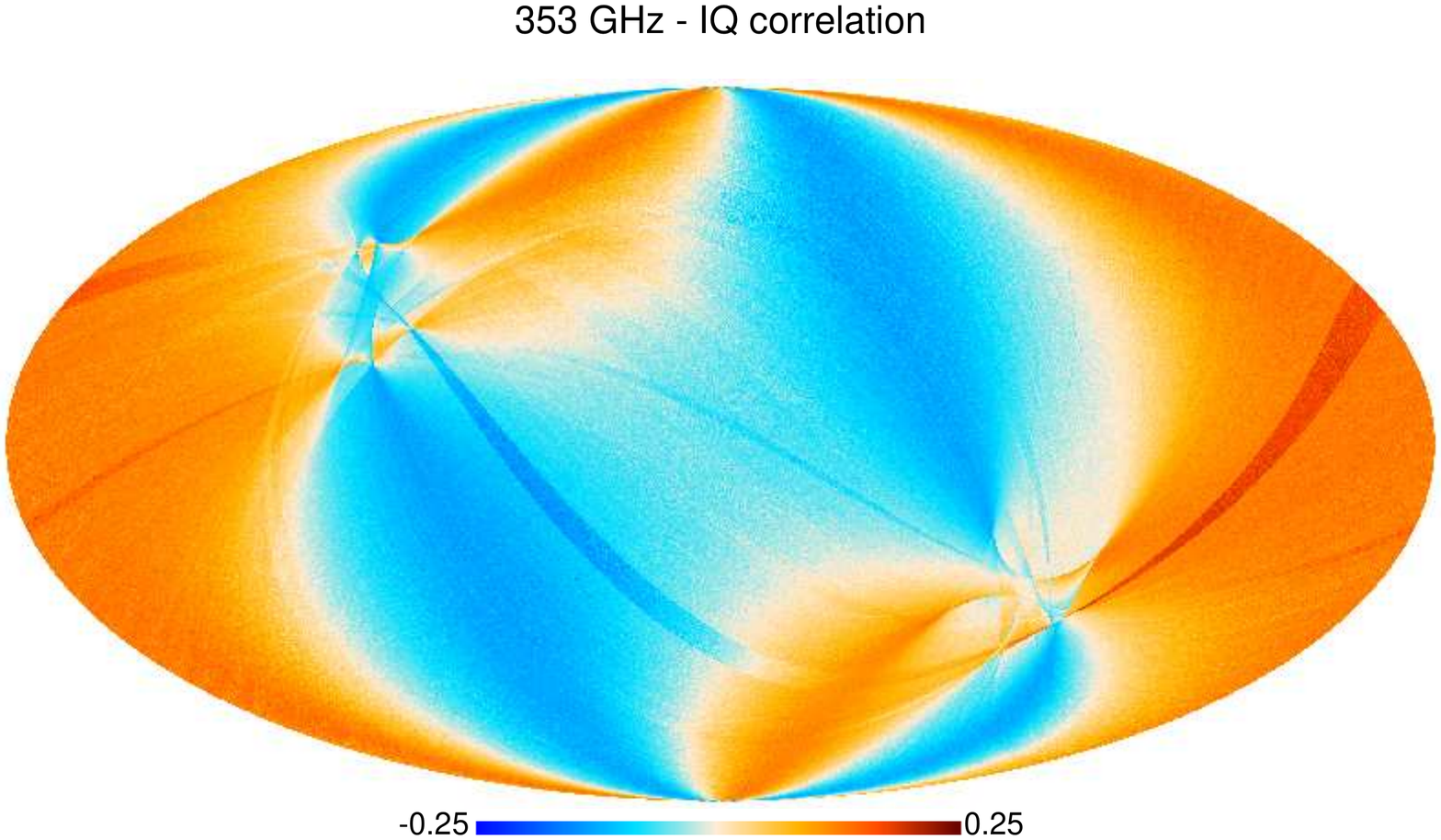} & \includegraphics[width=.32\textwidth,viewport=0 125 790 540,clip]{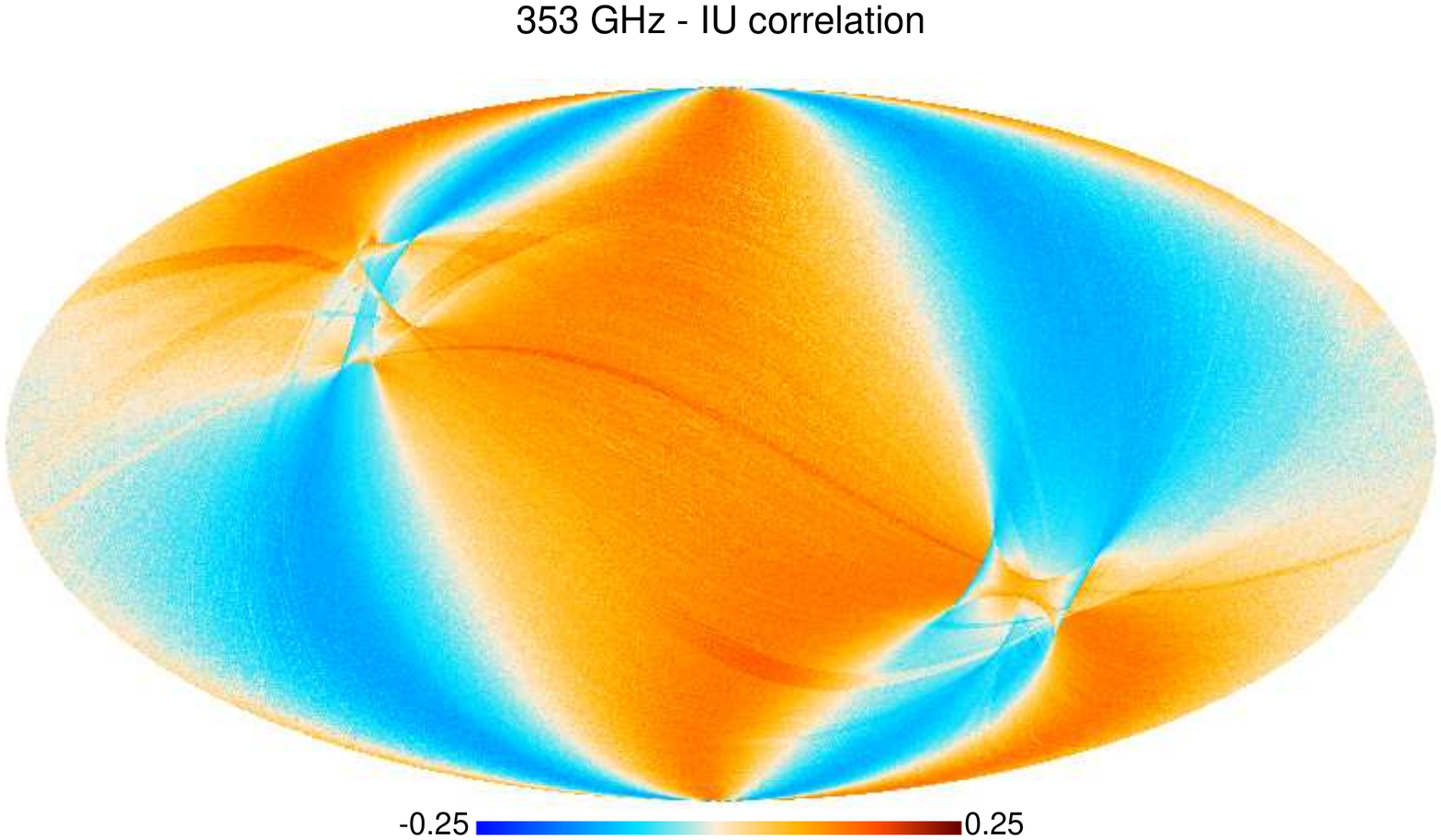} & \includegraphics[width=.32\textwidth,viewport=0 125 790 540,clip]{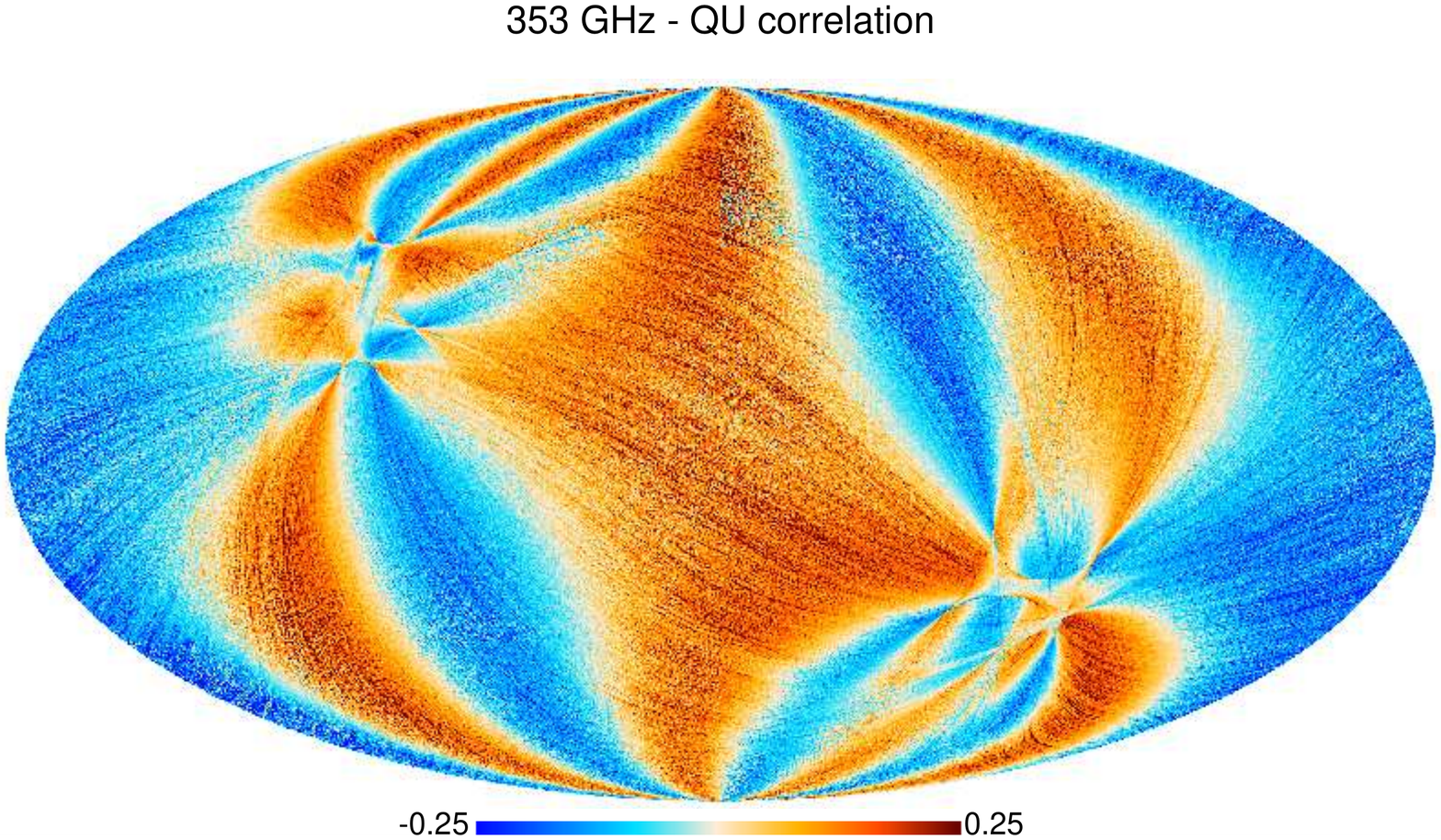} \\
        \end{tabular}
	\caption{Maps of the correlation $IQ$ ({\it left}), $IU$ ({\it middle}), $QU$ ({\it right}) between Stokes parameters for the 100, 143, 217 and 353\GHz\ ({\it from top to bottom}).}
	\label{fig:IQU_pix_correlation}
\end{figure*}

We present in Fig.~\ref{fig:IQU_pix_correlation} the distribution of the $I$, $Q$, and $U$ correlations in each pixel for the HFI frequencies where polarization is reconstructed.

\subsection{Leakage correction maps}
Section~\ref{sec:leakage} describes the origin and the formalism of the intensity-to-polarization leakage in HFI and how it is, in principle, possible to quantify and correct for these systematics effects. This is particularly important for any studies of the large angular scales in polarization, where dipole, monopole, and dust leakages are the main limiting factors. 

Although dust and CO bandpass leakage effects can be estimated from given intensity templates and electromagnetic spectra (Sect.~\ref{annex:bpm}), such is not the case for the calibration and monopole leakage levels, which have to be estimated directly from the maps (Sect.~\ref{annex:ggf}).

\subsubsection{Bandpass leakage correction maps}
\label{annex:bpm}
The small differences in the bandpasses of the different bolometers combined to produce polarization maps give rise to some leakage from intensity to polarization for the CO and the (Galactic) dust emission; this is estimated using the known band profiles and templates of the emission (see Sect.~\ref{sec:bpm}).

Bandpass leakage $Q$ and $U$ maps are provided for dust for all channels, using ground-based measurement of the bandpass integrated over the dust emission law. Each FITS file contains a single extension, with two columns containing the $Q$ and $U$ leakage maps in K$_\mathrm{CMB}$ units.

\subsubsection{The generalized global fit (GGF) approach for leakage correction}
\label{annex:ggf}
The generalized global fit (GGF) method is a template fitting approach that has been developed to consistently solve for calibration, monopole and bandpass mismatch (BPM) leakage effects, at the map level. Each $Q$ and $U$ map at 100, 143, 217, and 353\GHz\ is modelled as
\begin{eqnarray}
\label{eq:ggf}
\left[Q,U\right]^\nu &=& \left[Q,U\right]_{\mathrm{CMB}}+\left[Q,U\right]^\nu_{\mathrm{dust}}\\\nonumber
&+& \sum_{b\in\nu} \alpha^\nu_b\Gamma^b_{IQ,IU}+ \sum_{b\in\nu} \beta^\nu_b\Gamma^b_{IQ,IU} \times I^{\mathrm{dipole}}\\\nonumber
&+& \sum_{b\in\nu} \gamma^\nu_b\Gamma^b_{IQ,IU} \times I^{\mathrm{dust}}+ \sum_{b\in\nu} \delta^\nu_b\Gamma^b_{IQ,IU} \times I^{\mathrm{CO}}\;,
\end{eqnarray}
where:
\begin{itemize}
	\item the first line corresponds to the physical polarization signal coming from the CMB and the dust (synchrotron is assumed negligible at these frequencies);
	\item the second line corresponds to the monopole and calibration leakage terms, the templates of which make use of the leakage pattern maps $\Gamma$ described in Sect.~\ref{sec:leakage};
	\item the dust and CO BPM-induced leakage terms are gathered in the third line, again using the leakage patterns maps $\Gamma$;
	\item each summation is performed over the polarized bolometers of the frequency channel $\nu$.
\end{itemize}

We use the \Planck\ 353\GHz\ $Q$ and $U$ maps as polarized dust templates and perform the fitting procedure at $1\deg$ resolution and $N_{\mathrm{side}}=64$. Focussing on polarized dust and leakage effects only, Eq.~\eqref{eq:ggf} becomes
\begin{eqnarray}
	\label{eq:ggf2}
	\left[Q,U\right]^\nu &=& (1-\epsilon^\nu) \left[Q,U\right]_{CMB} + \epsilon^\nu \left[Q,U\right]^{353}\\\nonumber
	&+& \sum_{b\in\nu} \alpha^\nu_b\Gamma^b_{IQ,IU}+ \sum_{b\in\nu} \beta^\nu_b\Gamma^b_{IQ,IU} \times I^{\mathrm{dipole}}\\\nonumber
	&+& \sum_{b\in\nu} \gamma^\nu_b\Gamma^b_{IQ,IU} \times I^{\mathrm{dust}}+ \sum_{b\in\nu} \delta^\nu_b\Gamma^b_{IQ,IU} \times I^{\mathrm{CO}}\\\nonumber
	&-& \epsilon^\nu \times \left[\sum_{b\in 353}\alpha^{353}_b\Gamma^b_{IQ,IU}+ \sum_{b\in 353}\beta^{353}_b\Gamma^b_{IQ,IU} \times I^{\mathrm{dipole}}\right. \\\nonumber
	&+& \left. \sum_{b\in 353} \gamma^{353}_b\Gamma^b_{IQ,IU} \times I^{\mathrm{dust}}+ \sum_{b\in 353} \delta^{353}_b\Gamma^b_{IQ,IU} \times I^{\mathrm{CO}}\right]\;,
\end{eqnarray}
where $\epsilon^\nu$ is the overall factor scaling the dust from 353\GHz\ to the frequency channel $\nu$ and where $\alpha$, $\beta$, $\gamma$, and $\delta$ define the amplitude of the monopole, dipole, dust and CO leakage effects respectively. The last two lines of Eq.~\eqref{eq:ggf2} correct for the total leakage added when using $\left[Q,U\right]^{353}$ as dust templates.
This equation is at the core of the GGF method, which is then implemented as follows:
\begin{enumerate}
	\item A first fit is performed to solve for the coefficients of Eq.~\eqref{eq:ggf2}, for each channel ($\nu = 100$, $143$, and $217$\GHz) independently. There are strong degeneracies between the leakage templates at frequency $\nu$ and at 353\GHz, so that this first fit does not provide reliable $\alpha$, $\beta$, $\gamma$, and $\delta$ coefficients that can be used to compute leakage correction maps. However, it enables an accurate determination of the overall scaling factor of the dust $\epsilon^\nu$ between the channel under scrutiny and 353\GHz.
	\item These $\epsilon^\nu$ values are used as inputs to solve Eq.~\eqref{eq:ggf2}, simultaneously for all channels, in a consistent manner. At this stage, we also add some extra constraints (such as minimizing detector-set and survey differences) in order to lift further the leakage degeneracies between the four channels. This generalized global fit allows us to extract the set of $\alpha$, $\beta$, $\gamma$, and $\delta$ for all HFI polarized channels, including 353\GHz.
\end{enumerate}

\begin{figure}[htbp]
	\centering
	\includegraphics{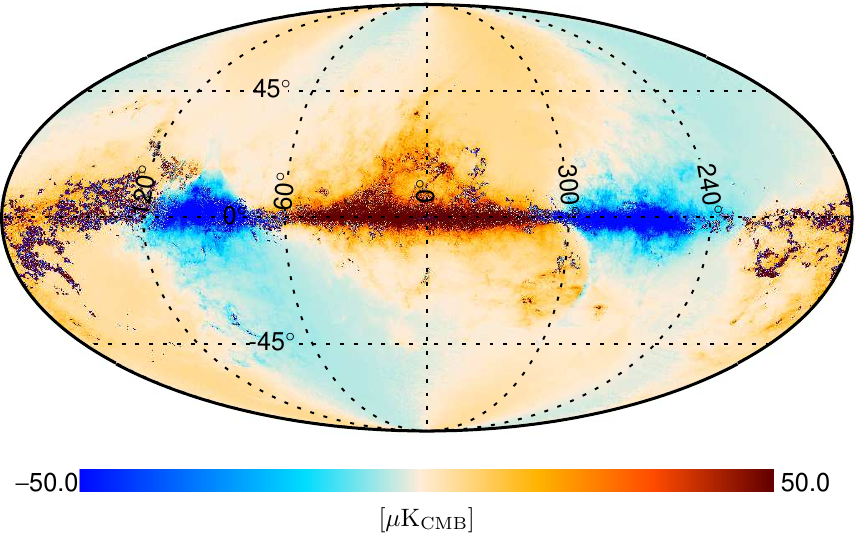} 
	\includegraphics{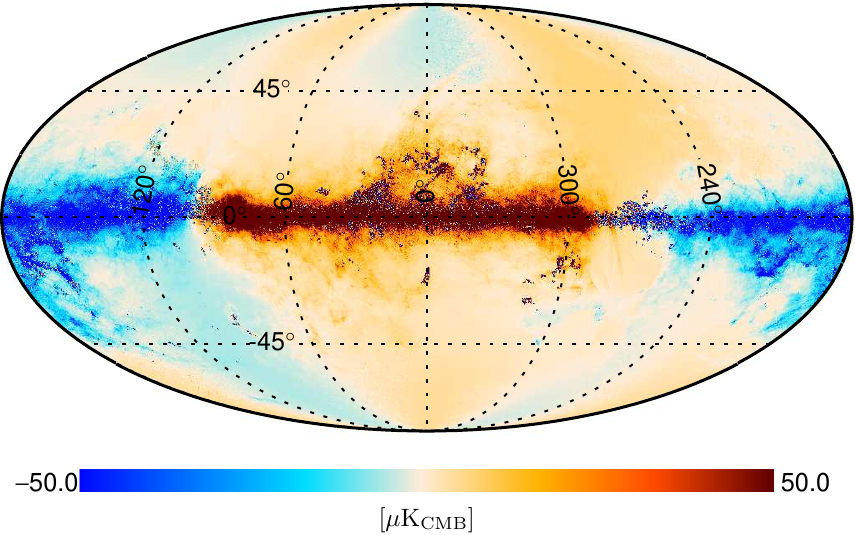}
	\caption{Total leakage correction maps in $Q$ ({\it top}) and $U$ ({\it bottom}) at 353\GHz\ computed with the generalized global fit (GGF) method. These include calibration, CO and dust leakage templates.}
	\label{fig:leak_ggf_maps_dust}
\end{figure}

Solving for survey differences in the second step is crucial. Although the leakage coefficients are independent of scan angle, the $\Gamma$ maps change because in different surveys pixels are scanned at different angles.
Total leakage maps, allowing a correction for monopole, calibration, and BPM leakage, are then computed using the coefficients extracted in the second step of the procedure. Figure~\ref{fig:leak_ggf_maps_dust} shows the total correction in $Q$ (top) and $U$ (bottom) for the 353\GHz\ channel. Dust and CO BPM leakage effects are clearly dominant near the Galactic plane, while the large patterns at high latitude are mainly due to the dipole leakage.

\end{document}